\documentclass[traditabstract]{aa} 
\usepackage[lofdepth,lotdepth]{subfig}
\usepackage{graphicx}
\usepackage{wrapfig}
\usepackage[varg]{txfonts}
\usepackage{longtable}
\usepackage{lscape}
\usepackage{natbib}
\usepackage{url}
\usepackage{color}
\usepackage{txfonts}
\usepackage{float}
\usepackage{color}
\usepackage{multirow}
\usepackage{caption}
\usepackage{appendix}
\usepackage{threeparttable}

\bibpunct{(}{)}{;}{a}{}{,} 

\newcommand\teff{T_{\rm eff}}

\usepackage[breaklinks,
        colorlinks=true,
        urlcolor=blue,          
        filecolor=blue,         
        citecolor=blue,         
        linkcolor=blue,         
        pdfpagemode=None,
        hypertexnames=false,
        bookmarksopen=true]{hyperref}

\begin{document}
   \title{The origin and evolution of r- and s-process elements\\ in the Milky Way stellar disk\thanks{This paper includes data gathered with the 6.5 meter Magellan Telescopes at the Las Campanas Observatory, Chile and the ESO 1.5-m, 2.2-m. and 3.6-m telescopes on La Silla, Chile (ESO Proposal ID 65.L-0019, 67.B-0108, 76.B-0416, 82.B-0610); and data from UVES Paranal Observatory Project (ESO DDT Program ID 266.D-5655).}\fnmsep\thanks{Tables 3 and 4 are only available at the CDS via anonymous ftp to \tt{cdsarc.u-strasbg.fr (130.79.128.5)} or via \tt{http://cdsarc.u-strasbg.fr/viz-bin/qcat?J/A+A/XXX/XXX}}}
   
     \titlerunning{r- and s- process elements in the Milky Way stellar disk}
     \authorrunning{Battistini, C. \& Bensby, T.}


   \author{Chiara Battistini \inst{1,2}
          \and
         Thomas Bensby \inst{1}}

   \institute{Lund Observatory, Department of Astronomy and Theoretical Physics, Box 43, SE-221\,00, Lund, Sweden
     \and
   current address: Zentrum f{\"u}r Astronomie der Universit{\"a}t Heidelberg, Landessternwarte, K{\"o}nigstuhl 12, 69117 Heidelberg, Germany
   \\
    \email{cbattistini@lsw.uni-heidelberg.de}}

   \date{Received 16 September 2015 / Accepted 2 November 2015}
 
 \abstract
 {Elements heavier than iron are produced through neutron-capture processes in the different phases of stellar evolution. Asymptotic giant branch (AGB) stars are believed to be mainly responsible for elements that form through the {\it slow} neutron-capture process, while the elements created in the {\it rapid} neutron-capture process have less understood production sites. Knowledge of abundance ratios as functions of metallicity can lead to insights on the origin and evolution of our Galaxy and its stellar populations.
  }
 {We aim to trace the chemical evolution of the neutron-capture elements Sr, Zr, La, Ce, Nd, Sm, and Eu in the Milky Way stellar disk. This will allow us to constrain the formation sites of these elements as well as to probe the evolution of the Galactic thin and thick disks.
  }
 {Using spectra of high resolution ($42\,000\lesssim R \lesssim 65\,000$) and high signal-to-noise ($S/N\gtrsim 200$) obtained with the MIKE and FEROS spectrographs, we determine Sr, Zr, La, Ce, Nd, Sm, and Eu abundances for a sample of 593 F and G dwarf stars in the Solar neighbourhood. The abundance analysis is based on spectral synthesis using one-dimensional, plane-parallel, local thermodynamic equilibrium (LTE) model stellar atmospheres calculated with the MARCS 2012 code. 
 }
 {We present abundance results for Sr (156 stars), Zr (311 stars), La (242 stars), Ce (365 stars), Nd (395 stars), Sm (280 stars) and Eu (378 stars). We find that Nd, Sm, and Eu show trends similar to what is observed for the $\alpha$-elements in the [$X$/Fe]-[Fe/H] abundance plane. For [Sr/Fe] and [Zr/Fe] we find decreasing abundance ratios for increasing metallicity, reaching sub-solar values at super-solar metallicities. [La/Fe] and [Ce/Fe] do not show any clear trend with metallicity, being close to solar values at all [Fe/H]. The trends of abundance ratios [$X$/Fe] as a function of stellar ages present different slopes before and after 8\,Gyr. 
 }
{The rapid neutron-capture process is active early in the Galaxy, mainly in type II supernovae from stars in the mass range  8-10\,M$_{\odot}$. Europium is almost completely produced by r-process but Nd and Sm show similar trends to Eu even if their s-process component is higher. Strontium and Zr are thought to be mainly produced by s-process, but show significant enrichment at low metallicity that requires extra r-process production, that probably is different from the classical r-process. Finally, La and Ce are mainly produced via s-process from AGB stars in mass range 2-4\,M$_{\odot}$, which can be seen by the decrease in [La/Eu] and [Ce/Eu] at $\mathrm{[Fe/H] \approx-0.5}$. The trend of [$X$/Fe] with age could be explained by considering that the decrease in [$X$/Fe] for the thick disk stars can be due to the decrease of type II supernovae with time meaning a reduced enrichment of r-process elements in the interstellar medium. In the thin disk the trends are flatter that probably is due to that the main production from s-process is balanced by Fe production from type Ia supernovae.
}
  


   \maketitle
%

\section{Introduction}
Elements with atomic numbers up to iron can be synthesised via nuclear fusion in the interiors of stars.  Heavier elements are formed by the addition of neutrons in the stellar interiors, and are called
neutron-capture elements. There are two main ways to add neutrons, through {\it slow} neutron-capture (s-process), or through {\it rapid} neutron-capture (r-process), depending on if the neutron-capture is slow or rapid compared to the timescale for $\beta^{-}$ decay  \citep{Burbidge1957}. The s-process requires a low neutron flux and the creation of the new elements moves along the valley of $\beta$ stability. In the case of the r-process, the neutron flux is intense, permitting the creation of elements outside the valley of stability.

An important step in the study of neutron-capture elements was the discovery that metal-poor stars show high relative abundances of certain neutron-capture elements compared to Fe, meaning that the r- and s-processes were already active at early times (e.g. \citealt{Frebel2013}). In particular for the r-process, this was by the significant number of stars with high levels of the r-process element Eu at very low metallicities (e.g. the neutron-capture review by \cite{Sneden2008}). However, the production sites for r-process elements are poorly understood and currently there are at least three possible scenarios. First, the classic scenario for r-process production is neutrino-induced winds from type II supernovae (SN\,II) \citep{Woosley1994}. Extremely energetic neutrinos are produced during the collapse of the SN\,II and they are potentially able to interact with the dense material that is falling onto the core of the star. This interaction can heat the material giving to it the additional energy necessary to recreate the energy output observed of $\approx 10^{51}$ ergs. Another scenario is the merging of neutron stars \citep{Freiburghaus1999}, or the merging of a neutron star with a black hole \citep{Surman2008}. \cite{Rosswog2014} shows that that dynamic ejecta from this merging produce r-process elements with A > 130 with a pattern independent from the kind of merging, while neutrino-driven winds are responsible for nucleosynthesis of elements from A = 50 to A = 130 but in this case the exact output depends on the merging parameters. Lastly, polar jets from SN\,II with the use of a pure magnetohydrodynamic explosion seems to lead to the right conditions to have r-process nucleosynthesis \citep{Nishimura2006}. Unfortunately there are still theoretical problems in the modelling of SN\,II explosions with neutrino wind that prevent a definitive confirmation of the latter production site. In addition to this, theoretical predictions for r-process production have not been entirely successful in synthesising the observed total abundance distribution of r-process nuclei \citep{Cescutti2006}.

The production of s-process elements, on the other hand, can occur both in massive stars, in the He-burning core, and in the convective C-burning shell \citep{Pignatari2010}, as well as in asymptotic giant branch (AGB) of lower mass stars at solar and lower metallicities \citep{Bisterzo2011}. Considering that  the s-process is responsible for the production of approximately half of the nuclides from Fe to Bi, in particular feeding the groups Sr-Y-Zr, Ba-La-Ce-Pr-Nd, and Pb, understanding how and where the s-process elements are produced is very important. However, our understanding is poor, especially in respect to the where, when, and how much of these elements is produced. Abundance surveys of metal-poor dwarf stars in the halo revealed the presence of very old stars that are extremely rich in s-process elements \citep{Sneden2008}. These high abundances are difficult to explain with stellar evolution and nucleosynthesis theories because dwarf stars on the main-sequence cannot produce these elements. However, if a dwarf star experienced accretion from a more massive giant companion the peculiar atmospheric abundance of the dwarf star can be explained. In fact, if the giant companion already experienced dredge-up, bringing s-process elements produced during its AGB phase to the surface, these elements can be transferred onto the dwarf companion \citep{Aoki2001}.

Most neutron-capture elements can be produced by both s- and r-processes, and it is not always easy to constrain which of the processes that are involved in the creation path(s) of each element. An important work on the production rates for neutron-capture elements was the one from \cite{Arlandini1999}. They claim that the s-process production is responsible for 85\,\% of Sr abundance, 83\,\% of Zr, 62\,\% of La, 77\,\% of Ce, 30\,\% of Sm, and only 7\,\% of Eu. Thanks to its very limited s-process component, Eu is considered a ``pure'' r-process element and is well suited for determining the corresponding r-process contribution for other elements \citep{Winckler2006}. For Nd, the production is equally divided between the s- and r-processes (56\,\% produced by s-process).  Since \cite{Arlandini1999} several works have derived new production rates and yields for the neutron-capture elements, including better cross-section measurements for the reactions and evolutionary model for stars of different metallicities and masses. However, large uncertainties are still present in the s- and r-process calculations \citep{Karakas2014}. At the same time, observations usually show spread in abundances at a given metallicity and this can be due to the dependence of neutron-capture process on metallicity \citep{Travaglio2004}.

\begin{table}
\caption{Analysed elements and spectral lines.\tablefootmark{$^\dagger$} 
\label{tab:list_of_elements}}
\setlength{\tabcolsep}{2.7mm}
\tiny
\centering
\begin{tabular}{lcc|l|l}
\hline
\hline
\noalign{\smallskip}
Element & $\mathrm{\lambda}$ & EP & Isotopic ratio & Reference \\
& $[\AA]$ & [eV] & & \\
\noalign{\smallskip}
\hline
\noalign{\smallskip}
\ion{Sr}{i} & 4607 & 0.00 & $^{84}$Sr 0.56\,\% & \cite{Bautista2002}\\
& & & $^{86}$Sr 9.86\,\%  & \\
& & & $^{87}$Sr 7.00\,\%  & \\
& & & $^{88}$Sr 82.58\,\% & \\
\noalign{\smallskip}
\hline
\noalign{\smallskip}
\ion{Zr}{ii} & 4208 & 0.71 & $^{90}$Zr 51.45\,\% & \cite{Ljung2006}\\
\ion{Zr}{i}  & 4687 & 0.73 & $^{91}$Zr 11.22\,\% & \\
\ion{Zr}{i}  & 4739 & 0.65 & $^{92}$Zr 17.15\,\% & \\
\ion{Zr}{ii} & 5112 & 1.66 & $^{94}$Zr 17.38\,\% & \\
& & & $^{96}$Zr  2.38\% &  \\
\noalign{\smallskip}
\hline
\noalign{\smallskip}
\ion{La}{ii} & 4662 & 0.00 & $^{138}$La 0.09\,\% & \cite{Ivans2006}\\
\ion{La}{ii} & 4748 & 0.93 & $^{139}$La 99.91\,\% & \\
\ion{La}{ii} & 5122 & 0.32 & & \\
\ion{La}{ii} & 6390 & 0.32 & & \\
\noalign{\smallskip}
\hline
\noalign{\smallskip}
\ion{Ce}{ii} & 4523 & 0.68 & $^{136}$Ce 0.19\,\% & \cite{Lawler2009} \\
\ion{Ce}{ii} & 4572 & 0.68 & $^{138}$Ce 0.25\,\% & \\
\ion{Ce}{ii} & 4628 & 0.52 & $^{140}$Ce 88.45\,\% & \\
\ion{Ce}{ii} & 5187 & 1.21 & $^{142}$Ce 11.11\,\% &  \\
\noalign{\smallskip}
\hline
\noalign{\smallskip}
\ion{Nd}{ii} & 4177 & 0.06 & $^{142}$Nd 27.20\,\% & \cite{Roederer2008}\\
\ion{Nd}{ii} & 4358 & 0.32 & $^{143}$Nd 12.20\,\% &\\
\ion{Nd}{ii} & 4446 & 0.20 & $^{144}$Nd 23.8\,\% &  \\
\ion{Nd}{ii} & 5130 & 1.30 & $^{145}$Nd 8.30\,\% & \\
\ion{Nd}{ii} & 5319 & 0.55 & $^{146}$Nd 17.20\,\% & \\
 & & & $^{148}$Nd 5.70\,\% & \\
 & & & $^{150}$Nd 5.60\,\% & \\
\noalign{\smallskip}
\hline
\noalign{\smallskip}
\ion{Sm}{ii} & 4467 & 0.66 & $^{144}$Sm 3.07\,\% &  \cite{Lawler2006}\\
\ion{Sm}{ii} & 4523 & 0.43 & $^{147}$Sm 14.99\,\% &  \cite{Roederer2008}\\
\ion{Sm}{ii} & 4577 & 0.25 & $^{148}$Sm 11.24\,\% & \\
\ion{Sm}{ii} & 4669 & 0.28 & $^{149}$Sm 13.82\,\% & \\
& & & $^{150}$Sm 7.38\,\% & \\
 & & & $^{152}$Sm 26.75\,\% & \\
& & & $^{154}$Sm 22.75\,\% &  \\
\noalign{\smallskip}
\hline
\noalign{\smallskip}
\ion{Eu}{ii} & 4129 & 0.00  & $^{151}$Eu 47.80\,\% & \cite{Lawler2001} \\
& &  & $^{153}$Eu 52.20\% &\\
\noalign{\smallskip}
\hline
\end{tabular}
\tablefoot{\tablefoottext{$\dagger$}{Column 1 shows the name of the element and the ionisation stage. The wavelength and the energy potential for the different lines are listed in columns 2 and 3, respectively. Column 4 shows the  isotopic ratio while column 5 give the references for the different elements. 
}}
\end{table}

The relative contribution from each of the processes to the abundance of an element and to the various abundance ratios change during the evolution of the Galaxy. This means that a complete overview of the stellar populations at different metallicities is needed for a better understanding of the evolution of these elements. Considering the possible production sites listed above, and that the first low-mass stars in the Universe reached their AGB phase about 500 millions of years after the Big Bang (see for example \citealt{Sneden2008}), the s-process enrichment occurs with some delay with respect to SN\,II, that start to explode after few millions years after the onset of star formation. The study of this delay and how the different elemental abundances relate to each other at different metallicities can help us to understand the evolution of the Galactic disk. For this purpose, dwarf stars are very suitable because they have very long lifetimes, comparable to the age of the Universe \citep{Sackmann1993} and their surface abundance can be considered to be representative of the chemical composition of the gas cloud they were born from \citep{Lambert1989,Freeman2002}. This means that the information that can be derived from these types of stars are directly related to the enrichment processes happened in the previous stellar generations, all the way back to the first billion years of the Milky Way history. 

In this paper we derive abundances of some neutron-capture elements (Sr, Zr, La, Ce, Nd, Sm, and Eu) in the Solar neighbourhood and in addition to this, in our investigation, we made use of the abundances of Ba derived by \cite{Bensby2014}. Even if the method used is not the same (for Ba was used equivalent width measurements), one of the scope of this paper is to analyse elements that were not studied in \cite{Bensby2014}.

The paper is organised as follow: In Sect.~\ref{sect:abundance_analysis} the stellar sample and the abundance analysis are described. Section~\ref{sect:results} give the abundance results for Sr, Zr, La, Ce, Nd, Sm, and Eu. Section~\ref{sect:evoution_s_r} discusses possible origins of these neutron-capture elements and their evolution in the Milky Way. Finally, Sect.~\ref{sect:summary} summarises our findings.

\section{Abundance analysis \label{sect:abundance_analysis}}

\subsection{Stellar sample and stellar parameters}

The stars in this study are a subset of the 714 F and G dwarf star sample of \cite{Bensby2014}, namely those 593 stars observed with the FEROS and MIKE high-resolution spectrographs that have complete wavelength coverage from about 3500 to above 9000\,{\AA}. This is important as many of the spectral lines that we utilise are located in the blue spectral region. The spectra have spectral resolutions between $R = 48\,000$ to $65\,000$ and the signal-to-noise ratios are generally greater than $S/N>200$. Further details are given in \cite{Bensby2014}.

Stellar parameters, ages, and elemental abundances for O, Na, Mg, Al, Si, Ca, Ti, Cr, Fe, Ni, Zn, Y, and Ba were determined for all 714 stars in \cite{Bensby2014} where the reader is directed for full details. Briefly, the determination of stellar parameters was performed by requiring excitation balance of abundances from \ion{Fe}{i} lines for the effective temperature ($\teff$), ionisation balance between  \ion{Fe}{i} and  \ion{Fe}{ii} lines for surface gravity ($\log g$), and that abundances from \ion{Fe}{i} lines are independent of reduced line strength to get microturbulance parameter ($\mathrm{\xi_{t}}$). Stellar ages were determined using a grid of $\alpha$-enhanced Yonsei-Yale isochrones by \cite{Demarque2004} using probability distribution functions as explained in \cite{Bensby2014}. In short, the age probability distribution of each star is constructed considering the errors in effective temperature, surface gravity and metallicity of that specific star, that permit to derive a most likely age for the stars (and these ages are the ones used in the paper) together with a lower and higher estimation (used to calculate the errors on the ages).
In addition, Sc, V, Mn, and Co abundances were determined in \cite{Battistini2015} for part of the sample.

\begin{table*}
\tiny
\centering
\caption{Elemental abundances for individual lines in the solar spectrum.\tablefootmark{$^\dagger$}\label{tab:solar_abundance}}
\setlength{\tabcolsep}{3mm}
\begin{tabular}{c c c c c c c c | c}
\hline
\hline
\noalign{\smallskip}
Line & FEROS & MIKE 1 & MIKE 2 & MIKE 3 & MIKE 4 & <MIKE> & $\sigma$ <MIKE> & Solar abundance\\
 & $R=48\,000$ & $R = 65\,000$ & $R = 65\,000$ & $R = 65\,000$ & $R = 42\,000$ & & & from  \cite{Asplund2009}\\
\noalign{\smallskip}
 \hline
\noalign{\smallskip}
 Sr 4607 & 2.92 & 2.76 & 2.62 & 2.78 & 2.61 & 2.69 & 0.09 & 2.87 $\pm$ 0.07\\
\noalign{\smallskip}
 \hline
\noalign{\smallskip}
 Zr 4208 & 2.39 & 2.42 & 2.44 & 2.42 & 2.35 & 2.44 & 0.06 & 2.58 $\pm$ 0.04\\
 Zr 4687 & 2.43 & 2.55 & 2.48 & 2.52 & 2.43 & 2.50 & 0.05 &\\
 Zr 4739 & 2.39 & 2.57 & 2.47 & 2.57 & 2.46 & 2.52 & 0.06 &\\
 Zr 5112 & 2.57 & 2.54 & 2.48 & 2.51 & 2.45 & 2.50 & 0.04 & \\
\noalign{\smallskip}
 \hline
\noalign{\smallskip}
 La 4662 & 1.31 & 1.21 & 1.29 & 1.36 & 1.38 & 1.31 & 0.08  & 1.10 $\pm$ 0.04\\
 La 4748 & 1.22 & 1.28 & 1.38 & 1.22 & 1.19 & 1.27 & 0.08 & \\
 La 5122 & 1.10 & 1.11 & 1.18 & 1.14 & 1.21 & 1.16 & 0.05 & \\
 La 6390 & 1.10 & 1.23 & 1.23 & 1.12 & 1.15 & 1.18 & 0.06 &\\
\noalign{\smallskip}
 \hline
\noalign{\smallskip}
 Ce 4523 & 1.55 & 1.65 & 1.70 & 1.70 & 1.62 & 1.67 & 0.07 & 1.58 $\pm$ 0.04\\
 Ce 4572 & 1.72 & 1.70 & 1.68 & 1.80 & 1.65 & 1.71 & 0.07 &\\
 Ce 4628 & 1.70 & 1.66 & 1.72 & 1.72 & 1.60 & 1.68 & 0.06 &\\
 Ce 5187 & 1.48 & 1.45 & 1.45 & 1.36 & 1.38 & 1.41 & 0.05 &\\
\noalign{\smallskip}
 \hline
\noalign{\smallskip}
 Nd 4177 & 1.33 & 1.45 & 1.49 & 1.45 & 1.41 & 1.45 & 0.03 & 1.42 $\pm$ 0.04\\
 Nd 4358 & 1.62 & 1.76 & 1.64 & 1.66 & 1.62 & 1.67 & 0.06 &\\
 Nd 4446 & 1.30 & 1.38 & 1.43 & 1.38 & 1.36 & 1.39 & 0.03 &\\
 Nd 5130 & 1.44 & 1.61 & 1.61 & 1.57 & 1.49 & 1.57 & 0.06 &\\
 Nd 5319 & 1.28 & 1.25 & 1.25 & 1.31 & 1.33 & 1.29 & 0.04 &\\
\noalign{\smallskip}
 \hline
\noalign{\smallskip}
 Sm 4467 & 0.83 & 0.85 & 0.93 & 0.92 & 0.84 & 0.89 & 0.05 & 0.96 $\pm$ 0.04\\
 Sm 4523 & 0.84 & 0.93 & 1.00 & 1.00 & 0.99 & 0.98 & 0.04 & \\
 Sm 4577 & 0.91 & 1.04 & 1.00 & 0.93 & 0.95 & 0.98 & 0.05 &\\
 Sm 4669 & 1.25 & 1.25 & 1.22 & 1.33 & 1.40 & 1.30 & 0.08 &\\
\noalign{\smallskip}
 \hline
\noalign{\smallskip}
 Eu 4129 & 0.34 & 0.34 & 0.48 & 0.41 & 0.33 & 0.39 & 0.07 & 0.52 $\pm$ 0.04\\
\noalign{\smallskip}
 \hline
\end{tabular}
\tablefoot{\tablefoottext{$\dagger$}{Columns 2-6 list the single abundances for the different instruments and resolutions. Columns 7 and 8 show the median abundance values for MIKE spectra for the different observation runs and the one $\sigma$ dispersion, respectively. The mean abundances are used as solar abundance for MIKE spectra observed in the run with $R = 55\,000$ (see Table 1 in \citealt{Bensby2014}) since no solar spectra is available in this setting. In col.~9 are listed the solar abundances values from \cite{Asplund2009}.}}
\end{table*}

\begin{table*}
\tiny
\centering
\caption{Abundance from single lines.\label{tab:abundances_single_lines}\tablefootmark{$^\dagger$}}
\setlength{\tabcolsep}{1mm}
\begin{tabular}{c | c | c c c c | c c c c | c c c c | c c c c c | c c c c | c   }
\hline
\hline
\noalign{\smallskip}
\multirow{2}{*} {HIP}& \multicolumn{1}{c|} {$\mathrm{\epsilon(Sr)}$} & \multicolumn{4}{c|}{$\mathrm{\epsilon(Zr)}$} & \multicolumn{4}{c|}{$\mathrm{\epsilon(La})$} & \multicolumn{4}{c|}{$\mathrm{\epsilon(Ce)}$} & \multicolumn{5}{c|}{$\mathrm{\epsilon(Nd)}$} & \multicolumn{4}{c|}{$\mathrm{\epsilon(Sm)}$} & \multicolumn{1}{c}{$\mathrm{\epsilon(Eu)}$}\\
\cline{2-24}
\noalign{\smallskip}
 & 4607 & 4208 & 4687 & 4739 & 5112 & 4662 & 4748 & 5122 & 6390 & 4523 & 4572 & 4628 & 5187 & 4177 & 4358 & 4446 & 5130 & 5319 & 4467 & 4523 & 4577 & 4669 & 4129 \\
\noalign{\smallskip}
\hline
\noalign{\smallskip}
80 & 2.06 & 1.95 & $-$ & $-$ &  1.97 & $-$ & $-$ & $-$ & $-$ & $-$ & 1.00 & 0.96 & $-$ & $-$ & $-$ & $-$ & 1.05 & 0.80 & 0.53 & $-$ & $-$ & $-$ & 0.12 \\
305 & $-$ & 2.51 & $-$ & $-$ & $-$ & 1.51 & $-$ & $-$ & $-$ & 1.75 & 1.95 & $-$ & 1.65 & $-$ & 1.72 & $-$ & $-$ & 1.38 & $-$ & $-$ & $-$ & $-$ & 0.52 \\
407 & $-$ & 2.56 & 2.75 & $-$ & $-$ & $-$ & 1.32 & $-$ & 1.26 & 1.71 & $-$ & 1.78 & 1.63 & $-$ & 1.71 & $-$ & 1.68 & 1.35 & $-$ & 1.07 & $-$ & $-$ & 0.47 \\
\vdots & \vdots & \vdots & \vdots & \vdots & \vdots & \vdots & \vdots & \vdots & \vdots & \vdots & \vdots & \vdots & \vdots & \vdots & \vdots & \vdots & \vdots & \vdots & \vdots & \vdots & \vdots & \vdots & \vdots\\
\noalign{\smallskip}
\hline
\end{tabular}
\tablefoot{\tablefoottext{$\dagger$}{Abundances from the single line analyzed in this work. The table is only available in electronic form at the CDS.}}
\end{table*}

\begin{table*}
\tiny
\caption{Stellar parameters and abundance results for the entire sample.\label{tab:abundances}\tablefootmark{$^\dagger$}}
\centering
\begin{tabular}{c c c c c c c c c c c c}
\hline
\hline
\\
HIP & $\teff$ & $\log g$ & $\mathrm{[Fe/H]}$ & $\mathrm{\xi_{t}}$ & [Sr/H] & [Zr/H] & [La/H] & [Ce/H] & [Nd/H] & [Sm/H] & [Eu/H] \\
\\
\hline
\\
80 & 5856 & 4.1 & $-0.59$ & $1.12$ & $-0.72$ & $-0.53$ & $-$ & $-0.73$ & $-0.52$ & $-0.39$ & $-0.29$\\
\vdots & \vdots & \vdots & \vdots & \vdots & \vdots & \vdots & \vdots & \vdots & \vdots & \vdots & \vdots\\
\\
\hline
\end{tabular}
\tablefoot{\tablefoottext{$\dagger$}{The table is only available in electronic form at the CDS}\label{tab:abundances_H}}
\end{table*}

\subsection{Spectral line synthesis}

The spectral lines from the heavy neutron-capture elements are usually located in the blue regions of the visible spectrum. The blue region is very crowded with spectral lines, and especially so for metal-rich disk stars, meaning that blends from other species are frequent. The crowdedness of the blue spectral regions also make the placement of the continuum difficult. In addition, heavy elements usually consist of several isotopes and their spectral lines can be affected by hyperfine splitting. Therefore, to determine abundances from these lines one has to use spectral line synthesis in order to accurately account for wavelength shifts of the different isotopes, hyperfine components, and other blending features. 

Table~\ref{tab:list_of_elements} lists the spectral lines that we use in this study. It shows the analysed lines with the isotopic ratios of the elements, ionisation stages, wavelengths, and excitation potential together with reference work where these details were gathered. For Sm and Nd, that suffer in large part by isotopic shifts and also hfs, we use data from \cite{Roederer2008}, for La we used hfs only components from \cite{Ivans2006} since La has only one naturally-occurring isotope, and for Eu we used shifts correction for isotopic and hfs substructures from \cite{Lawler2001}. 
However, some of the lines of the elements affected by hfs are anyway treated as single lines because the wavelength splitting is small enough to be ignored. In our linelists the lines treated as single lines are \ion{La}{ii} at 4748\,{\AA}, \ion{Nd}{ii} at 5130\,{\AA} and at 5319\,{\AA}, and \ion{Sm}{ii} at 4523\,{\AA}, 4577\,{\AA}, and 4669\,{\AA}. All lines from \ion{Sr}{ii}, \ion{Zr}{ii} and \ion{Ce}{ii} lines are treated as single lines since there is no information in literature on their hyperfine structure and isotopic shifts or because they do not suffer of these problems. Atomic data for blending lines and other nearby lines were taken from the compilation available in the Vienna Atomic Line Database (VALD, \citealt{Piskunov1995,Ryabchikova1997,Kupka1999,Kupka2000}). The complete linelists with hyperfine components can be found in Appendix~\ref{sec:linelists}.


\begin{figure}
\centering
\resizebox{\hsize}{!}{
\includegraphics[viewport=15 60 812 640,clip]{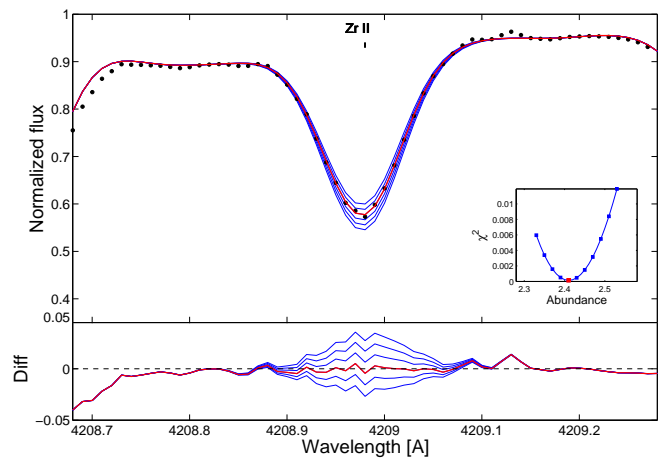}}
\caption{Synthesis of the \ion{Zr}{ii} line at 4208\,{\AA} in the solar spectrum (in this case Vesta, observed with the MIKE spectrograph in January 2006). The upper panel shows synthetic spectra with different Zr abundances in steps of 0.04\,dex (blue lines). The difference between the observed spectrum and the synthetic spectra is plotted in the lower panel with the best abundance spectrum plotted as a red line in both the upper and lower panel. The inset shows the $\chi^{2}$ fit to determine the best abundance value expressed as $\log \epsilon ({\rm Zr})$ (shown as a red dot).\label{fig:example_sun}}
\end{figure}

\subsection{Abundance determination}\label{sect:syn_creation}

The methodology for abundance determination is the same as in \cite{Battistini2015} where we analysed Sc, V, Mn, and Co for the same sample of stars. It is based on comparisons between observed spectra and synthetic spectra to find the best fitting abundance. The synthetic spectra were calculated with the MARCS2012 code \citep{Gustafsson2008}, under the assumption of local thermodynamic equilibrium (LTE), and one-dimensional plane-parallel model atmospheres. 

The synthetic spectra were created using SME (Spectroscopy Made Easy, \cite{Valenti1996,Valenti2005}) with stellar parameters from \cite{Bensby2014} as input. The abundance analysis is done through a minimisation routine of an un-normalised $\chi^{2}$ function based on the difference between the observed and the synthetic spectra. An example of the comparison between observed and synthetic spectra can be seen in Fig.~\ref{fig:example_sun} for the \ion{Zr}{ii} at 4208 $\AA$ in one of the available solar spectra. In Table~\ref{tab:solar_abundance} the abundances from individual lines are listed for the different solar spectra that we analysed. The fewer abundances usually available for La are due to the fact that the lines are generally weak so more similar to the spectral noise, while for Sr the line is in most cases not well synthetized.

\begin{table*}[ht]
\tiny
\caption{Comparisons of stellar parameters and abundances for stars in common to other studies.
\label{tab:error_literature}\tablefootmark{$^\dagger$} }
\setlength{\tabcolsep}{2mm}
\centering
\begin{tabular}{lrrrrr}
\hline
\hline
\noalign{\smallskip}
& Reddy03 & Reddy06 & Mashonkina04 & Mashonkina07 & Mishenina13 \\
\noalign{\smallskip}
\hline
\noalign{\smallskip}
$\Delta \teff$ [K] & 132 $\pm$ 52 (13) & 114 $\pm$ 114 (55) & 4 $\pm$ 69 (19) & 20 $\pm$ 101 (26) & $-$9 $\pm$ 54 (8)\\
$\Delta \log g$    & 0.05 $\pm$ 0.09 (13) & 0.01 $\pm$ 0.17 (55) & 0.01 $\pm$ 0.12 (19)  & 0.03 $\pm$ 0.13 (26) & 0.04 $\pm$ 0.11 (8)\\
$\Delta$[Sr/H]    &  --          & --           & --          & --           & --         \\
$\Delta$[Zr/H]    & 0.19 (1)    & --           & --          & --0.02 (4)  & $0.08$ (3) \\
$\Delta$[La/H]    & --          & --           & --          & --          & 0.18 (1)   \\
$\Delta$[Ce/H]    & 0.23 (3)    & $-0.04$ (14)  & --          & $-0.18$ (4) & $0.12$ (1) \\
$\Delta$[Nd/H]    & 0.16 (1)    & $-0.04$ (15) & $-0.12$ (4) & --          & $0.12$ (1) \\
$\Delta$[Sm/H]    & --          & --           & --          & --          & 0.21 (2)   \\
$\Delta$[Eu/H]    & $-0.06$ (1) & 0.05 (21)    & --          & --          & 0.11 (2)   \\
\noalign{\smallskip}
\hline
\end{tabular}
\tablefoot{\tablefoottext{$\dagger$}{Differences in stellar parameters from \cite{Bensby2014} and abundances from this work with the stars in common with \cite{Reddy2003,Reddy2006}, \cite{Mashonkina2004,Mashonkina2007} and \cite{Mishenina2013}. The differences are given as this work minus the other studies. In parenthesis, the number of common stars used for the different comparisons.}}
\end{table*}

The analysis was done strictly differentially with respect to the Sun. The solar spectra for each observation run is listed in Table 1 in \cite{Bensby2014} and were used to determine abundances for the different lines using the same methodology as for stars in the sample. \cite{Bedell2014} noted that the derived abundances from different solar spectra observed with different instruments can differ by up to 0.04\,dex. Hence, we decided to normalise the abundances on a line-by-line basis, using the correspondent solar spectrum from the different runs, when available. For the observation runs that do not have a solar observation an average abundance based on abundances from all solar spectra was used. Table~\ref{tab:solar_abundance} gives the abundances we derived for individual lines in the different solar spectra and Figs.~\ref{fig:solar_spectrum1}--\ref{fig:solar_spectrum5} in Appendix~\ref{sec:solarabundanceplots} show all line fits in one of the solar spectra, and an example is shown in Fig.~\ref{fig:example_sun}. 

A difficulty in the analysis for these elements is that most of the lines are weak, meaning that even if the signal-to-noise of the spectrum is high, in some cases the noise is comparable to the strength of the lines (this is visible in . In the solar spectrum these lines show equivalent widths ranging from few m$\AA$  (Ce at 5187\,{\AA}, 4 m$\AA$, and La at 4748\,{\AA}, 5.3 m$\AA$) to several tens of m$\AA$ (Eu at 4129\,{\AA}, 35.6 m$\AA$, Sr at 4607\,{\AA}, 44.4 m$\AA$, and Zr at 4208\;{\AA}, 78 m$\AA$). So, at the end of the fitting procedure of the each spectral line a visual inspection was performed to evaluate if the value for the best abundance from each line actually accurately reproduced the shape of the observed line. On a total of 597 stars from the FEROS and MIKE spectra, we have Sr abundances for 156 stars, Zr for 311 stars, La for 242 stars, Ce for 365 stars, Nd for 395 stars, Sm for 280 stars, and Eu for 378 stars. All abundances from individual lines are given in Table~\ref{tab:abundances_single_lines}, while the solar-normalized averaged values are given in Table~\ref{tab:abundances_H}. When results from more than two lines are available, we use median value because it is less affected by outliers.


\subsection{Systematic and random error estimation}\label{sect:error}

Performing a differential analysis relative to the Sun means that systematic errors arising from uncertainties in atomic data and the analysis methods largely cancel out. To further check for systematic uncertainties we search the literature for other studies that have analysed the same stars to investigate possible offset in stellar parameters and derived abundances. Table~\ref{tab:error_literature} lists the differences in stellar parameters and abundances for the stars in common with \cite{Reddy2003,Reddy2006}, \cite{Mashonkina2004}, \cite{Mashonkina2007} and \cite{Mishenina2013}. Among the different works, we share the highest number of stars with neutron-capture abundances determination with \cite{Reddy2006} while for the other works, like \cite{Reddy2003} and \cite{Mishenina2013}, we only have one or two stars in common, giving no real information on possible offsets. Only the comparison with \cite{Reddy2006} contains enough stars (more than ten) to show that in this case there is good agreement.

The estimation of the random errors for our abundances was done by deriving how much the errors in the stellar parameters would affect the final abundances. We follow the same procedure as in \cite{Battistini2015}, selecting a random subsample of stars probing different part of the stellar parameters space and calculating the new abundances when the the random errors from \cite{Bensby2014} are applied to each star. First we calculate the difference between the abundances without and with the errors on stellar parameters applied, then all the differences for each element are used to calculate the final square mean error. We consider an error on the abundance determination of 0.05\,dex for wrong continuum placement and from not perfectly fitted lines. The list of the abundance errors together with the mean standard random errors on stellar parameters are listed in Table~\ref{tab:median_error}. 

As it is visible in Table~\ref{tab:median_error}, the error on Eu is the smallest between the elements we analysed and is comparable with the errors on [Fe/H]. On average, our errors for the elements are around 0.1 dex  but for Sr is higher, precisely of 0.15 dex, probably due to the fact that Sr presents high spread (as it will be explained in Sect.~\ref{sect:results}).

\begin{table}
\tiny
\caption{Random errors in stellar parameters and abundances.\tablefootmark{$^\dagger$} \label{tab:median_error} }
\setlength{\tabcolsep}{2 mm}
\centering
\begin{tabular}{c c}
\hline
\hline
\noalign{\smallskip}
 Parameter & Random error (1-$\sigma$) \\
\noalign{\smallskip}
\hline
\noalign{\smallskip}
$\teff$  &  51\,K\\
$\log g$ &  0.07\\
$\rm [Fe/H]$   &  0.05\\
$\mathrm{\xi_{t}}$ & 0.08\\
$\rm [Sr/H]$   &  0.15\\
$\rm [Zr/H]$   &  0.12\\
$\rm [La/H]$   &  0.11\\
$\rm [Ce/H]$   &  0.12\\
$\rm [Nd/H]$   &  0.10\\
$\rm [Sm/H]$   &  0.11 \\
$\rm [Eu/H]$   &  0.08\\
\noalign{\smallskip}
\hline
\end{tabular}
\tablefoot{\tablefoottext{$\dagger$}{Mean standard errors for stellar parameters from \cite{Bensby2014}. Errors for the neutron-capture elements are from this study.}}
\end{table}

\section{Results}\label{sect:results}

\begin{figure*}
\centering
\resizebox{\hsize}{!}{
\includegraphics[viewport=0 0 650 400,clip]{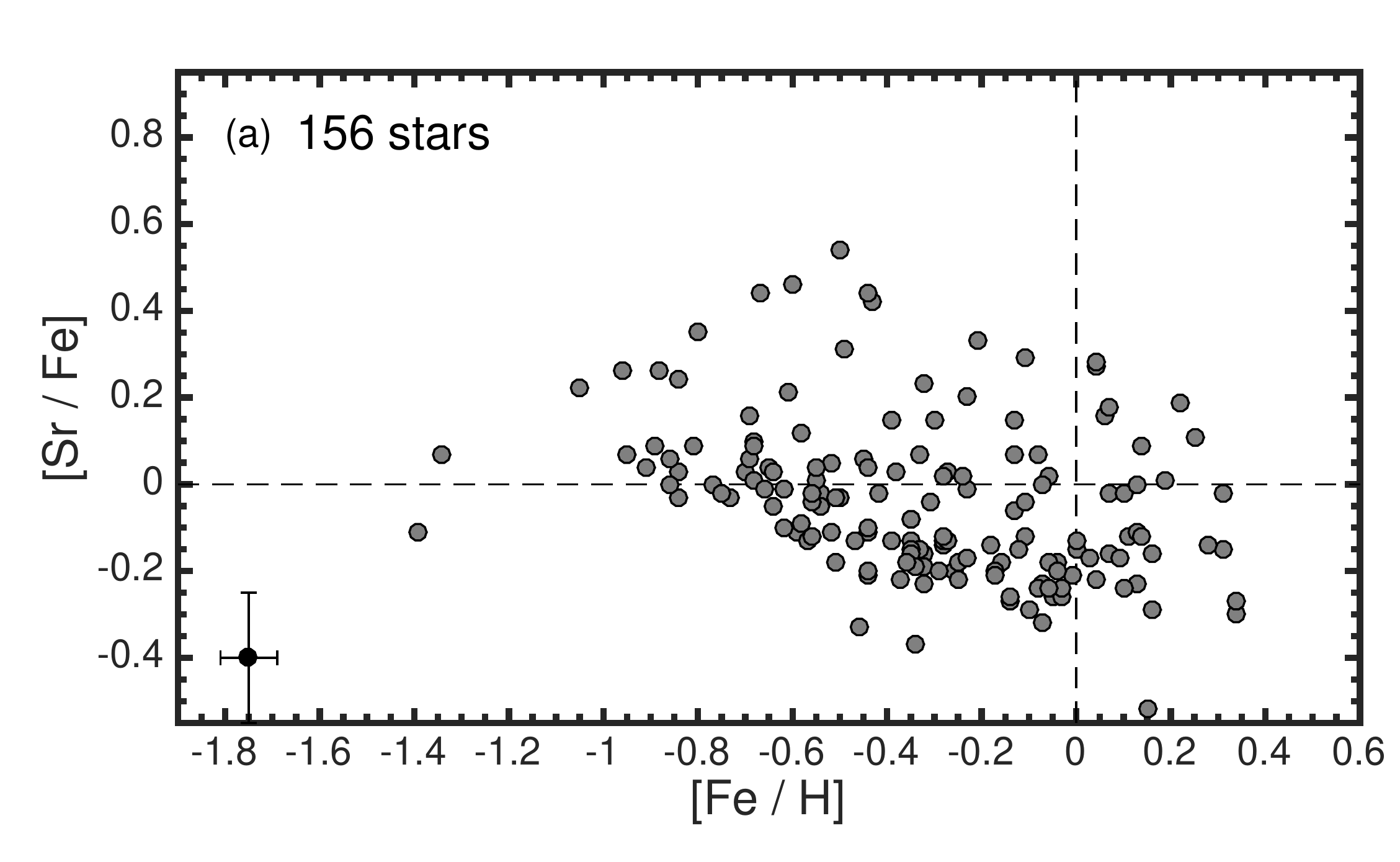}
\includegraphics[viewport=0 0 670 400,clip]{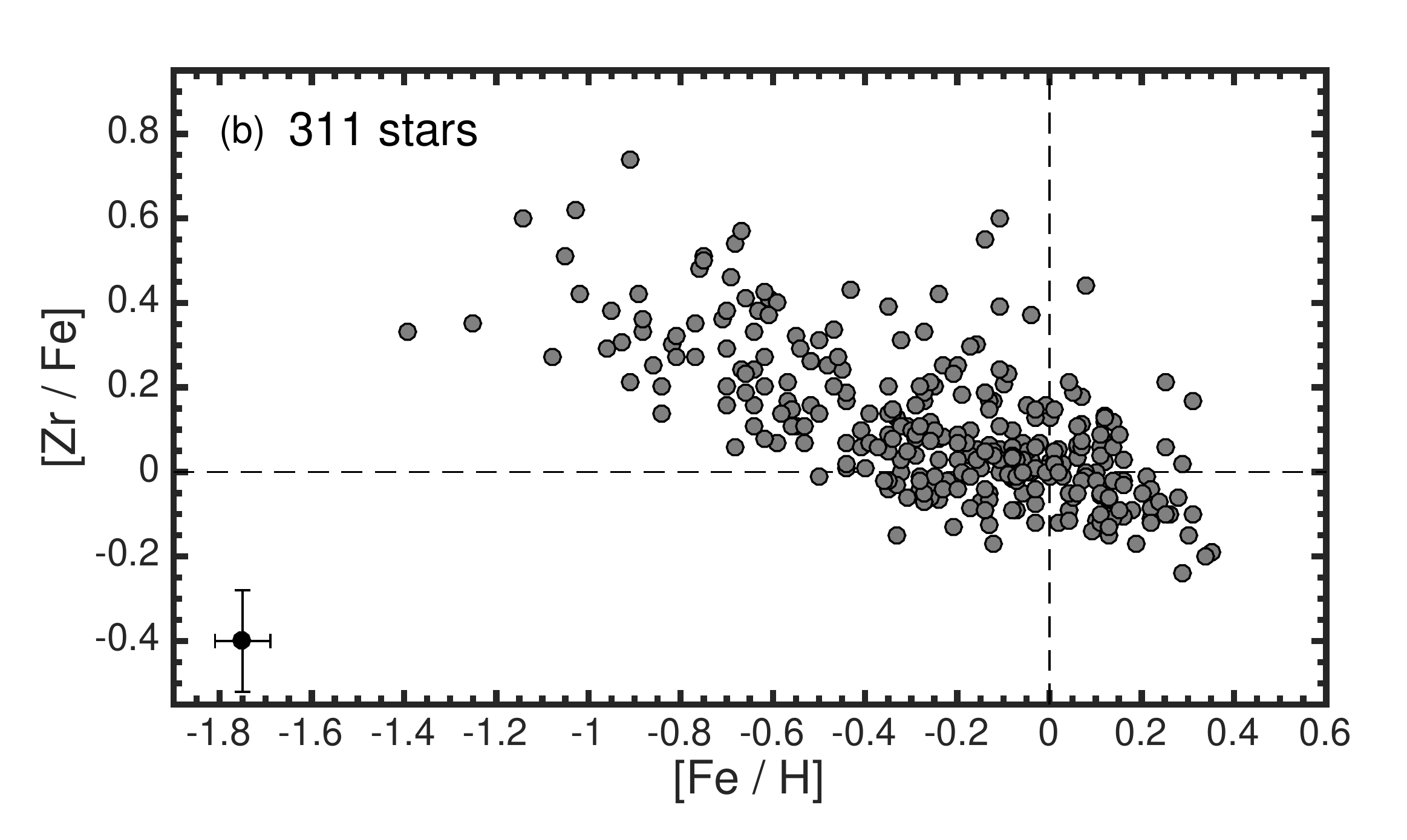}}
\resizebox{\hsize}{!}{
\includegraphics[viewport=0 0 650 400,clip]{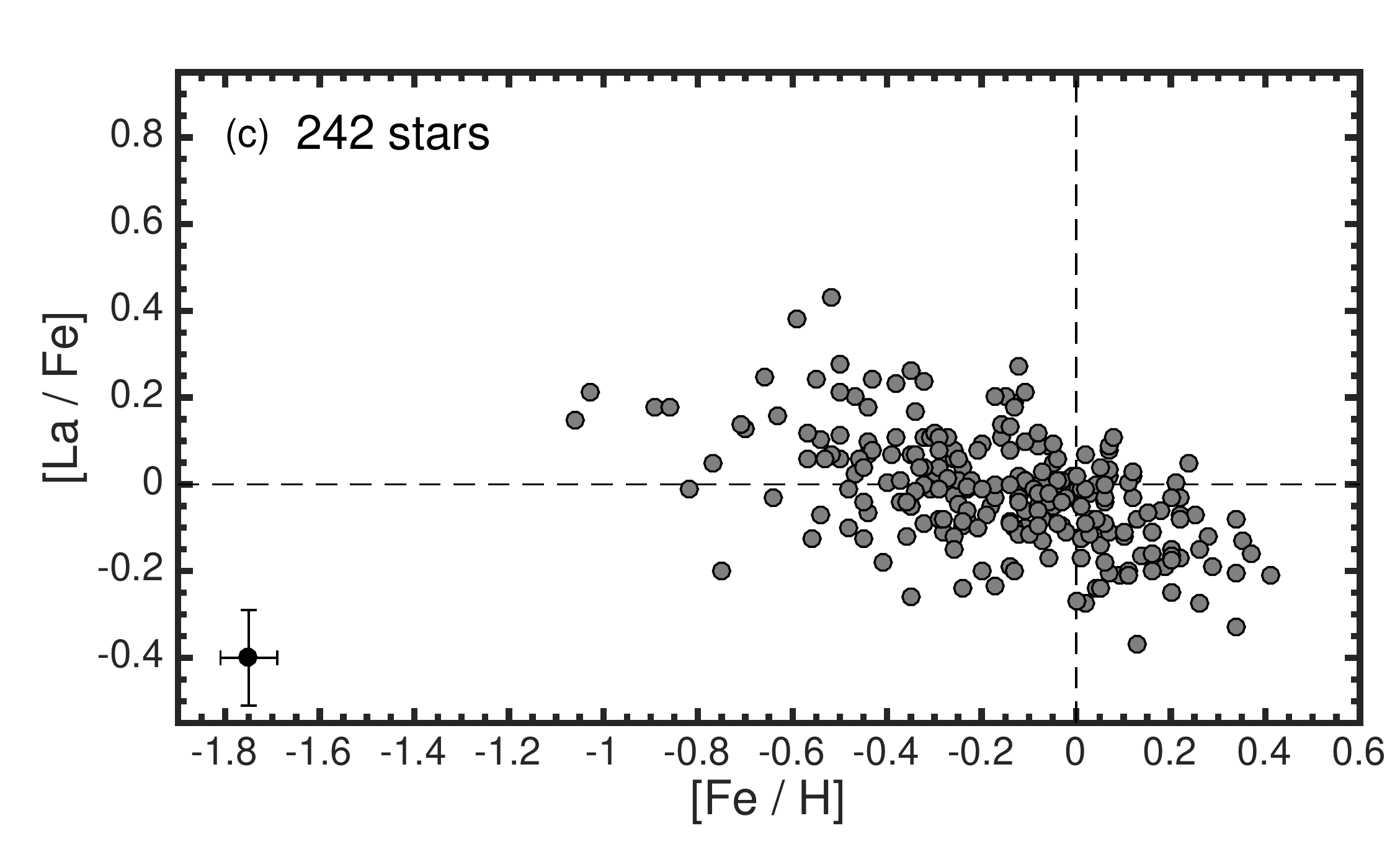}
\includegraphics[viewport=0 0 670 400,clip]{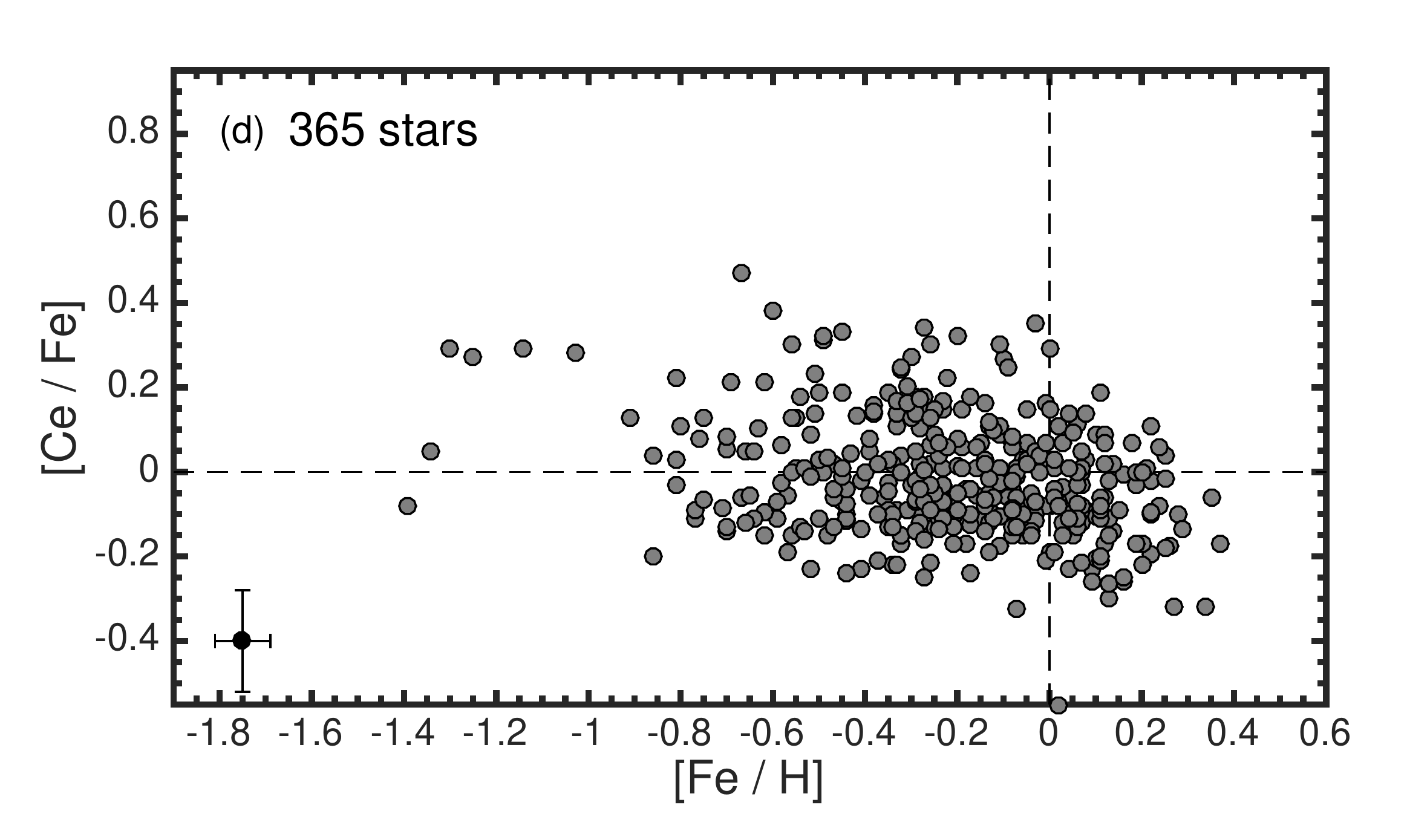}}
\resizebox{\hsize}{!}{
\includegraphics[viewport=0 0 650 400,clip]{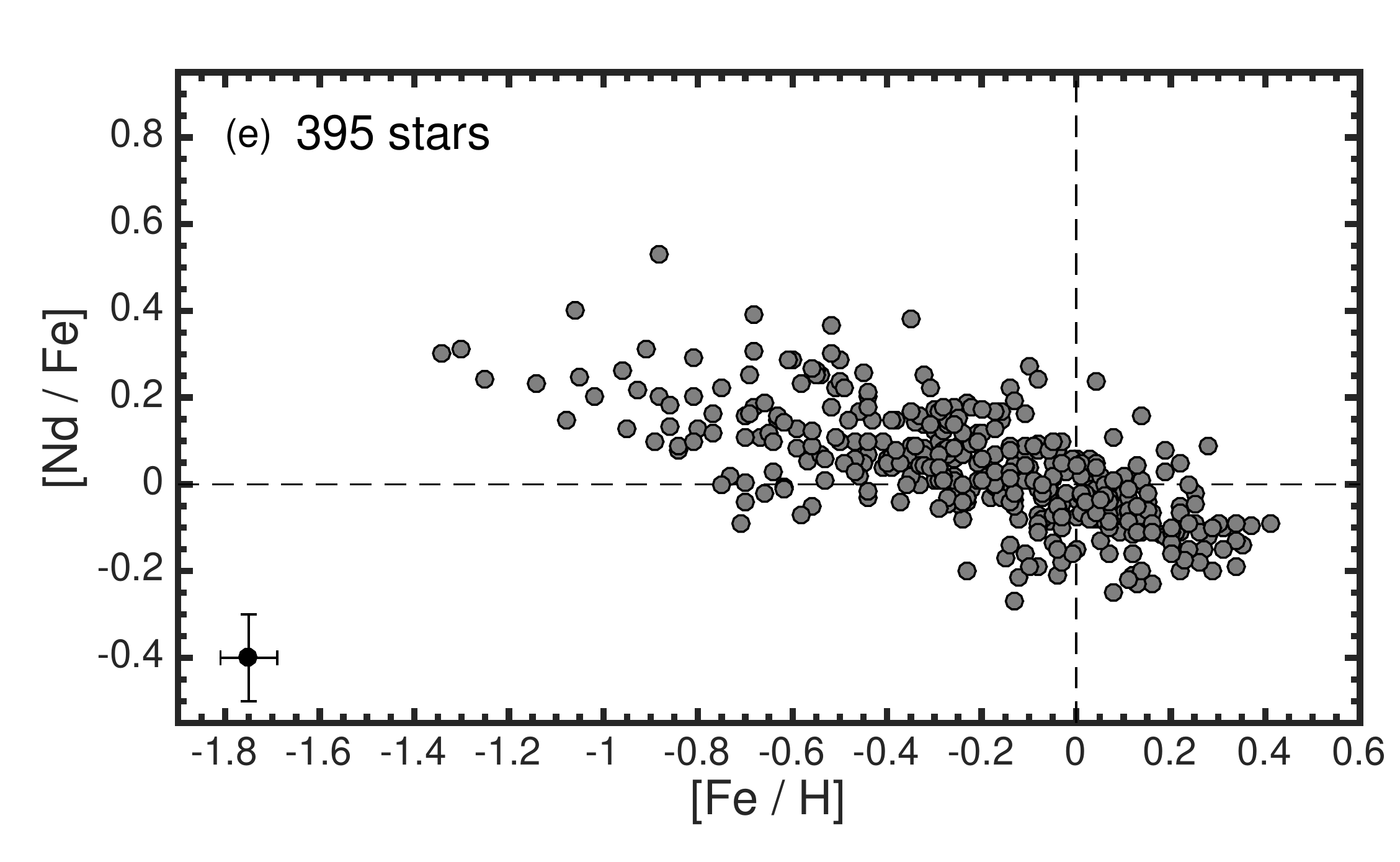}
\includegraphics[viewport=0 0 670 400,clip]{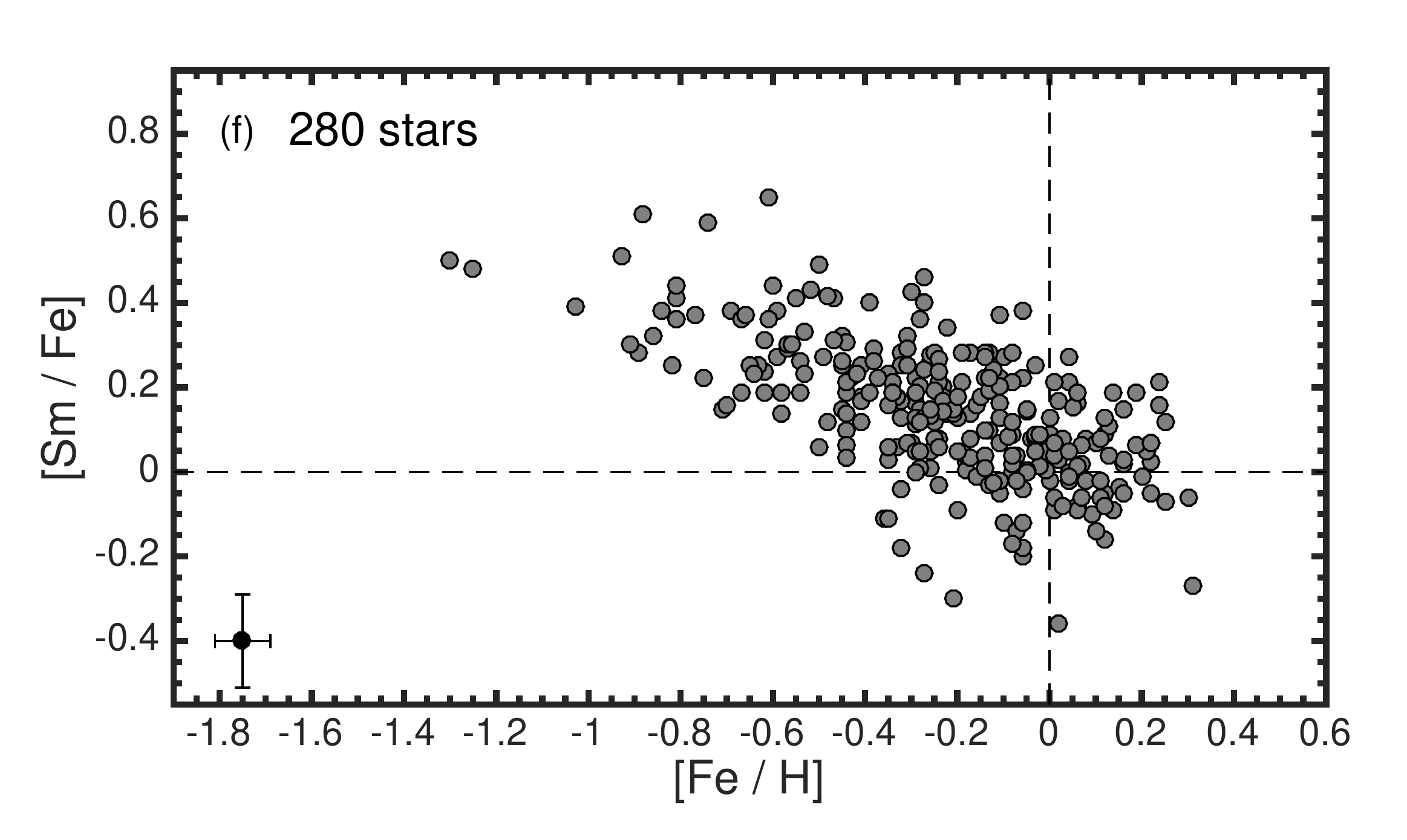}}
\resizebox{\hsize}{!}{
\includegraphics[viewport=-325 0 975 400,clip]{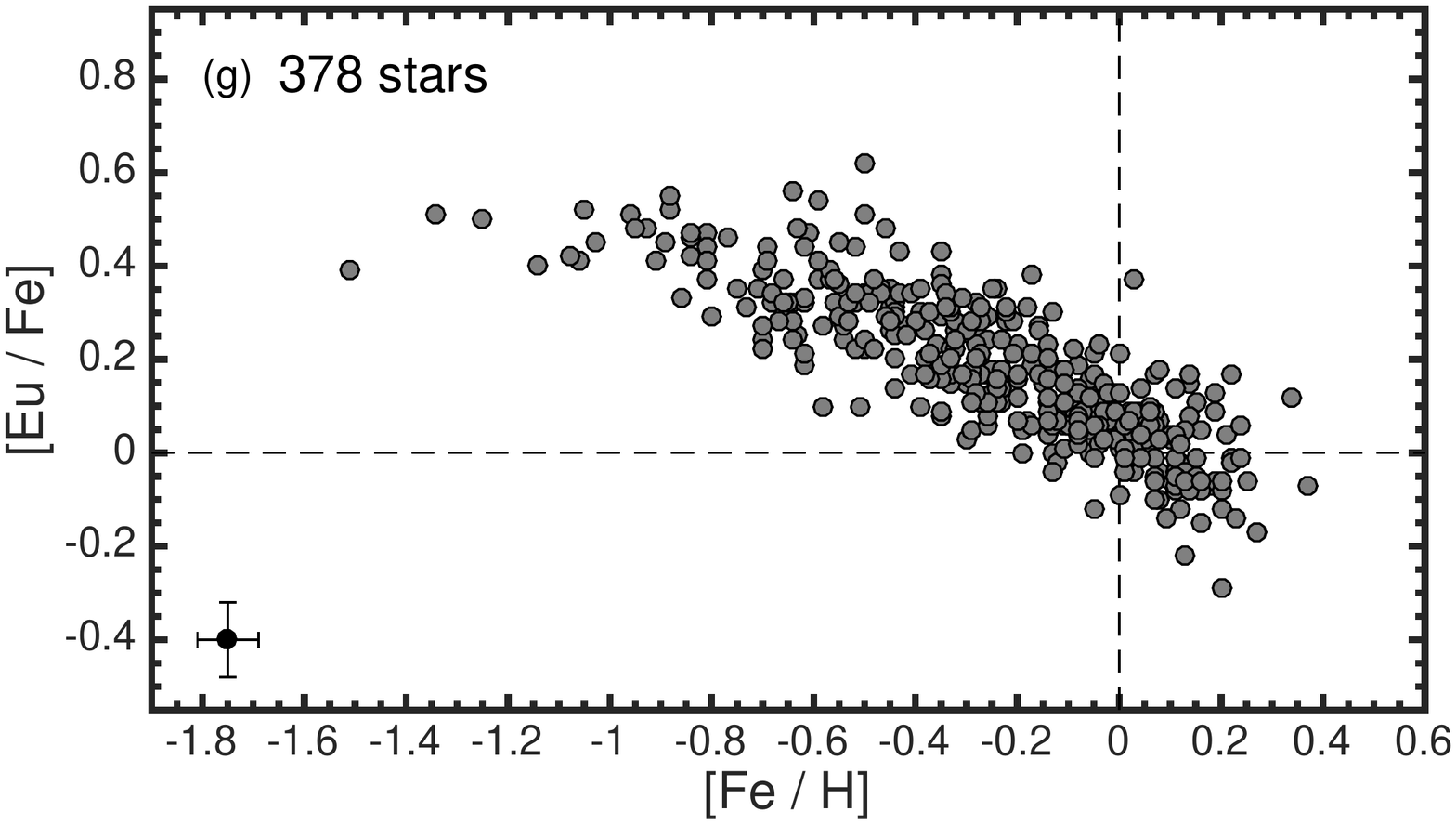}}
\caption{Elemental abundance trends with Fe as the reference element. The number for stars for which the abundance has been derived is indicated in each plot. \label{fig:abundance_full}}
\end{figure*}

\subsection{General abundance trends}

In this section we present the abundance results for the neutron-capture elements that we have analysed and Fig.~\ref{fig:abundance_full} shows the [$X$/Fe] versus [Fe/H] abundance trends.

\subsubsection{Strontium}

Strontium in Fig.~\ref{fig:abundance_full}a shows increasing [Sr/Fe] abundance ratios with decreasing metallicity: $\rm[Sr/Fe] \approx -0.2$ for solar metallicity stars that reaches $\rm[Sr/Fe]\approx 0$ at $\rm [Fe/H] \approx -1$. There are a few stars that show high [Sr/Fe] ratios, which could be due to the fact that only one Sr line was used. A similar spread is found in giant stars from \cite{Burris2000} and dwarf stars from \cite{Mashonkina2001}, \cite{Reddy2003} and \cite{Brewer2006} (even though they derived abundances from \ion{Sr}{ii} lines) for stars in a metallicity range similar to ours ($\rm -2.0 \lesssim [Fe/H] \lesssim 0.2$), where the trend is on average solar. [Sr/Fe] for very metal-poor stars ($\rm[Fe/H] < -2.5$), as for giants in \cite{Burris2000} and \cite{Andrievsky2011}, presents a down-turn. We notice that \cite{Andrievsky2011} used NLTE values for Sr but the same trend is still present even when compared to the LTE case. 
In addition to literature results, more recently \cite{Ishigaki2013} observed stars in thick disk and halo and found solar $\rm [Sr/Fe]$ ratios for $\rm [Fe/H] \lesssim -1$.
In addition to literature results, more recently \cite{Ishigaki2013} observed stars in thick disk and halo and found solar $\rm [Sr/Fe]$ ratios for $\rm [Fe/H] \lesssim -1$.

\subsubsection{Zirconium}
Zirconium in Fig.~\ref{fig:abundance_full}b shows an increasing trend with decreasing metallicity, from $\rm [Zr/Fe] \approx -0.1$ at $\rm [Fe/H] \approx 0.3$ to $\rm [Zr/Fe] \approx 0.3$ for the most metal-poor stars in the sample at $\rm [Fe/H] \approx -1$. 
\cite{Burris2000} found a [Zr/Fe] trend that seems to be only slightly super-solar with a large spread for stars with $\rm [Fe/H] < -1.5$, and \cite{Brewer2006} found basically solar [Zr/Fe] values for all their stars. The same was seen by \cite{Reddy2006} even if there seems to be a small trend from slight super-solar $\rm [Zr/Fe] $ at solar metallicity to slight sub-solar [Zr/Fe] for $\rm [Fe/H] < -0.6$. On the other hand, the trend found by \cite{Mashonkina2007} is similar to what we find with an increasing trend from solar [Zr/Fe] value for solar metallicity stars to $\rm [Zr/Fe] \approx 0.3$ for $\rm [Fe/H] \approx -1.5$. Recently \cite{Mishenina2013} observed F, G, and K dwarf stars in the Galactic disk finding a similar trend to ours even though the stars at super-solar metallicities show solar or slightly enhanced [Zr/Fe] values. Also \cite{Ishigaki2013} studied Zr finding a similar [Zr/Fe] trend to ours for their thick disk stars while for halo stars there is a decreasing trend to negative [Zr/Fe] values at low metallicities.

\subsubsection{Lanthanum}
As is shown in Fig.~\ref{fig:abundance_full}c, La presents mostly solar [La/Fe] values, with a possible trend from super-solar metallicity with $\rm [La/Fe] \approx -0.3$ to roughly solar [La/Fe] for $\rm[Fe/H] \approx -0.6$. \cite{Burris2000} observed average solar [La/Fe] for $\rm [Fe/H] \lesssim -$1 with increasing spread as metallicity decreases. \cite{Simmerer2004} derived abundances of La for giants and dwarfs in the metallicity range $-$3 < [Fe/H] < 0.3: for stars with the metallicity as in our sample, the agreement is good even if an offset of $\sim$ 0.2 dex seems to be present. \cite{Brewer2006} found a trend similar to what we found, with sub-solar [La/Fe] for solar metallicity stars that then increases to reach around solar [La/Fe] for $\rm[Fe/H] \approx -0.6$. More recently \cite{Mishenina2013} found results in agreement with our trend as well as the thick disk sample of \cite{Ishigaki2013}.

\subsubsection{Cerium}
In Fig.~\ref{fig:abundance_full}d Ce presents a basically flat and slightly sub-solar [Ce/Fe] abundance trend. This behaviour with $\rm [Ce/Fe] \approx 0$ is observed in \cite{Reddy2006} while a flat trend but slightly super-solar is found in \cite{Brewer2006}. However, in \cite{Mashonkina2007},  the [Ce/Fe] trend is clearly increasing as metallicity decreases. Similar to our result is \cite{Mishenina2013}, where for solar metallicity and sub-solar metallicities, [Ce/Fe] shows a large spread around the solar [Ce/Fe] value.

\subsubsection{Neodymium}
Neodymium shows a tighter trend compared to the previous elements, as it is visible in Fig.~\ref{fig:abundance_full}e. The trend presents similarities to Zr, with sub-solar [Nd/Fe] for super-solar metallicity stars that then raises to $\rm [Nd/Fe] \approx 0.2$ at $\rm [Fe/H] \approx - 0.6$ and for even lower metallicity the trend seems to be flat. \cite{Burris2000} presents a significant spread in $\rm [Nd/Fe]$, while \cite{Mashonkina2004} found a tight trend similar to what we found, with increasing $\rm [Nd/Fe] $ for decreasing metallicity up to $\rm [Nd/Fe] \approx 0.3$ for $\rm [Fe/H] \approx - 1$ and then a plateau down to lower metallicity. The same general trend was found in \cite{Brewer2006} while, on the other hand, the trend found in \cite{Reddy2006} is basically flat with $\rm [Nd/Fe] \approx 0$. The result from \cite{Mishenina2013} is similar to ours, while \cite{Ishigaki2013} found flat trend with $\rm [Nd/Fe] = 0$ for thick disk stars and $\rm [Nd/Fe] = 1$ at $\rm [Fe/H] = -0.6$, that decrease to reach solar $\rm [Nd/Fe] $ for $\rm [Fe/H] - 3$ for halo stars.

\subsubsection{Samarium}
In Fig.~\ref{fig:abundance_full}f it is possible to distinguish a rising trend from solar $\rm [Sm/Fe]$ at solar metallicity up to $\rm [Sm/Fe] \approx 0.5$ at $\rm [Fe/H] \approx -1$. Some of the previous listed works studied Sm as well. For example \cite{Mishenina2013} found a similar trend but with an offset of $\rm \approx 0.2$ dex, meaning that solar and super-solar metallicity stars have $\rm [Sm/Fe] \approx -0.2$. Also \cite{Ishigaki2013} studied $\rm [Sm/Fe] $ finding basically solar value.

\subsubsection{Europium}
Europium (Fig.~\ref{fig:abundance_full}g) clearly shows a typical $\alpha$-element trend, as is expected for an almost pure r-process element, since rapid neutron-capture is believed to happen in SN\,II. \cite{Prochaska2000} derived Eu abundances for four stars in metallicity range $\rm -0.7 < [Fe/H] < -0.4$  showing high $\rm [Eu/Fe]$. \cite{Burris2000} shows raising $\rm [Eu/Fe]$ as in \cite{Mashonkina2001} and \cite{Koch2002} where Eu shows a raising trend as we observed, with a steady increase from $\rm [Eu/Fe] \approx -0.2$ at $\rm [Fe/H] \approx 0$ up to $\rm [Eu/Fe] \approx 0.4$ for $\rm [Fe/H] \approx -1.0$.  A good agreement can be found in \cite{Simmerer2004}, \cite{Bensby2005}, \cite{Brewer2006} and in \cite{Reddy2006} even if in this case the spread is higher. The same trend with good agreement in values and shape is found in \cite{Mishenina2013} while in \cite{Ishigaki2013} the same trend is shifted to lower metallicity, as $\rm [Eu/Fe] \approx 0$ for $\rm [Fe/H] \approx -1$ and then $\rm [Eu/Fe]$ raises to $\rm +0.2$ at $\rm [Fe/H] \approx -1.8$.

\begin{figure}
\resizebox{\hsize}{!}{
\includegraphics[viewport=0 0 670 400,clip]{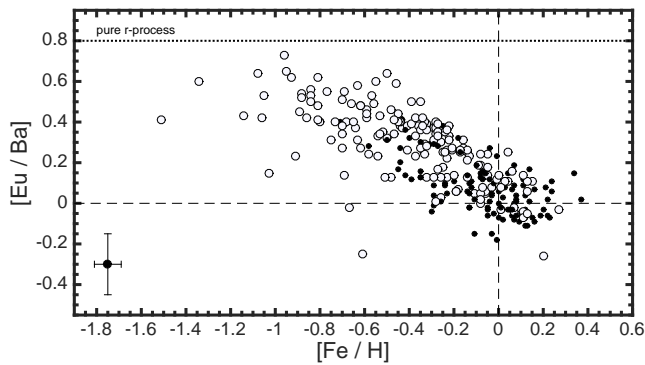}}
\caption{[Eu/Ba] as a function of [Fe/H]. The full sample is divided in thin and thick disk according to our age selection criterion. The dotted line represent pure r- process ratio derived from \cite{Bisterzo2014}. The average error is also presented.}\label{fig:EuBa_vs_FeH}
\end{figure}

\section{Origin of s- and r- elements}
\label{sect:evoution_s_r}

Most of the neutron-capture elements are produced from a mixture of r- and s-processes, and only in few cases only one of the two processes is the main responsible for the production. The comparison between such ``prototype'' elements of either the r- or the s-process, and other neutron-capture elements can help to constrain the production sites for the neutron-capture elements with uncertain origins.
Barium, for instance, is often used in comparisons between neutron-capture elements due to its high s-process component. For example, \cite{Arlandini1999} derived the contributions of the s- and r-process to the solar Ba abundance being 81\,\% and 19\,\% respectively. However, since the work by \cite{Arlandini1999} several studies have been published with updated r- and s-process rates. For our comparisons we will use the values from \cite{Bisterzo2014}. 

The Ba abundances that we use in our study are derived with equivalent width from \cite{Bensby2014}.  Since Ba is known to suffer from NLTE for $\teff$ > 6100 (i.e. \citealt{Korotin2011}) and for this reason, in the comparisons with Ba we discarded the stars with high temperature. In Fig. 16 in \cite{Bensby2014} is visible the trend of [Ba/Fe] as a function of [Fe/H] together with the typical error on [Ba/Fe] of about 0.1 dex. Once the stars with high temperature are removed, the [Ba/Fe] trend is basically flat.

Several studies of the abundance structure in the Galactic disk have revealed the presence of two different stellar components that differ in age, kinematic and $\alpha$-element abundances. The thin disk contains mostly young and kinematically cold stars with low $\alpha$-abundance for a given metallicity while, on the other hand, the thick disk contains mostly old and kinematically hot stars with high $\alpha$-abundance for given metallicity \citep{Fuhrmann1998, Bensby2003, Bensby2005, Reddy2003, Reddy2006, Adibekyan2012, Bensby2014}. As suggested by \cite{Haywood2013} and \cite{Bensby2014}, stellar ages seem to be a better separator between thin and thick disk stellar populations, than kinematic properties than largely overlap between the two. Following \cite{Bensby2014} we consider stars younger than 7\,Gyr  thin disk stars and stars older than 9\,Gyr thick disk stars. In all the following figures, if not specified otherwise, black dots are thin disk stars while white dots are thick disk stars as divided followed our criterion.

\subsection{Eu}

The comparison between Ba and Eu can be used as diagnostic of the neutron-capture process \citep{Mashonkina2001} due to the different production of the two elements (Eu almost completely r-process, Ba mostly s-process). In Fig.~\ref{fig:EuBa_vs_FeH} the [Eu/Ba] ratio  is close to the pure r-process line for metal-poor stars, meaning that the r-process was the only neutron-capture process active at the beginning of the formation of the Milky Way. As mentioned before, Eu is almost completely produced by the r-process, but Fig.~\ref{fig:EuBa_vs_FeH} shows that also Ba was initially produced in this way, as already found for example by \cite{Burris2000}. As soon as AGB stars start to be present and enrich the ISM with s-process elements, the [Eu/Ba] ratio decreases until it reaches the solar value at solar metallicity. Comparing [Zr/Ba] in Fig.~\ref{fig:SrZr_vs_FeH} with [Eu/Ba] it is possible to see that the rise in [Eu/Ba] is steeper due to the almost complete production of Eu by r-process, meaning that as soon as enrichment from AGB stars become predominant, the [Eu/Ba] ratio decreases quickly. 

\subsection{Sr and Zr}
\begin{figure*}
\resizebox{\hsize}{!}{
\includegraphics[viewport=0 15 650 400,clip]{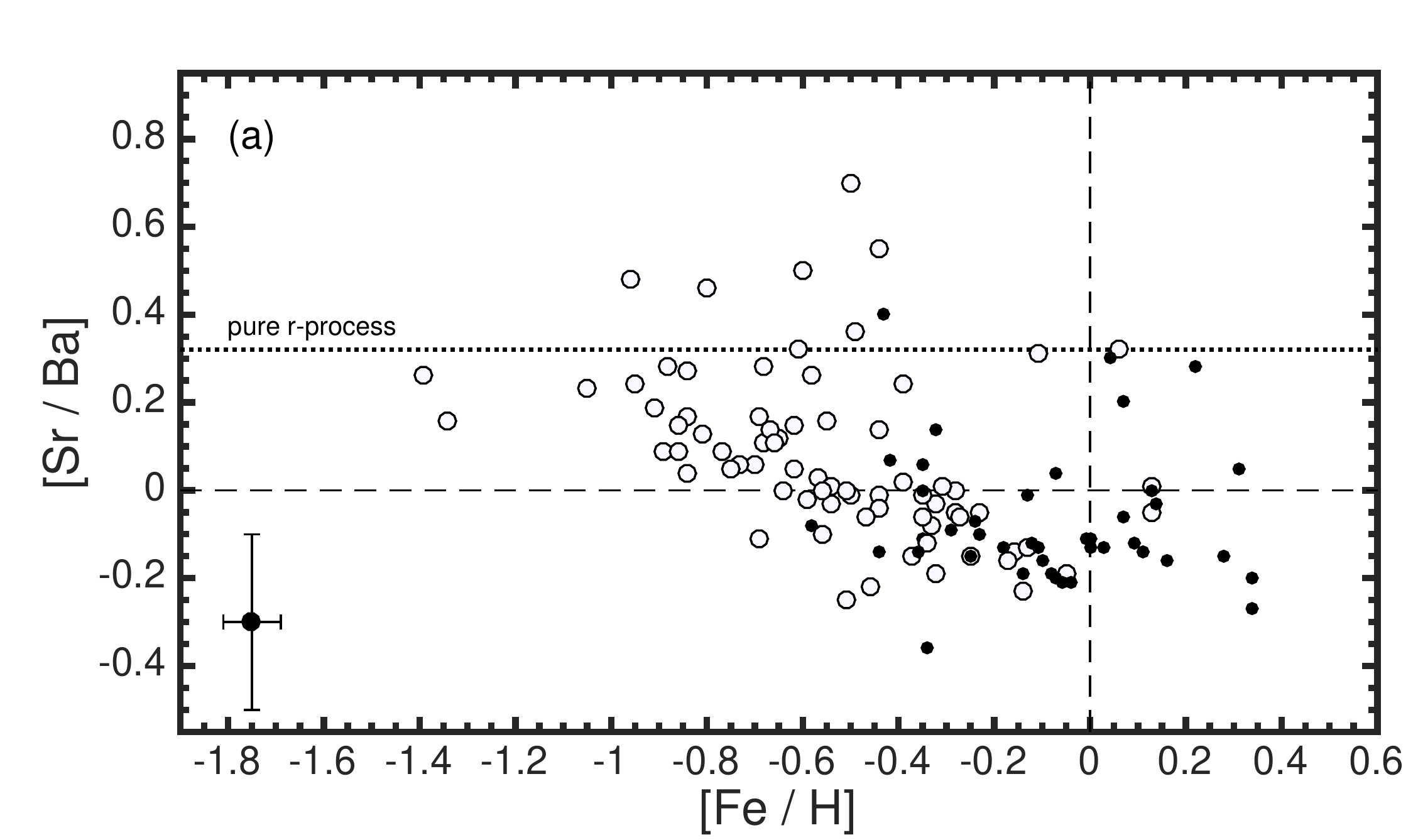}
\includegraphics[viewport=0 15 670 400,clip]{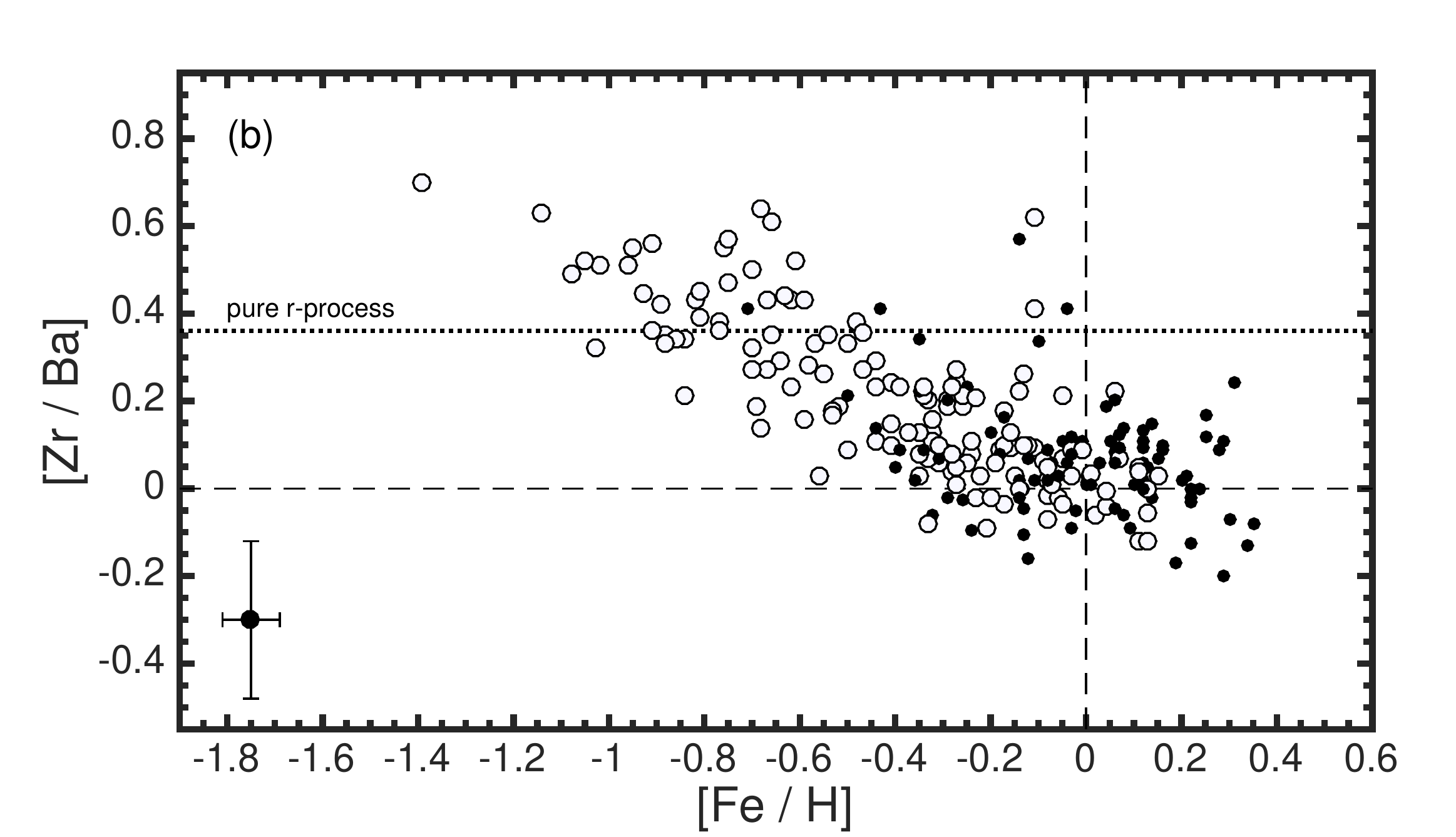}}
\resizebox{\hsize}{!}{
\includegraphics[viewport=0 15 650 380,clip]{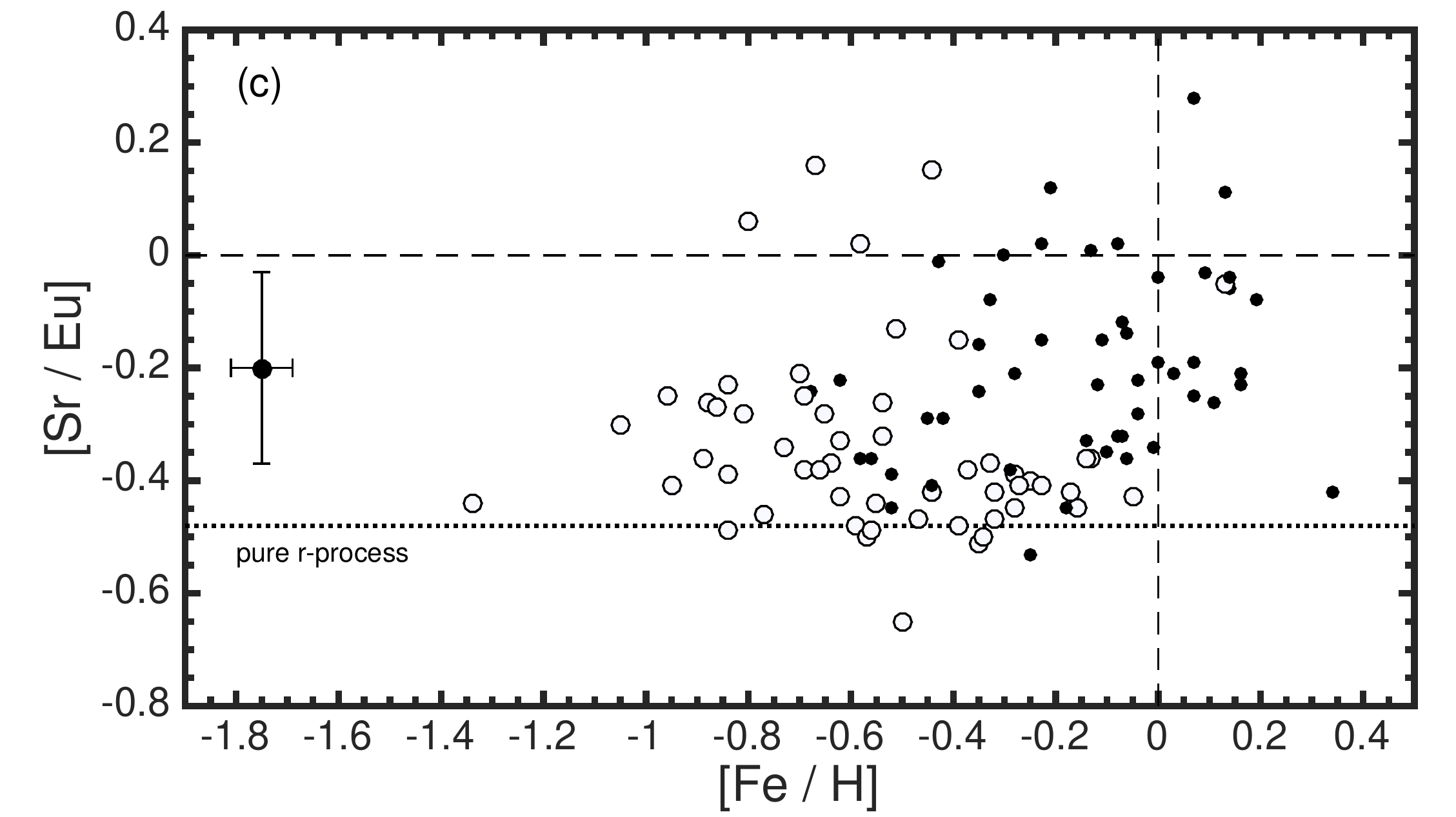}
\includegraphics[viewport=0 15 670 380,clip]{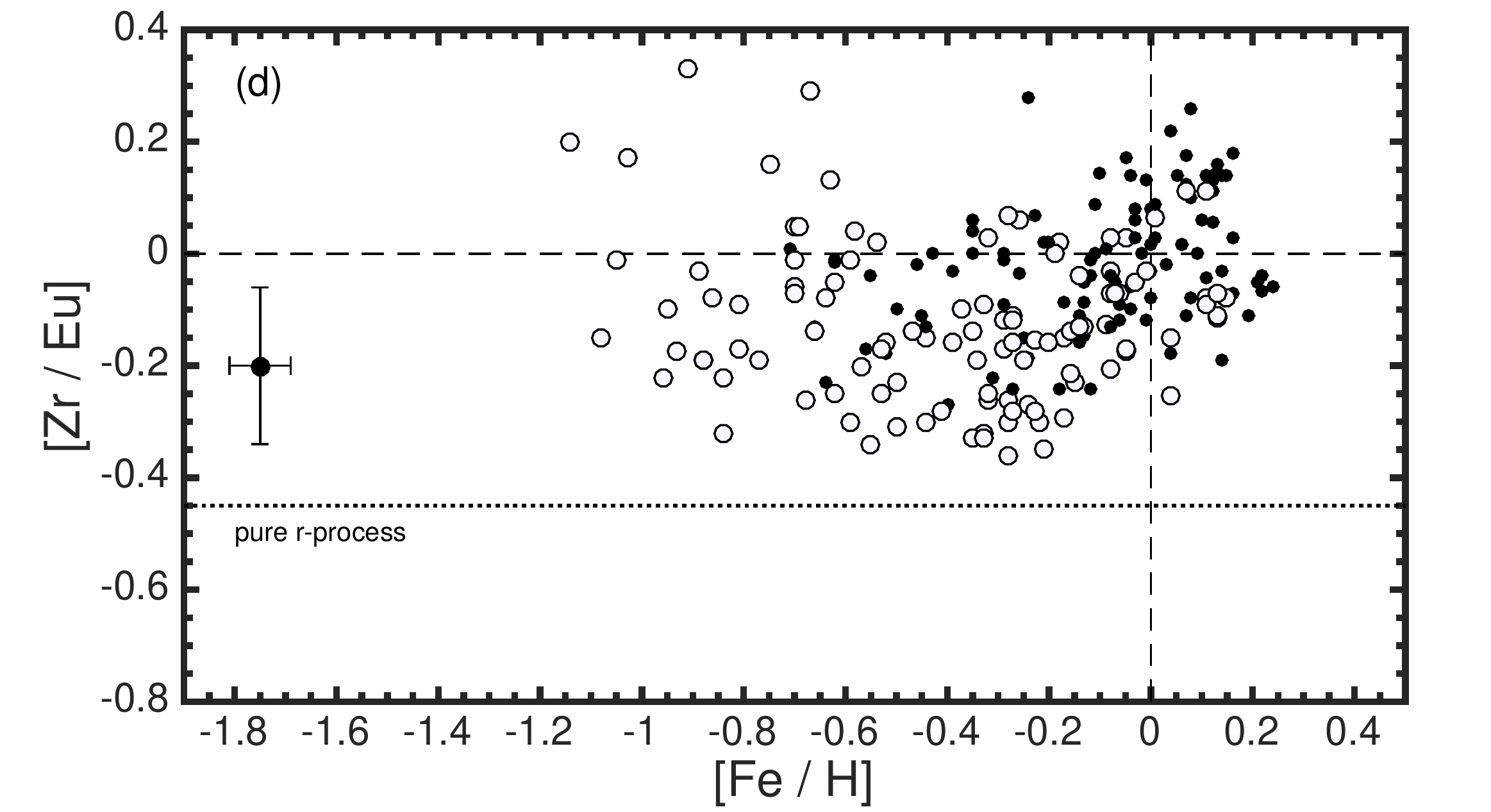}}
\caption{[Sr/Ba] (panel (a)), [Zr/Ba] (panel (b)) as a function of [Fe/H] and [Sr/Eu] (panel (c)) and [Zr/Eu] (panel(d)) as a function of [Ba/H]. The full sample is divided in thin and thick disk according to our age selection criterion. The dotted line represent pure r- process ratio derived from \cite{Bisterzo2014}. The average error is also indicated.}\label{fig:SrZr_vs_FeH}
\end{figure*}

Barium can also be compared to Sr and Zr abundances to diagnose the processes that formed the Sr-Y-Zr peak elements \citep{Travaglio2004}.  Figures~\ref{fig:SrZr_vs_FeH}a-b show the trends for [Sr/Ba] and [Zr/Ba] as a function of [Fe/H]. Even if the number of stars in the Sr case is lower than for Zr, the increases in [Sr/Ba] and [Zr/Ba] with decreasing metallicity are quite similar. Compared to [Eu/Ba], [Zr/Ba] shows a less steep rise with decreasing metallicity, probably because Zr has an higher s-process component than Eu. In addition to this, flat [Zr/Ba] and [Sr/Ba] trends are seen for stars around solar metallicity. 
Moreover, the abundances for thick disk stars at low-metallicity is particularly high, similar to what is visible in Fig.~\ref{fig:EuBa_vs_FeH} for Eu, and is difficult to explain only with the $\approx$15\,\% and $\approx$20\,\% of r-production of Sr and Zr, respectively, from \cite{Arlandini1999}. On the other hand, a shift of $\approx$\,0.2 dex is visible between [Sr/Ba] and [Zr/Ba].

\begin{figure*}
\resizebox{\hsize}{!}{
\includegraphics[viewport=0 15 650 400,clip]{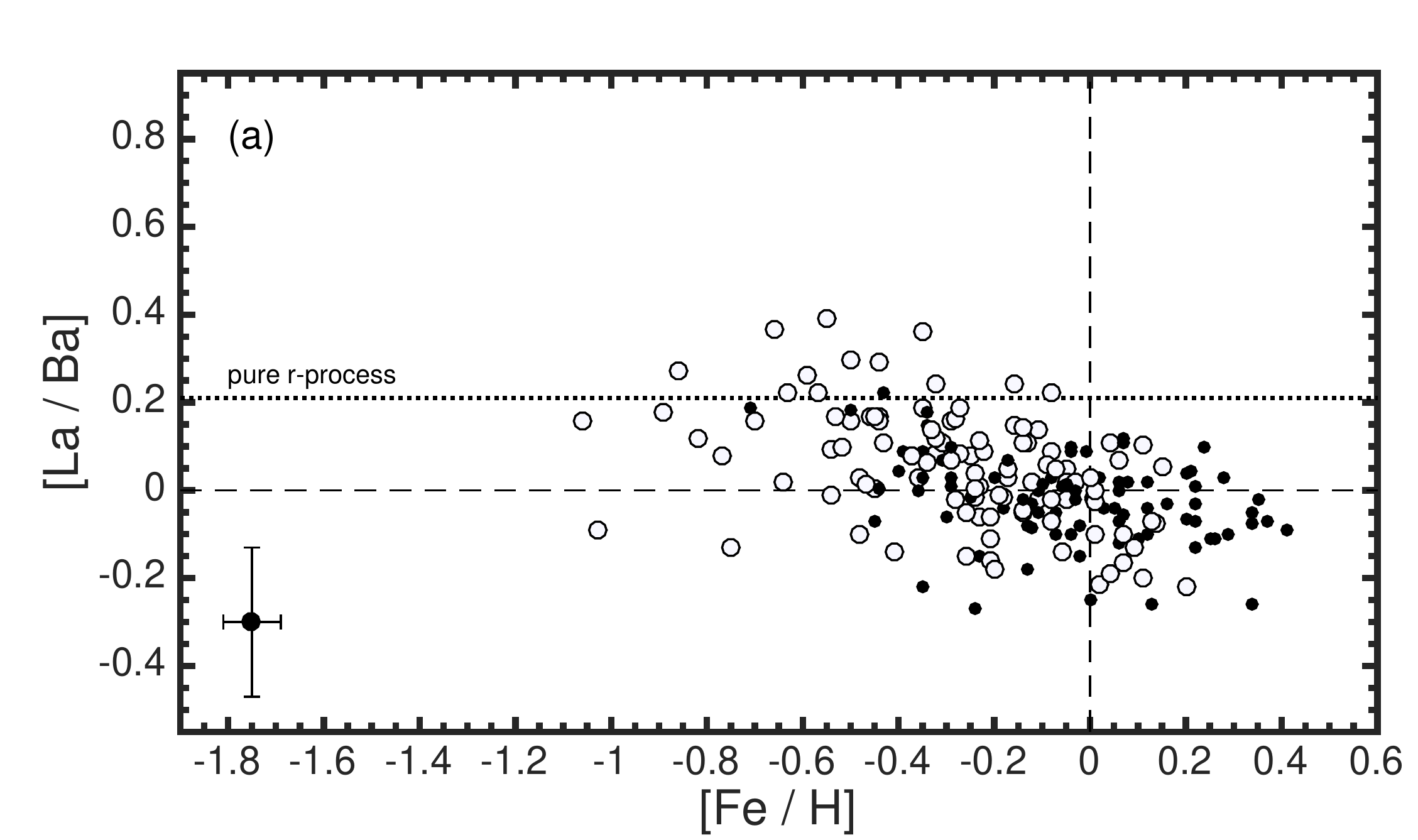}
\includegraphics[viewport=0 15 670 400,clip]{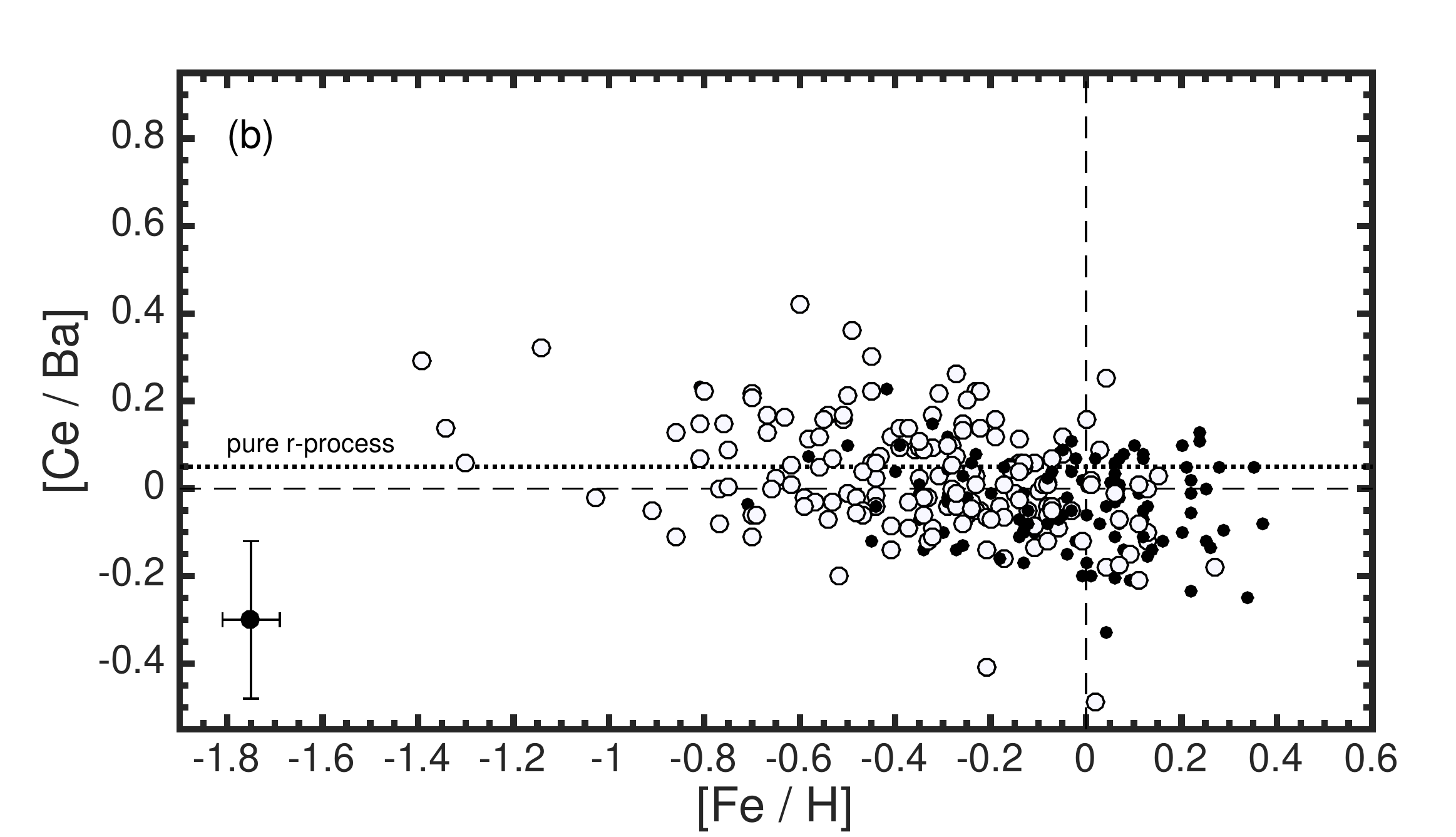}}
\resizebox{\hsize}{!}{
\includegraphics[viewport=0 15 650 380,clip]{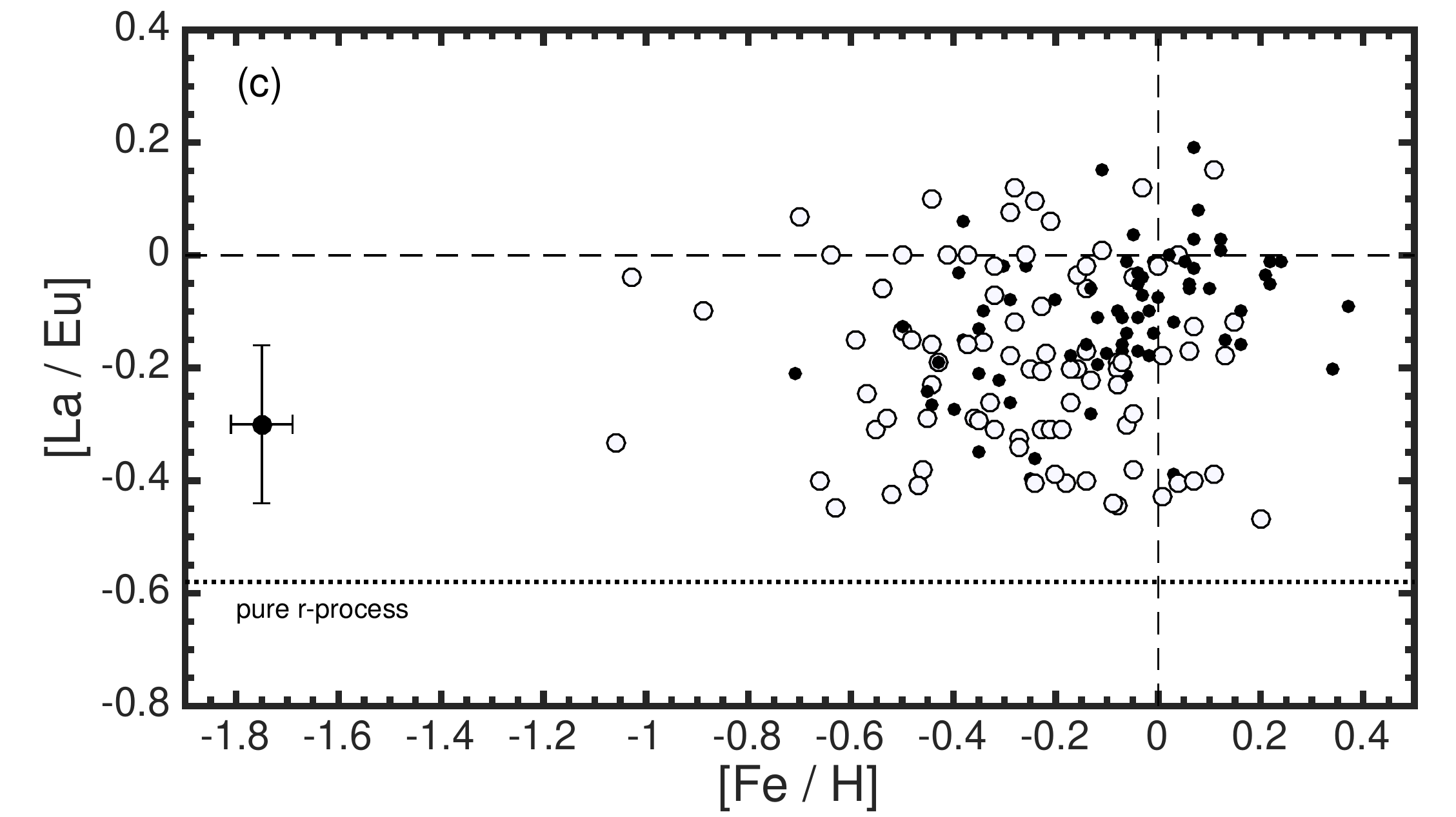}
\includegraphics[viewport=0 15 670 380,clip]{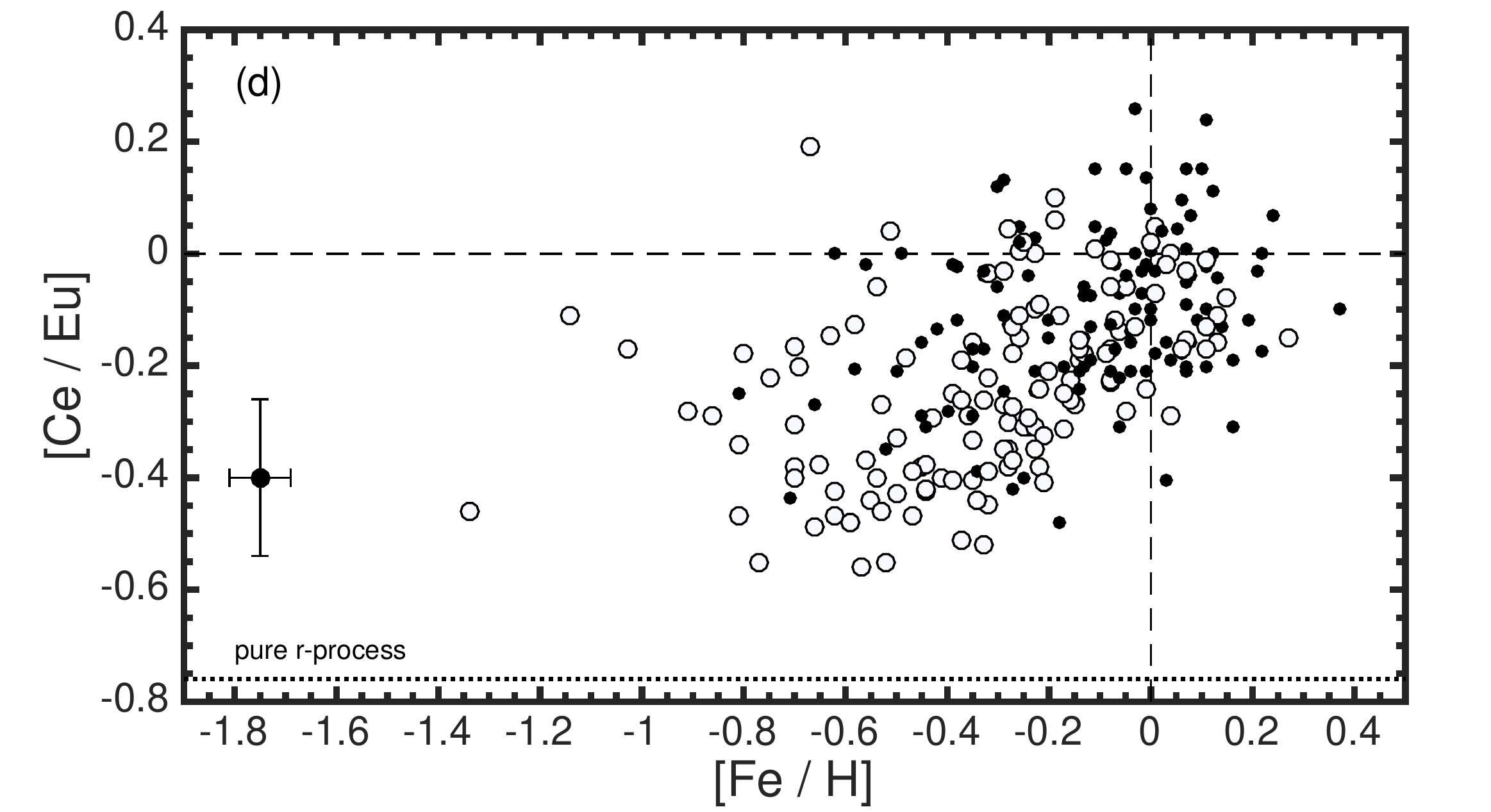}}
\caption{[La/Ba] and [Ce/Ba] (panels (a) and (b) respectively) and [La/Eu] and [Ce/Eu] (panels (c) and (d) respectively) as a function of [Fe/H]. The full sample is divided in thin and thick disk according to our age selection criterion. The dotted line represent pure r- process ratio derived from \cite{Bisterzo2014}. The average error is also indicated.}\label{fig:LaCe_vs_FeH}
\end{figure*}

\begin{figure*}
\resizebox{\hsize}{!}{
\includegraphics[viewport=0 15 650 400,clip]{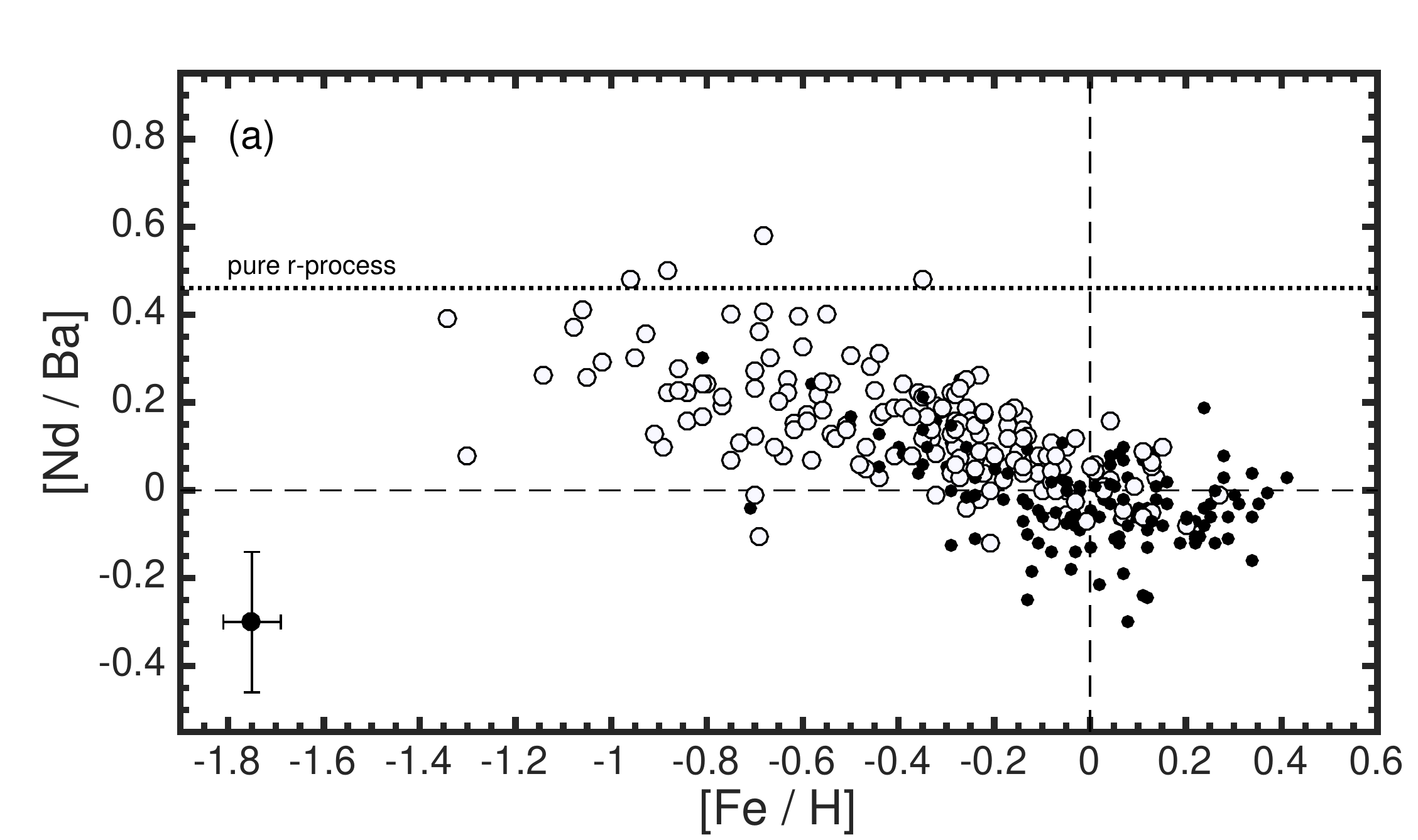}
\includegraphics[viewport=0 15 670 400,clip]{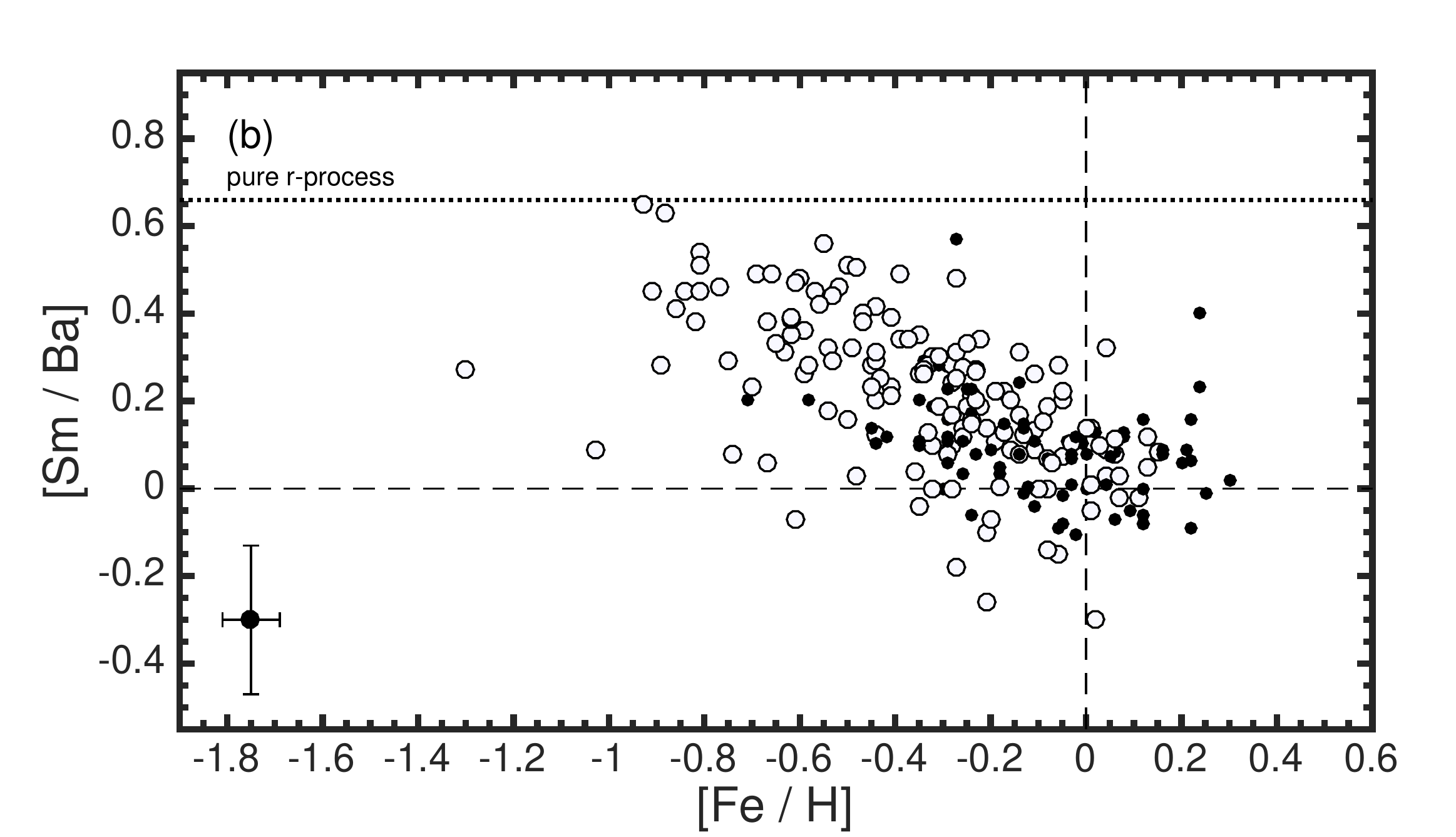}}
\resizebox{\hsize}{!}{
\includegraphics[viewport=0 15 650 380,clip]{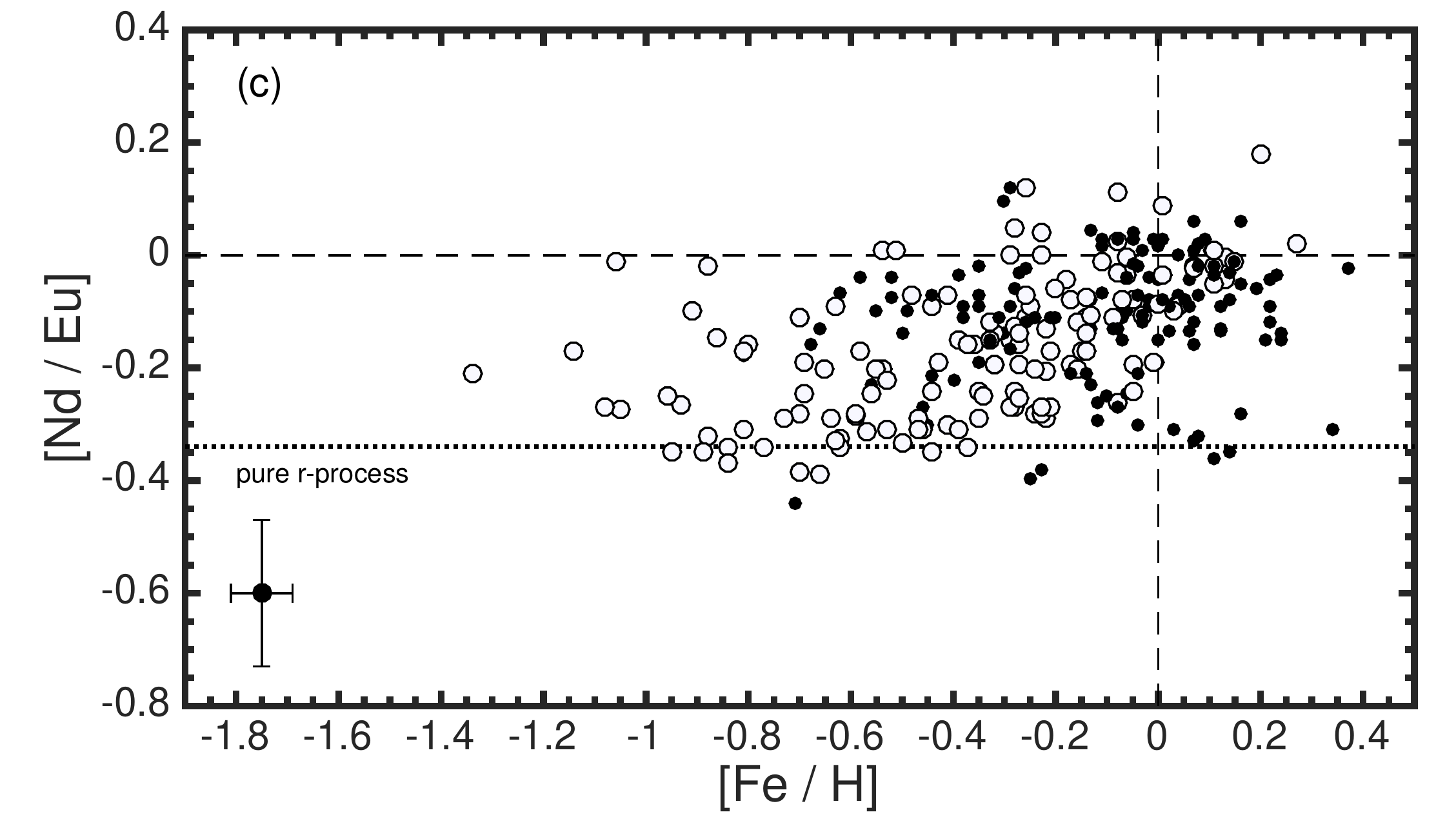}
\includegraphics[viewport=0 15 670 380,clip]{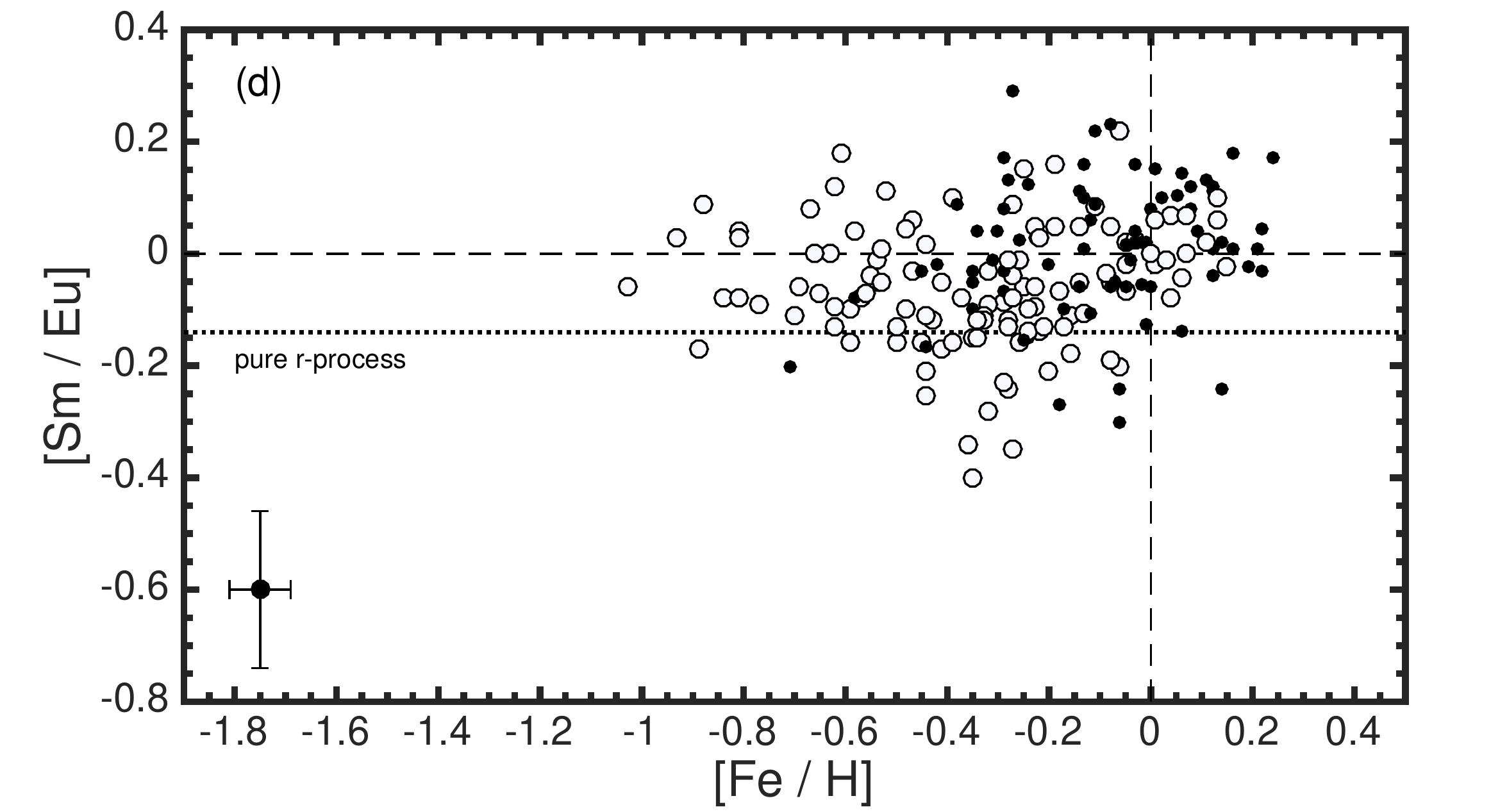}}
\caption{[Nd/Ba] and [Sm/Ba] (panels (a) and (b) respectively) and [Nd/Eu] and [Sm/Eu] (panels (c) and (d) respectively) as a function of [Fe/H]. The full sample is divided in thin and thick disk according to our age selection criterion. The dotted line represent pure r- process ratio derived from \cite{Bisterzo2014}. The average error is also indicated.}\label{fig:NdSm_vs_FeH}
\end{figure*}

\begin{figure*}
\centering
\resizebox{\hsize}{!}{
\includegraphics[viewport=0 250 650 600,clip]{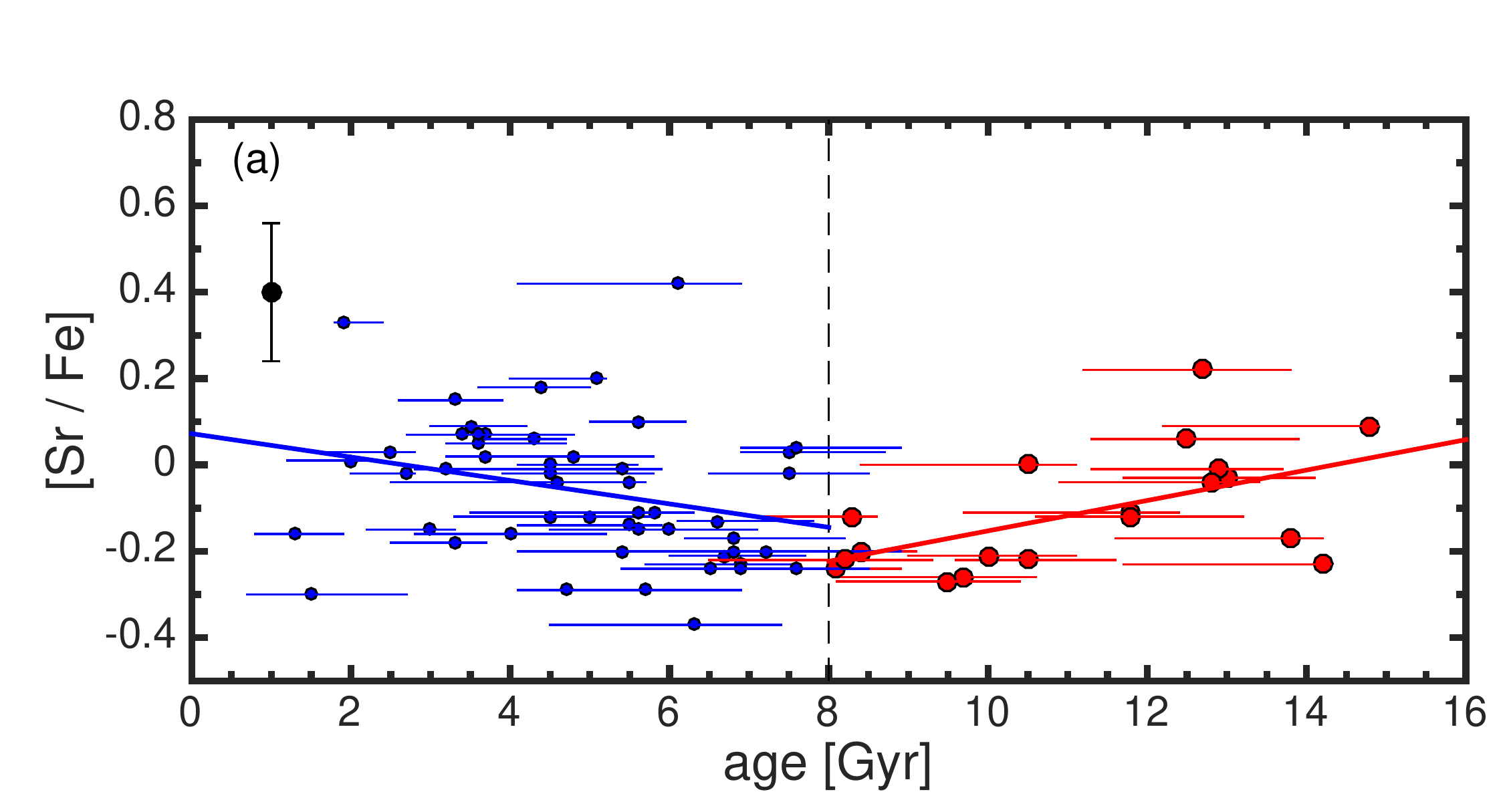}
\includegraphics[viewport=0 250 670 600,clip]{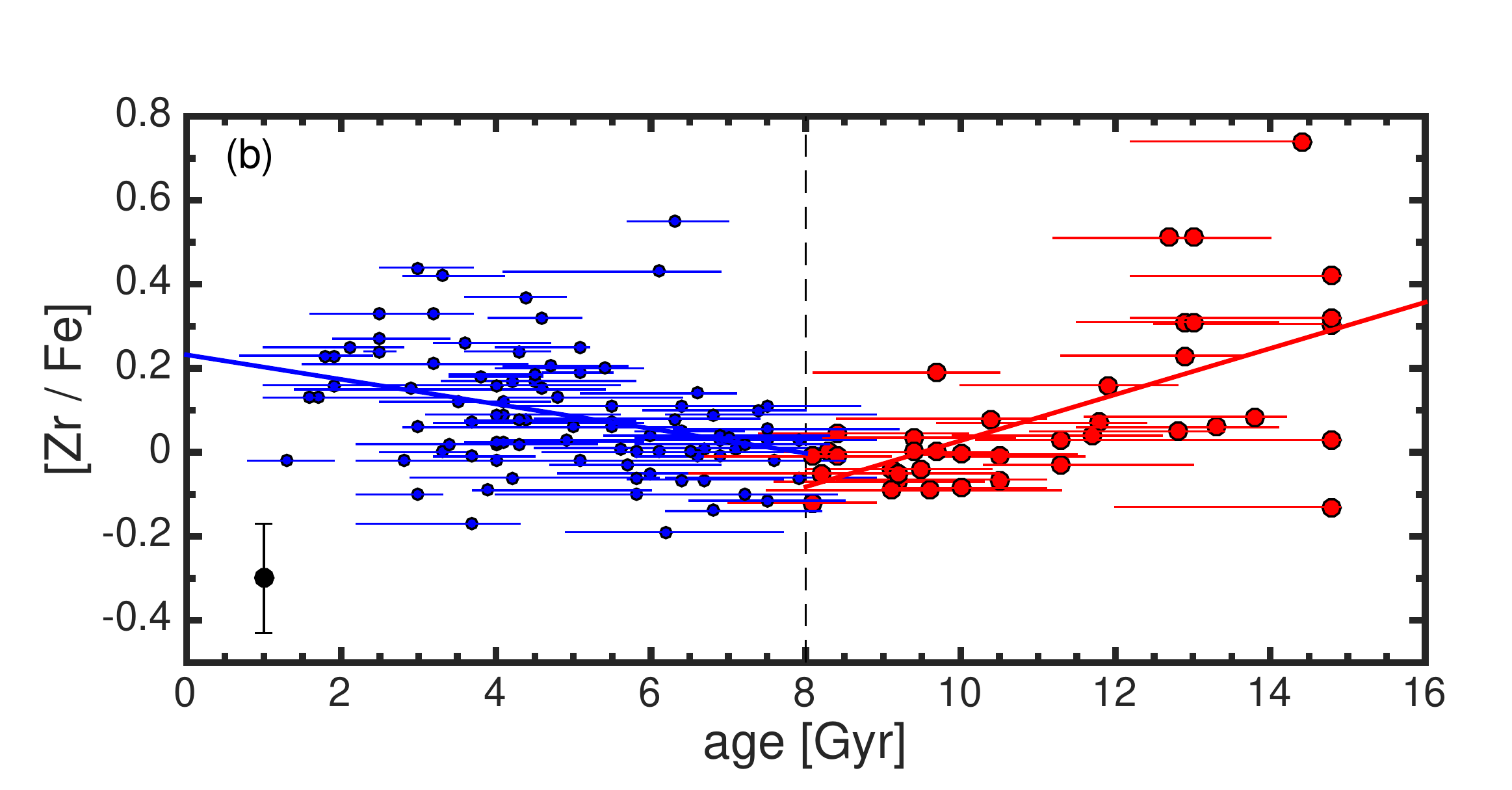}}
\resizebox{\hsize}{!}{
\includegraphics[viewport=0 250 650 560,clip]{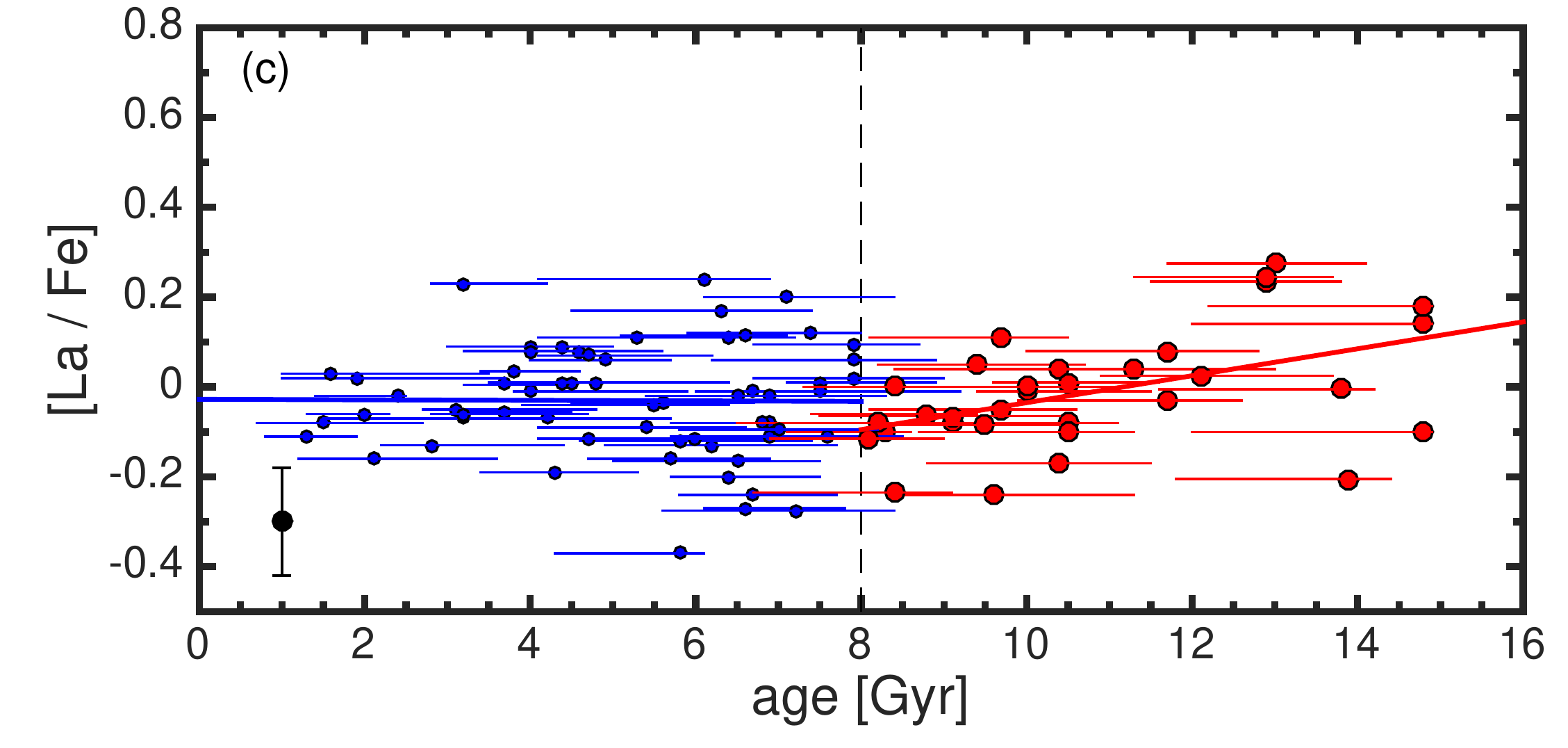}
\includegraphics[viewport=0 250 670 560,clip]{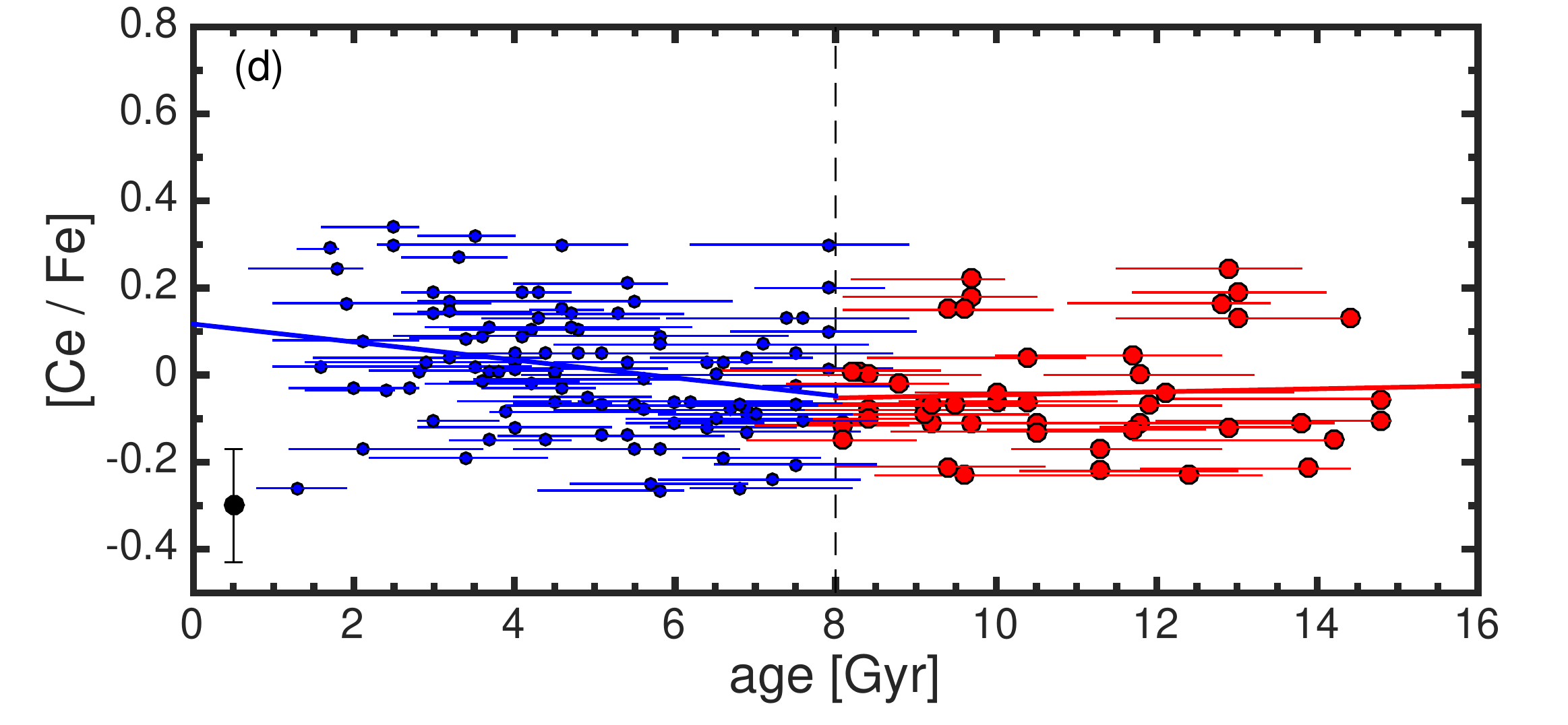}}
\resizebox{\hsize}{!}{
\includegraphics[viewport=0 250 650 560,clip]{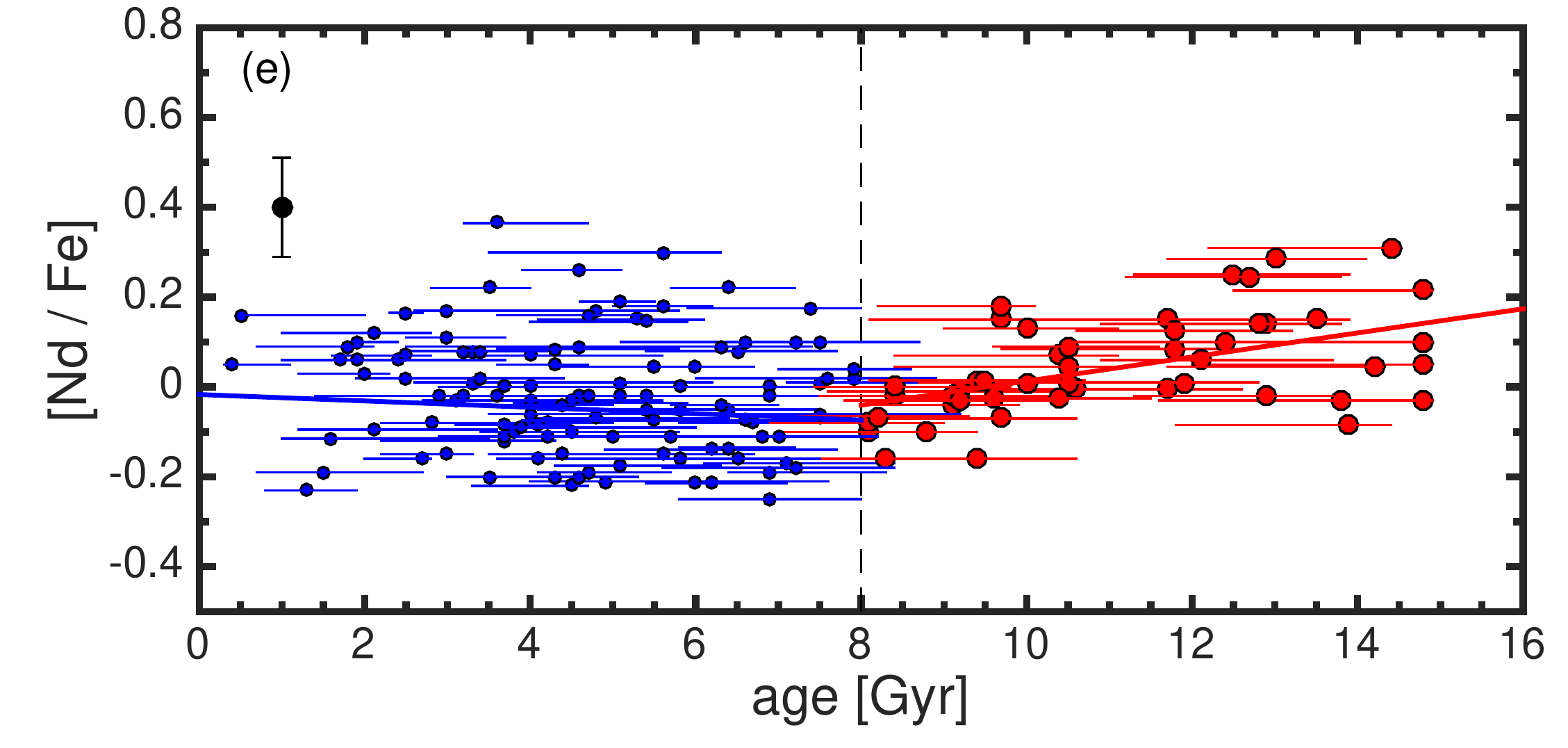}
\includegraphics[viewport=0 250 670 560,clip]{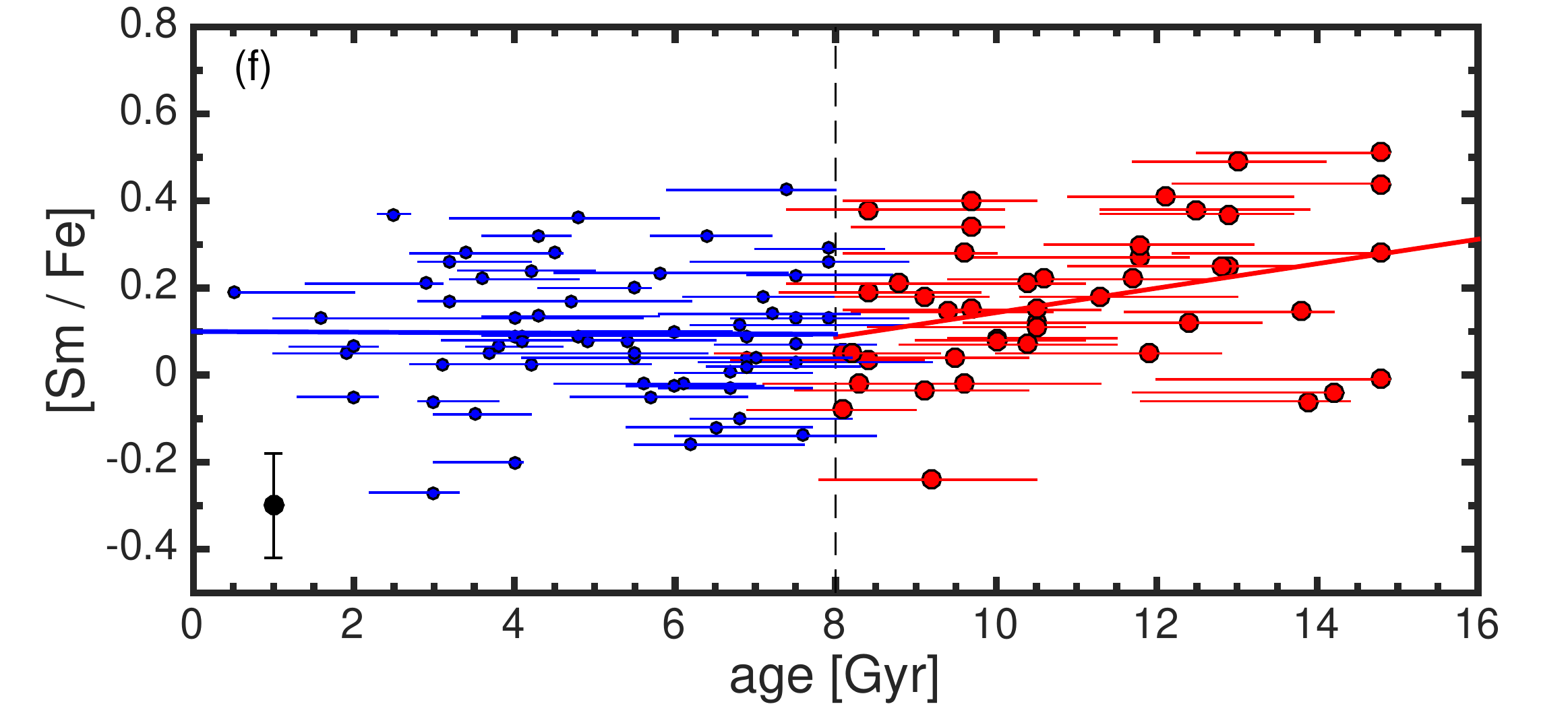}}
\resizebox{\hsize}{!}{
\includegraphics[viewport=-325 250 975 560,clip]{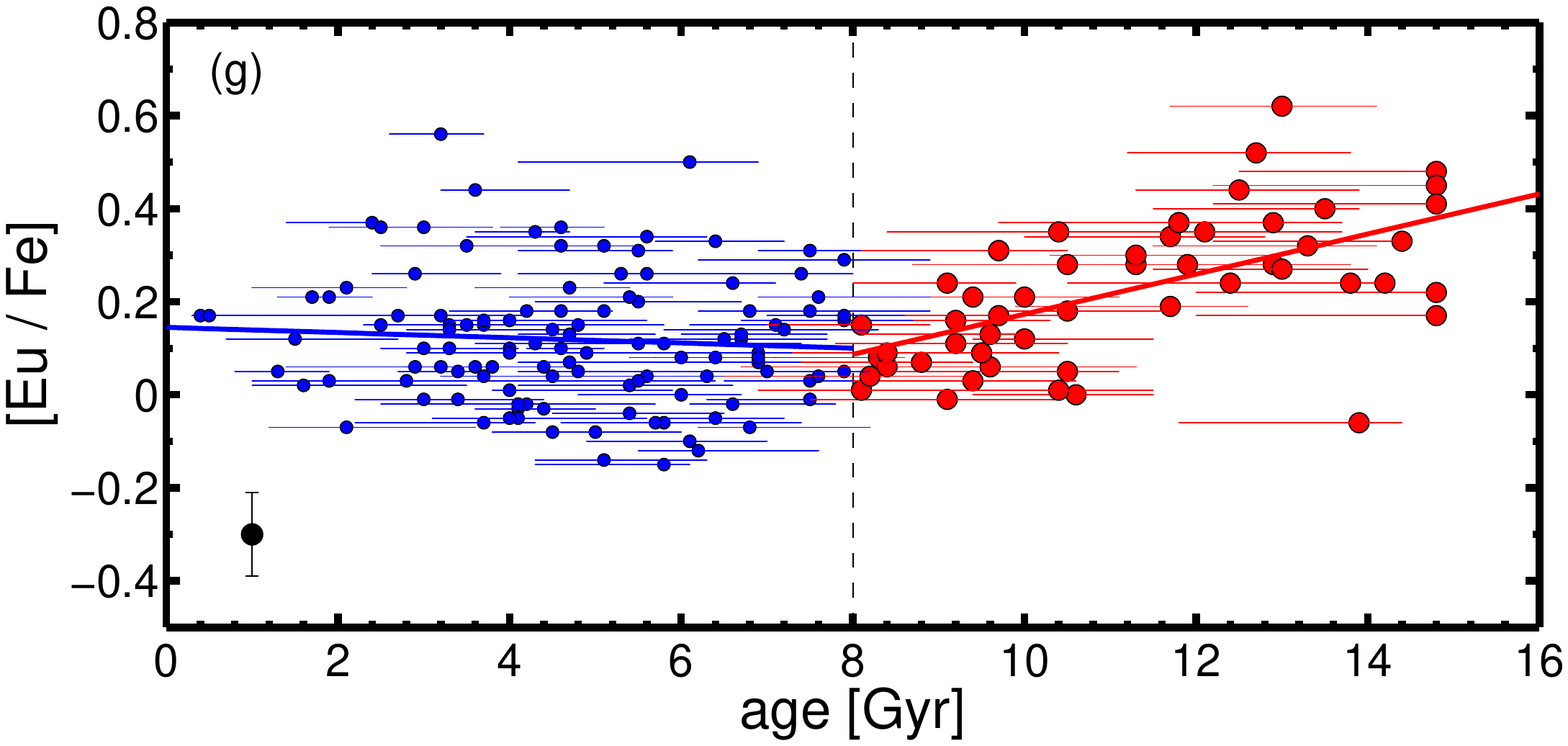}}
\caption{[$X$/Fe] for neutron-capture elements compared to age. Only stars with age uncertainties less than 3\,Gyr are plotted. The vertical dot line at 8 Gyr indicate an approximate age separation between thin and thick disk. Blue dots represent young thin disk stars while red dots are for old thick disk stars. The blue and red lines indicate the best fit for thin and thick disk stars. The errors on the ages are from \cite{Bensby2014}. The average error on the abundance ratio is indicated in black. \label{fig:abundance_age}}
\end{figure*}

\cite{Travaglio2004} compared their galactic chemical evolution model with observations down to $\rm [Fe/H]\approx -4$. The model consists of Sr produced of 71\,\% from low-intermediate mass AGB stars (1-8 M$_{\odot}$) defined as {\it main} s-process, and for 9\,\% from advanced evolution phases of massive stars defined as {\it weak} s-process. For Zr they derive 65\,\% {\it main} s-process contribution and an almost negligible contribution from the {\it weak} s-process of about 2\,\%. From their results, \cite{Travaglio2004} claim an r-process contribution of $\approx 20$\,\% and $\approx 30$\,\% that is higher than what derived by \cite{Arlandini1999}. However, in their comparison with r-process rich and very metal-poor stars, they derive that pure r-process production for Sr and Zr is $\approx 10$\,\% (more precisely 12\,\% for Sr and 15\,\% for Zr). Summing up the last derived r-contribution with the s- contribution mentioned above, it can be noticed that 8\,\% of Sr and 18\,\% of Zr is missing and it is of ``primary'' origin from massive stars at low metallicity. Unfortunately the real process for this LEPP (lighter element primary process) is still not clear because detailed supernovae model calculations for massive stars at low metallicity are still not available \citep{Travaglio2004}. More recently, \cite{Bisterzo2014} re-calculated the contribution for Sr and Zr, and found similar results as \cite{Travaglio2004}, still requiring LEPP as well. The different LEPP contribution for Sr and Zr could be the explanation of the $\approx$\,0.2 dex between Sr and Zr, where the higher plateau of [Zr/Ba] could be explained with the higher contribution from LEPP, i.e. more enrichment of Zr from high massive stars at low metallicity.

The comparison with the \cite{Travaglio2004} model, however, is unable to match our data since it predicts a solar [Sr/Ba] and [Zr/Ba] down to $\rm [Fe/H] \approx -1$. A possible explanation for this discrepancy can come from the large uncertainties that still exist for the yields of heavy elements from AGB stars due to missing models for stars in the range 1-8 M$_{\odot}$ at different metallicities \citep{Karakas2014}.

\begin{figure*}
\centering
\resizebox{\hsize}{!}{
\includegraphics[viewport=0 0 650 350,clip]{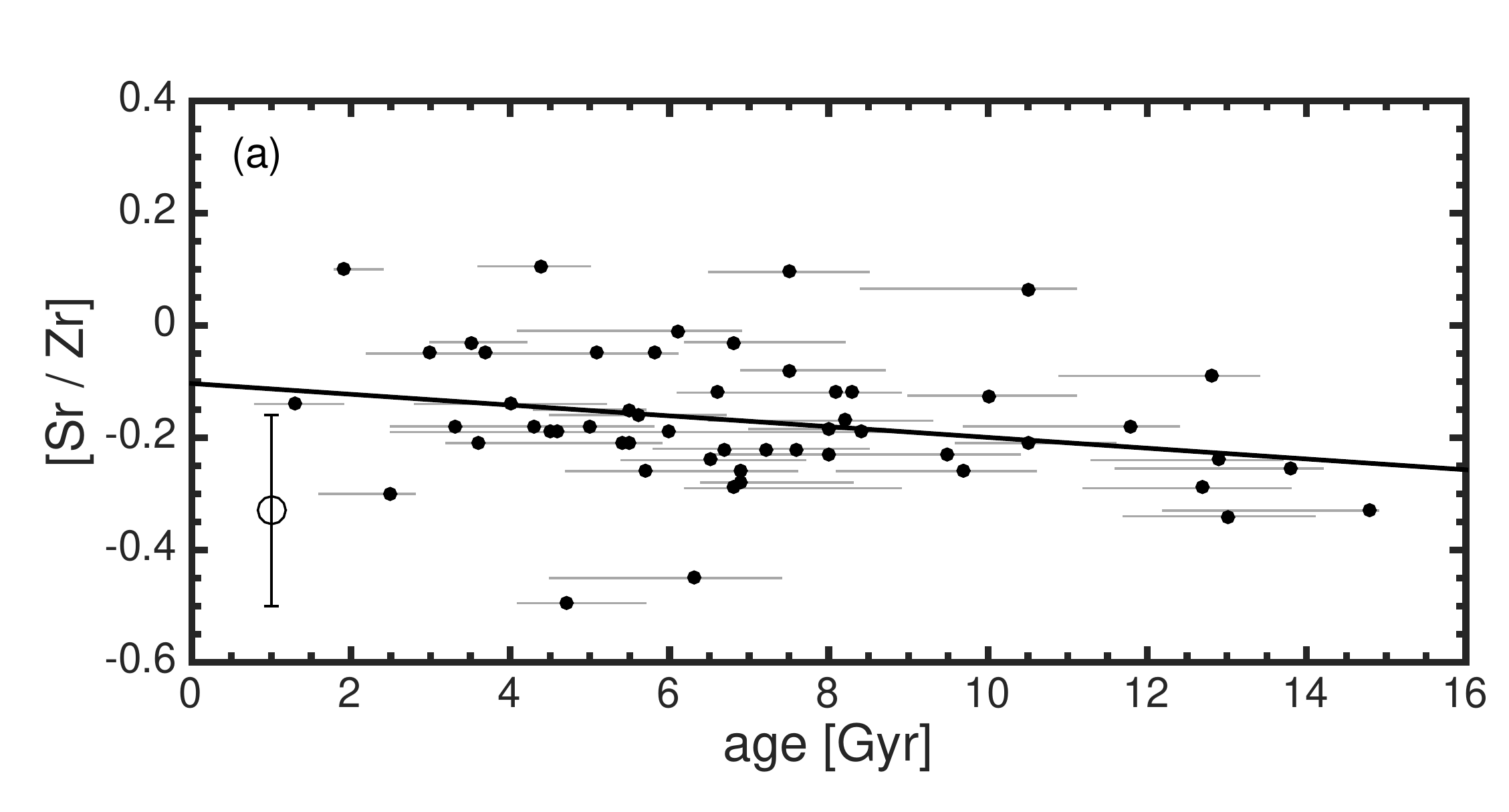}
\includegraphics[viewport=0 0 670 350,clip]{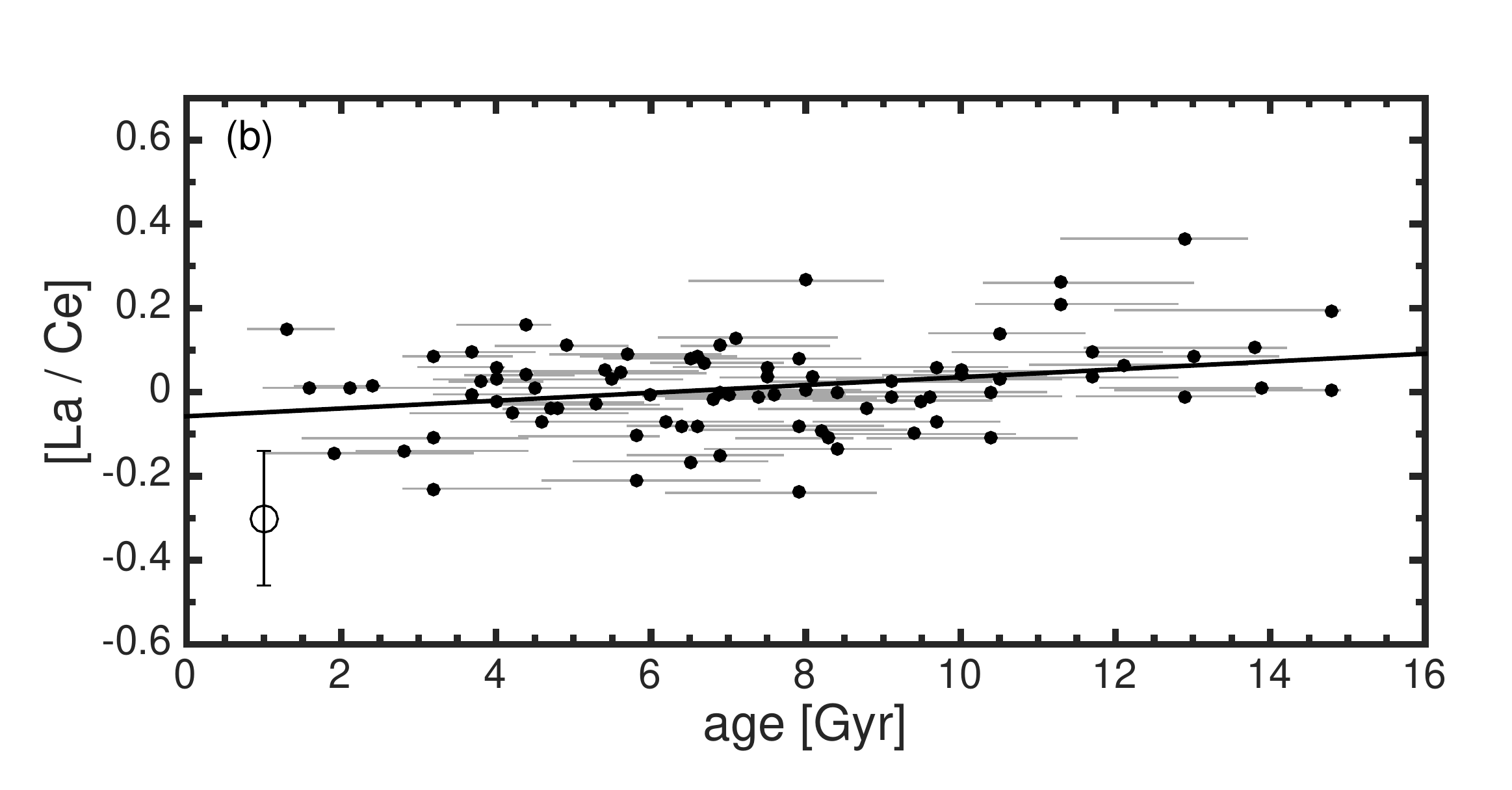}}
\resizebox{\hsize}{!}{
\includegraphics[viewport=0 0 650 315,clip]{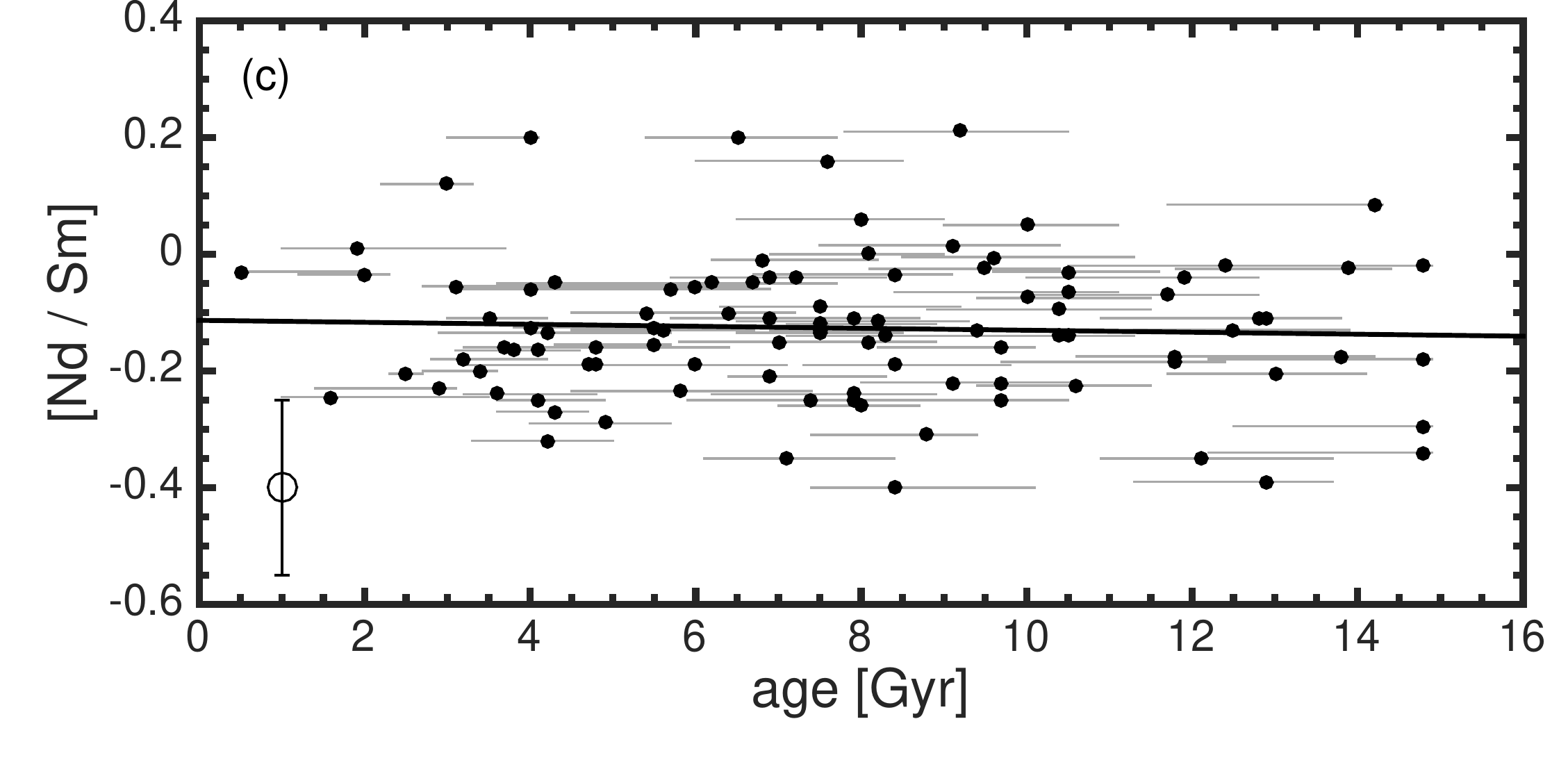}
\includegraphics[viewport=0 0 670 315,clip]{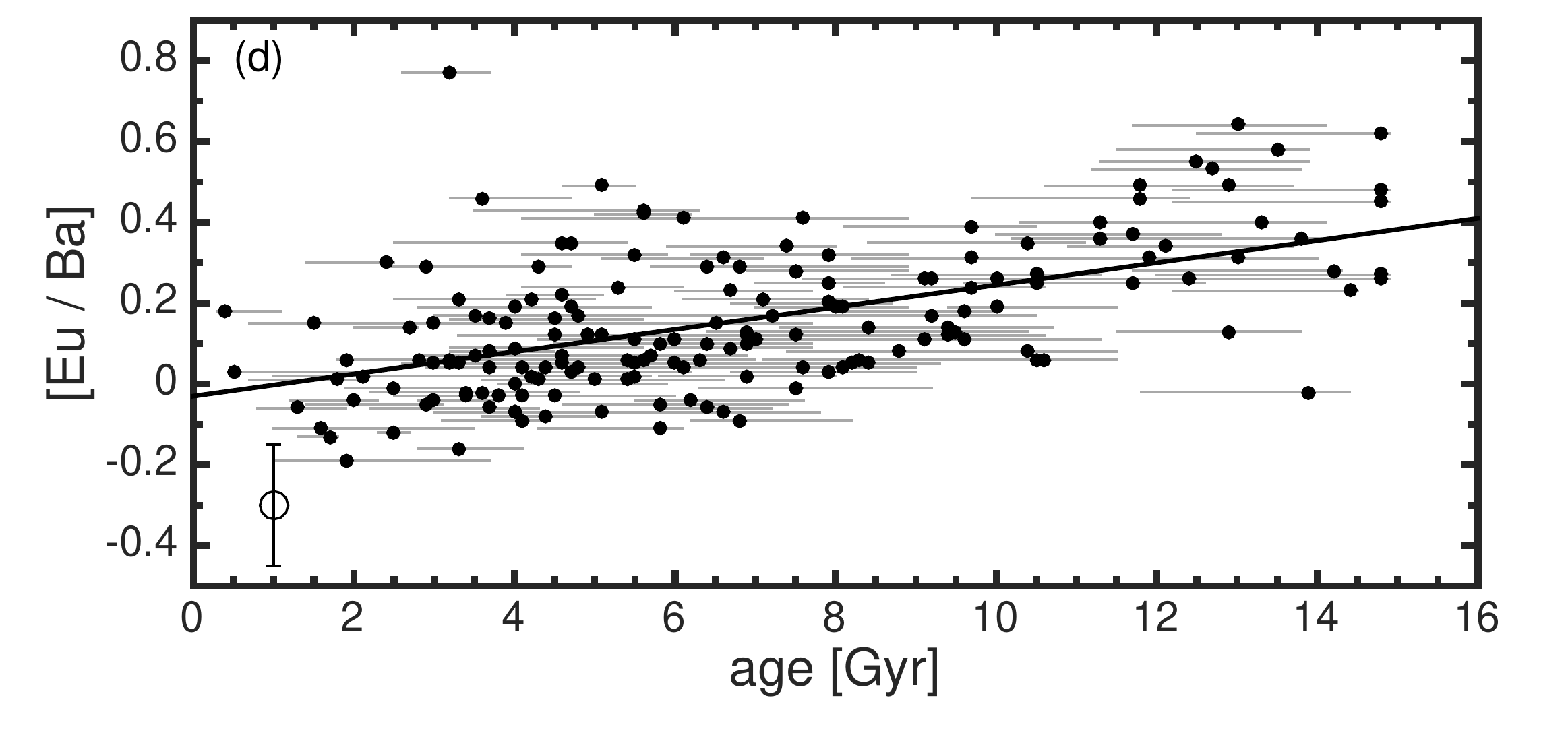}}
\caption{[Sr/Zr], [La/Ce], [Nd/Sm] and [Eu/Ba] compared to age. Only stars with age uncertainties less than 3\,Gyr are plotted. The black line in each plot represent the best fit. The errors on the ages are taken from \cite{Bensby2014}. The average error on the abundance ratio is indicated in the lower left part of each plot. \label{fig:abundance_age_ratio}}
\end{figure*}

\subsection{La, Ce, Nd and Sm}

Figures.~\ref{fig:LaCe_vs_FeH}a-d show [La/Ba], [Ce/Ba], [La/Eu], and [Ce/Eu] versus [Fe/H]. La and Ce are part of the second  peak of the magic neutron number 82 together with Sm and Eu. From Fig.~\ref{fig:abundance_full}c and Fig.~\ref{fig:abundance_full}d La and Ce do not show any particular trend with metallicity and this is also reflected in the upper panels of Figs.~\ref{fig:LaCe_vs_FeH}. In the lower panels of Fig.~\ref{fig:LaCe_vs_FeH}, [La/Eu] and [Ce/Eu] are plotted over [Fe/H] to evaluate their small r-process enrichment branch and the trends look similar. Compared to Sr and Zr, for example, it is possible to see that the increases in [La/Eu] and [Ce/Eu] are fast and happens when AGB stars start to contribute. From Fig.~\ref{fig:LaCe_vs_FeH}d it is visible a change in slope at [Fe/H] $\approx$\,$-$0.5. This could be the moment where the enrichment of ISM from AGB stars start to dominate, represented as a steady increase in [Ce/Eu] as [Fe/H] increases toward solar values.

A similar investigation was performed for Nd that is produced in almost equal parts by s- and r-processes. In both panels of Fig.~\ref{fig:NdSm_vs_FeH}, Nd shows smooth trends, with high Nd abundance for metal-poor stars that decreases as metallicity increases. Interestingly the most metal-poor stars in both panels are close to the theoretical abundance ratio for r-process derived using \cite{Bisterzo2014} values. Very similar results are derived in Nd work by \cite{Mashonkina2004}.
The decrease that is present in [Nd/Ba] is due to the higher rate of s-production of Ba compared to Nd (81\,\% for Ba and 56\,\% for Nd), while the decrease in [Nd/Eu] is produced by the higher rate (93\,\%) of r-production for Eu compared to Nd. Similar results can be found for Sm due to its high percentage of r-process production. In Fig.~\ref{fig:NdSm_vs_FeH}a, thin disk stars have on average smaller [Nd/Ba] compared to thick disk stars, and are more concentrated along solar [Nd/Ba] values. This means that during thick disk formation and evolution, the r-process contribution was higher than in the thin disk, probably because the thick disk formed in a rapid way with high star formation rate that produced more massive stars responsible for r-process enrichment.

We also investigated [Sm/Ba] and [Sm/Eu] versus [Fe/H] and the results can be seen in Figs.~\ref{fig:NdSm_vs_FeH}b and d.  From Fig.~\ref{fig:NdSm_vs_FeH}b it is clear that the most metal-poor stars in our sample are close to the pure solar r-process meaning that r-process was the main active channel for the production of Sm. This is expected, considering that Sm is produced for almost 70\,\% via rapid neutron-capture. Since Eu is an r-process, it is interesting to investigate [Sm/Eu] (Fig.~\ref{fig:NdSm_vs_FeH}d). The high percentage of production via rapid neutron-capture for both Sm and Eu, is responsible for the $\rm [Sm/Eu] \approx 0$ until solar metallicity. For super-solar metallicity, [Sm/Eu] seems to be on average higher than solar value, probably because at this metallicities regime the s-process production of Sm is more important.

The formation of the elements La, Ce, Nd and Eu was studied by \cite{Travaglio1999} and it is interesting since these elements are part of the magic neutron number N = 82. \cite{Travaglio1999} found that the main s-process contribution to these elements come from AGB stars in the range 2-4 M$_{\odot}$ while the r-process contribution is mainly due to stars in the rage 8-10 M$_{\odot}$. In their calculation, AGB contribution starts to be important for $\rm [Fe/H] \gtrsim -1.5$, meaning that after this point, the Eu abundance decreases while the abundances for the other elements increase, especially for Ce and La because of their larger s-process contribution. 

The models of \cite{Travaglio1999} are in agreement with our data, as can be seen comparing our Figs.~\ref{fig:abundance_full}c, d, e, f with their Figs.~7, 8, 10, and 11. Their model of Eu enrichment is in agreement with what we found in Fig.~\ref{fig:abundance_full}g and it can be explained by r-process derived from SN\,II from 8-10 M$_{\odot}$ stars.

\begin{figure*}
\centering
\resizebox{\hsize}{!}{
\includegraphics[viewport=0 250 650 600,clip]{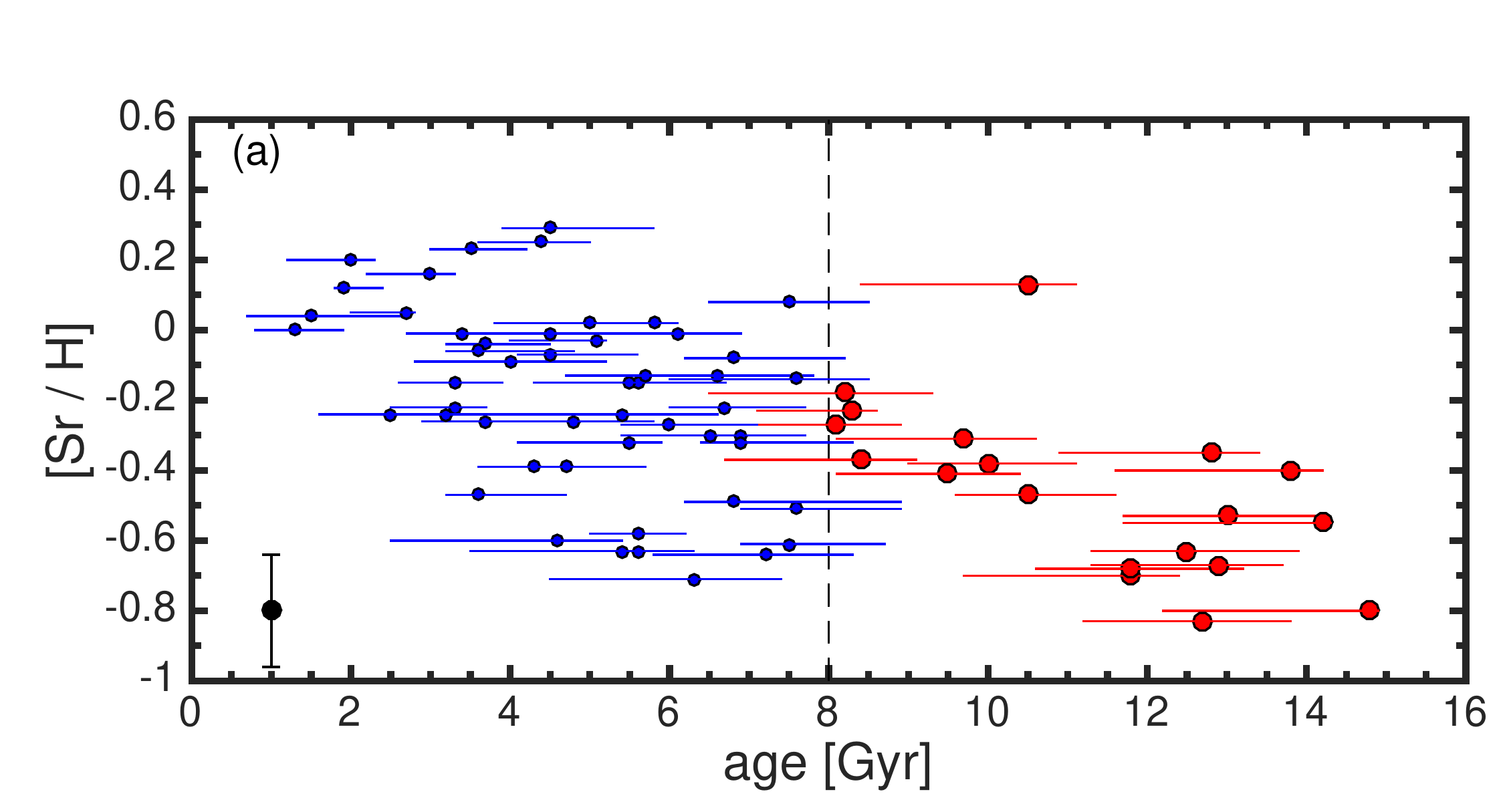}
\includegraphics[viewport=0 250 670 600,clip]{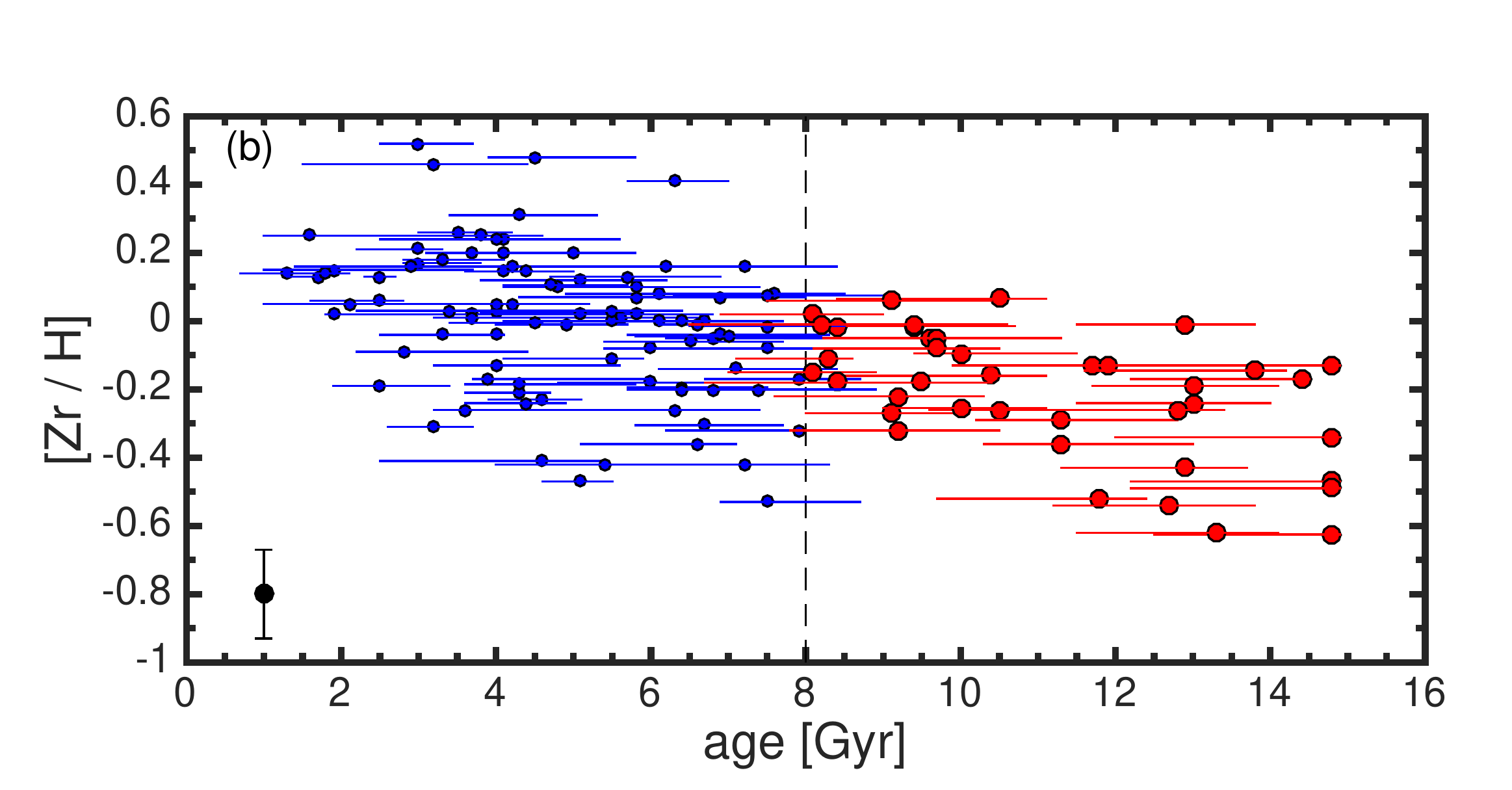}}
\resizebox{\hsize}{!}{
\includegraphics[viewport=0 250 650 560,clip]{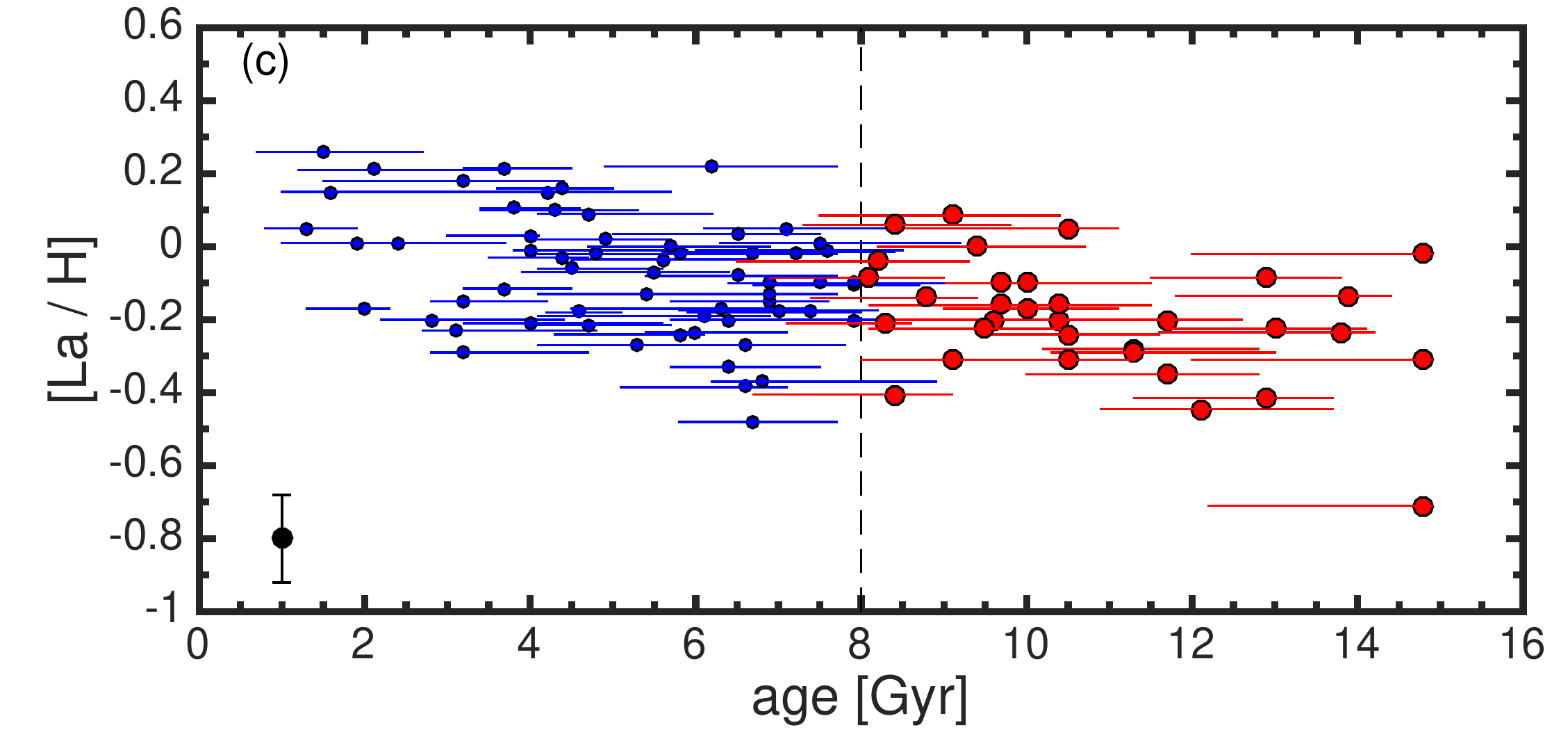}
\includegraphics[viewport=0 250 670 560,clip]{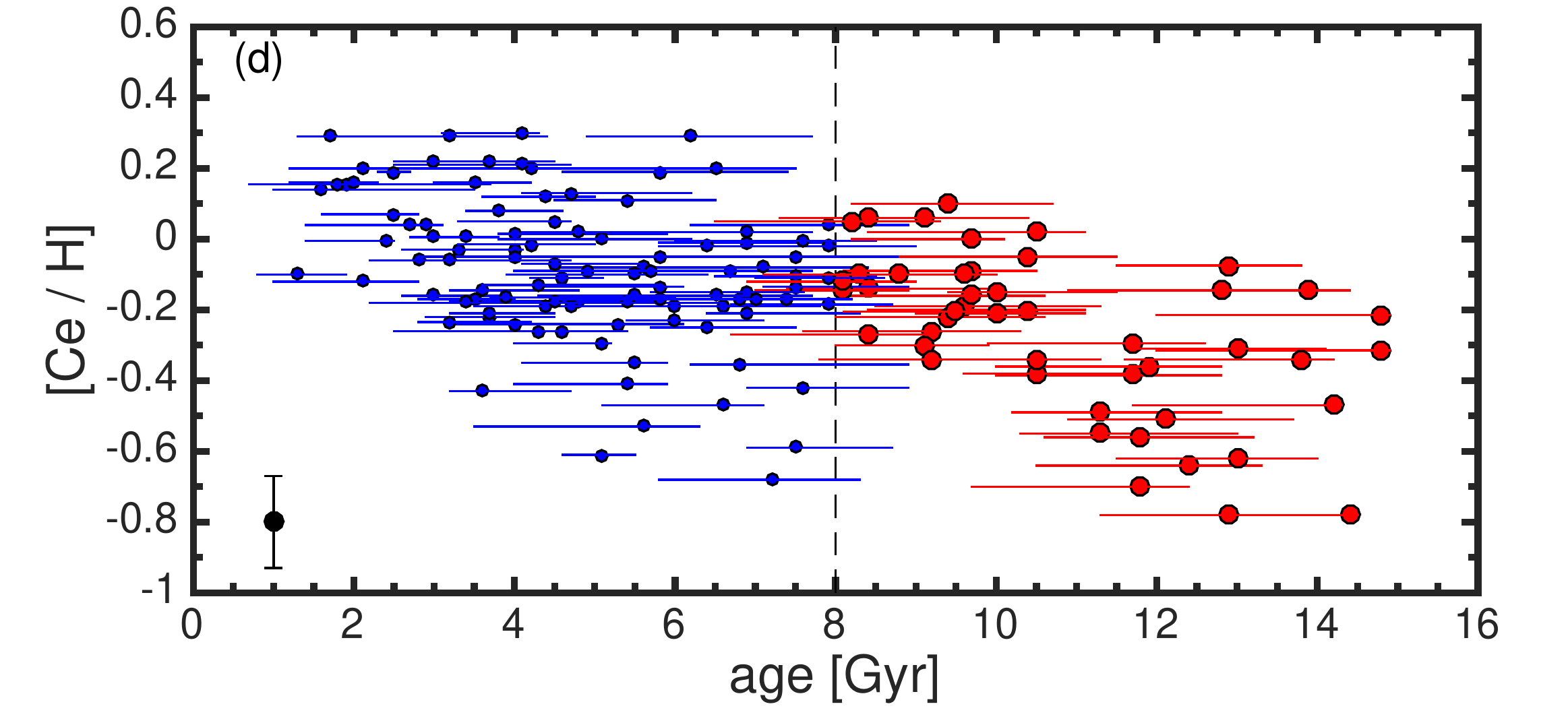}}
\resizebox{\hsize}{!}{
\includegraphics[viewport=0 250 650 560,clip]{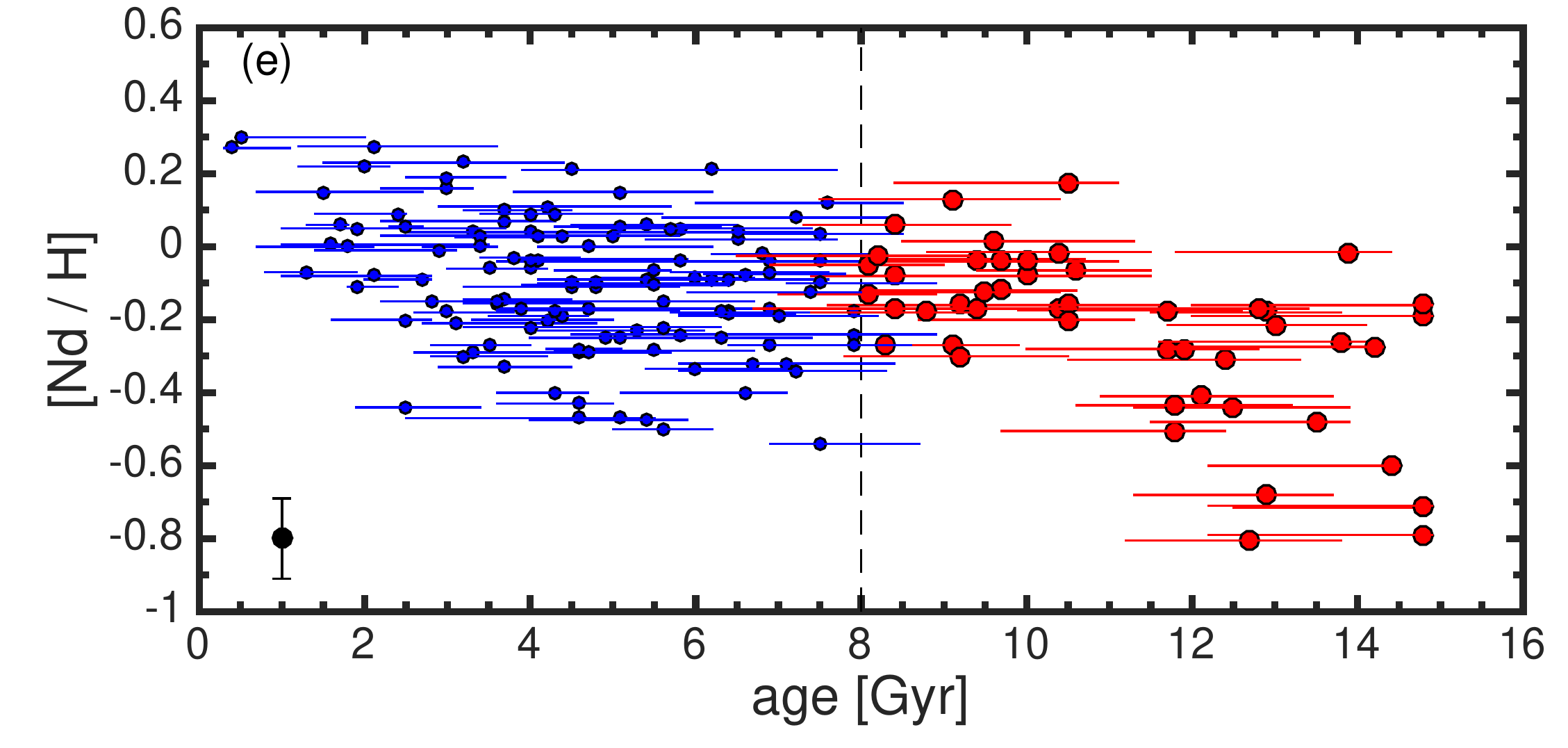}
\includegraphics[viewport=0 250 670 560,clip]{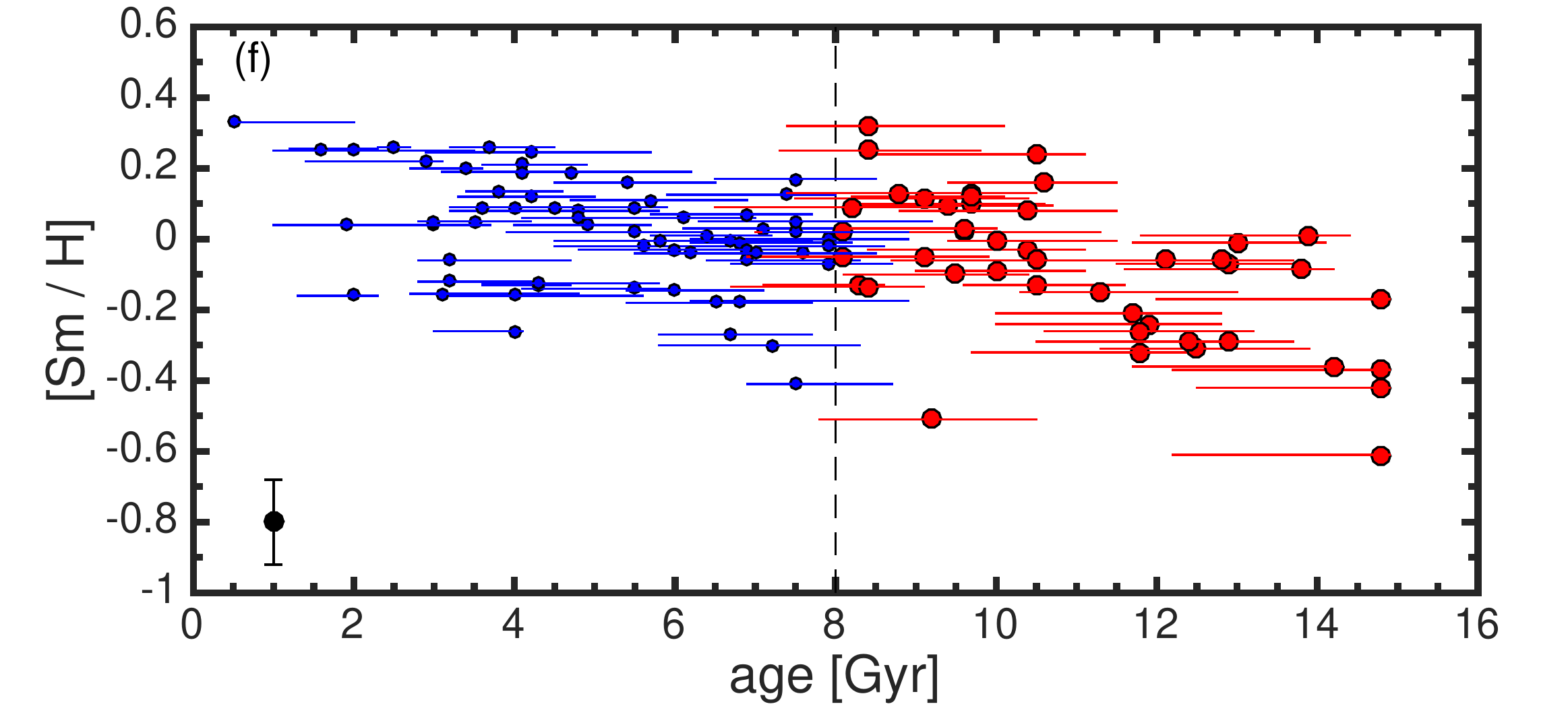}}
\resizebox{\hsize}{!}{
\includegraphics[viewport=-325 0 950 320,clip]{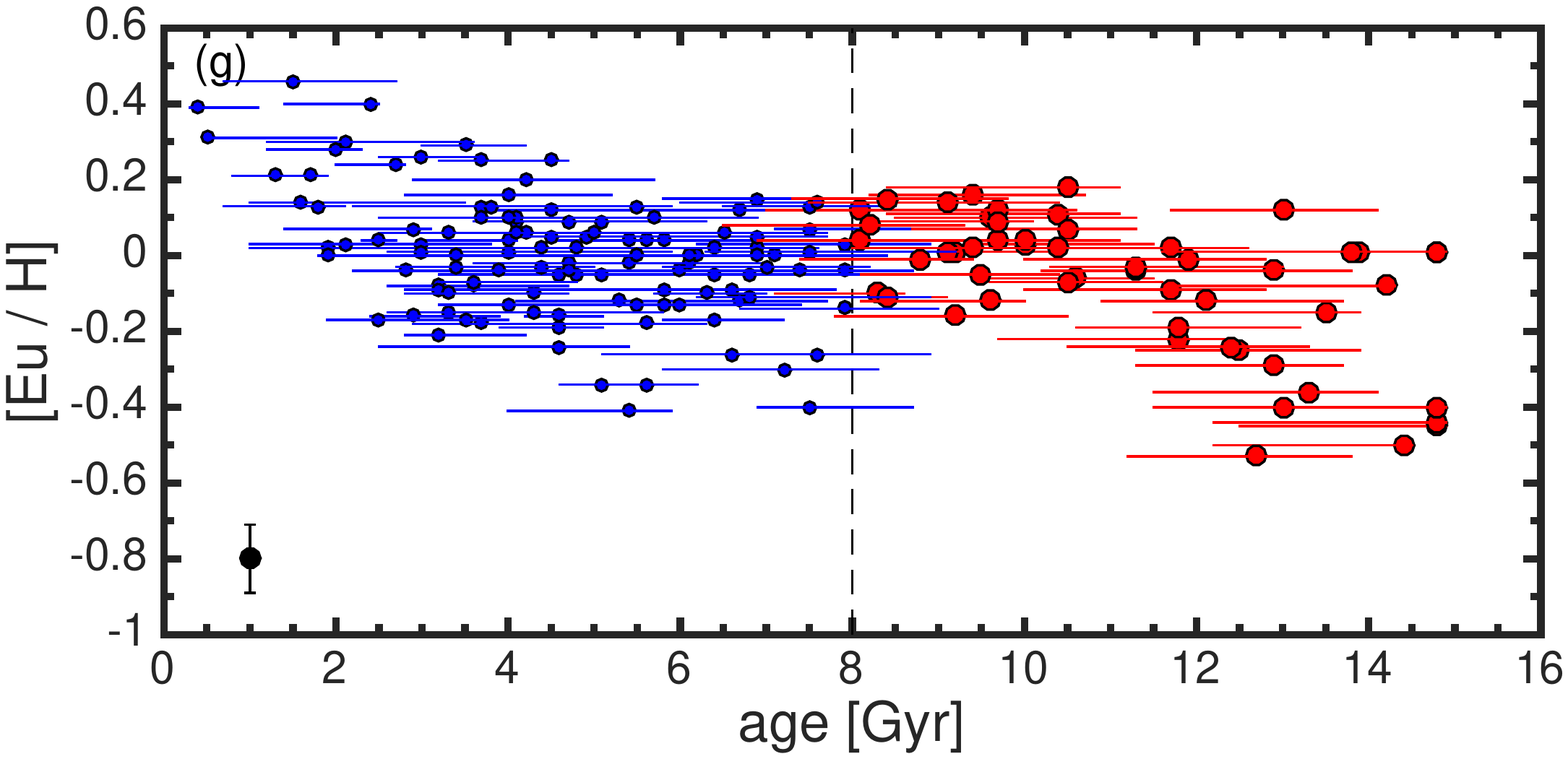}}
\caption{[$X$/H] for neutron-capture elements compared to age. Only stars with age uncertainties less than 3\,Gyr are plotted. The vertical dot line at 8 Gyr indicate an approximate age separation between thin and thick disk. Blue dots represent young thin disk stars while red dots are for old thick disk stars. The errors on the ages are from \cite{Bensby2014}. The average error on the abundance ratio is indicated in black. \label{fig:el_H_age}}
\end{figure*}

\begin{figure*}
\centering
\resizebox{\hsize}{!}{
\includegraphics[viewport=0 250 650 600,clip]{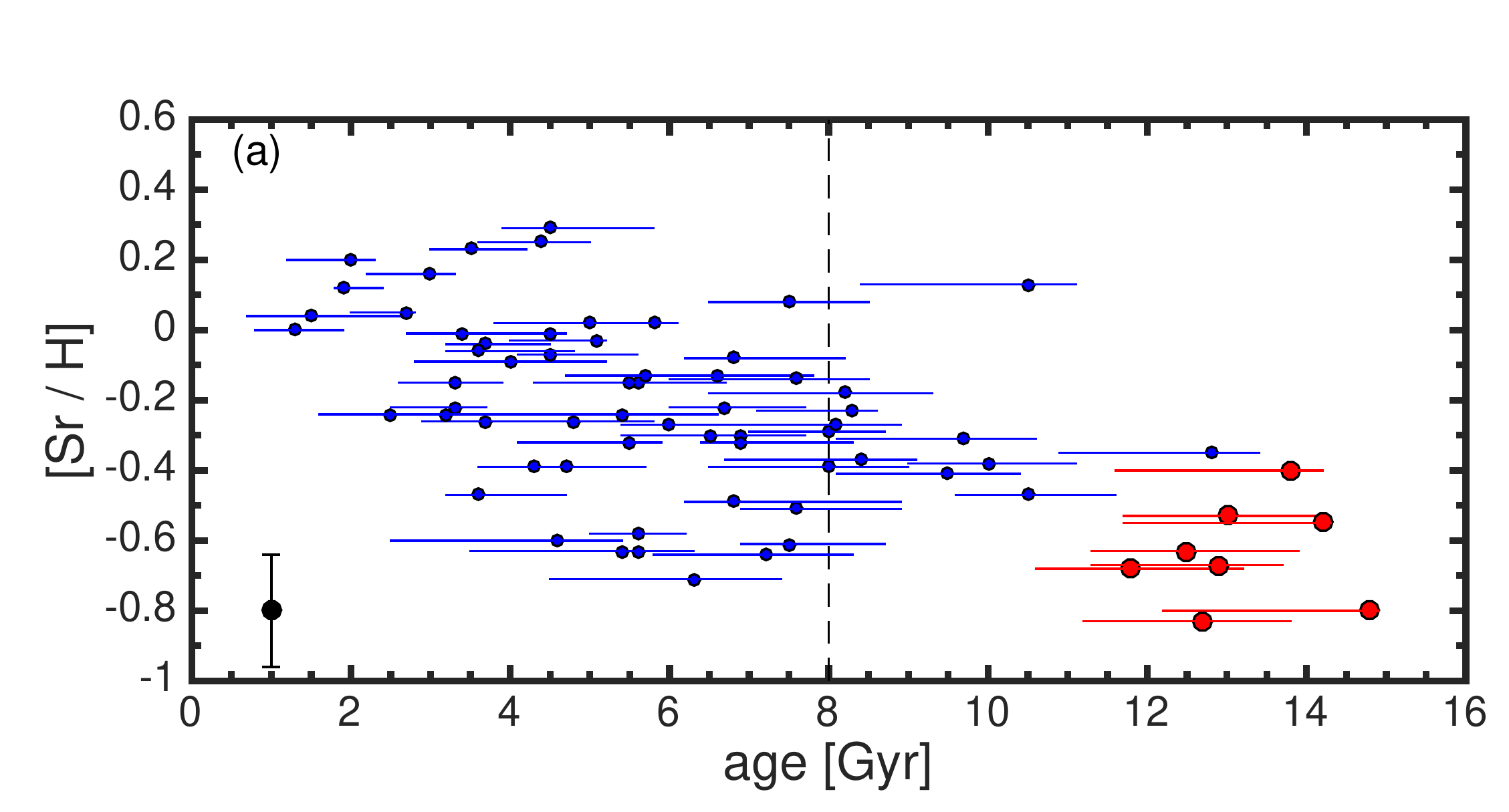}
\includegraphics[viewport=0 250 670 600,clip]{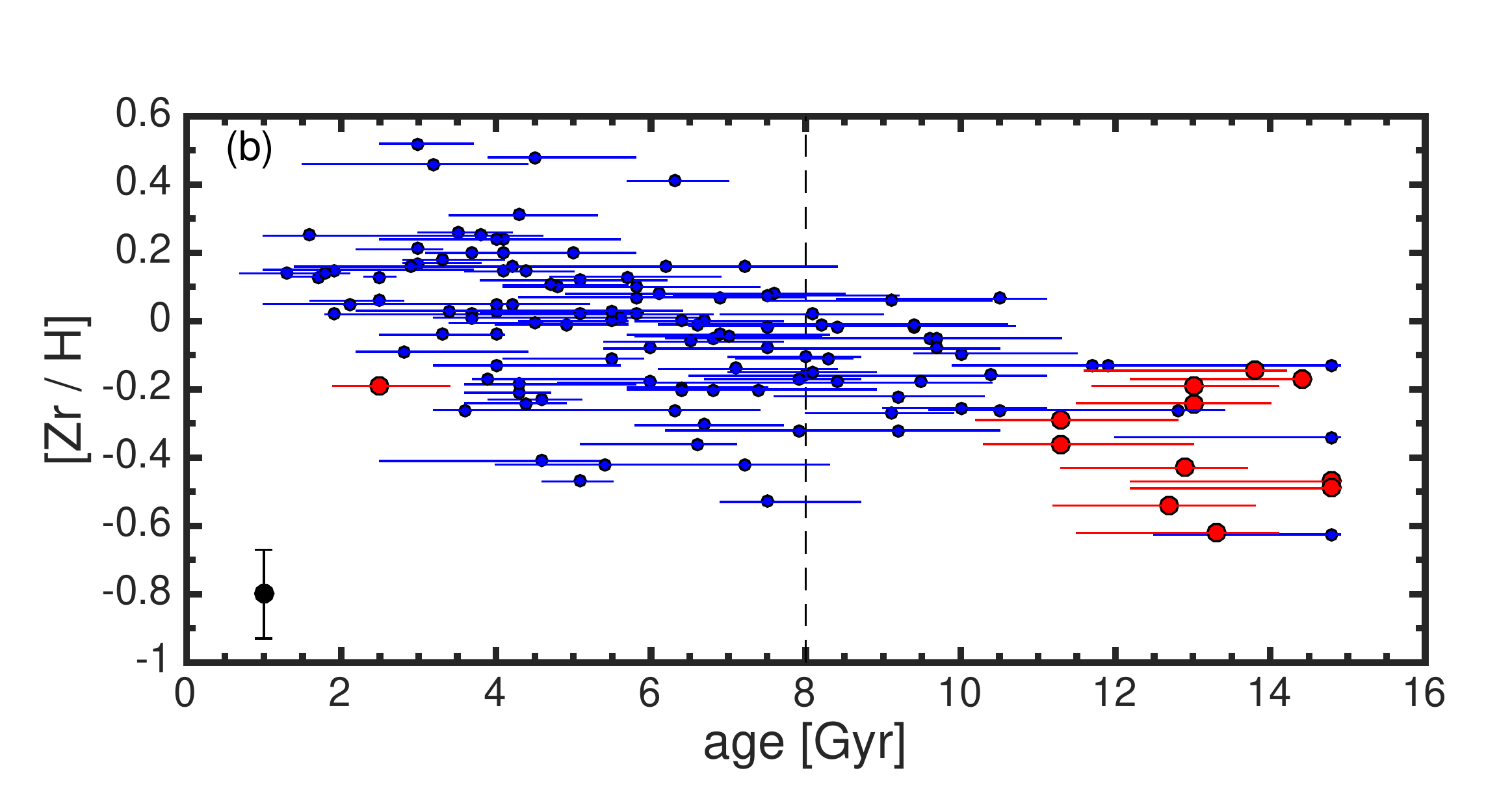}}
\resizebox{\hsize}{!}{
\includegraphics[viewport=0 250 650 560,clip]{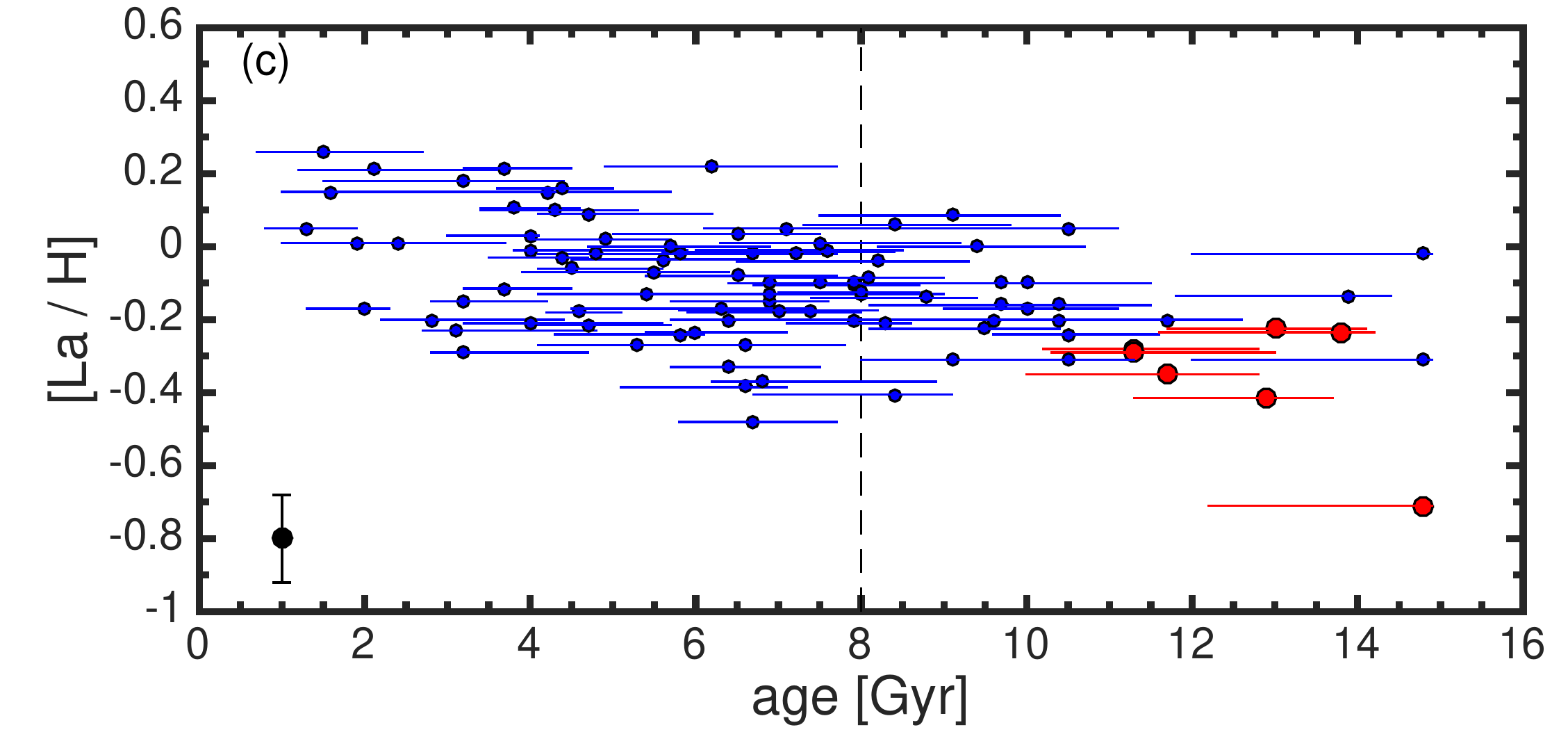}
\includegraphics[viewport=0 250 670 560,clip]{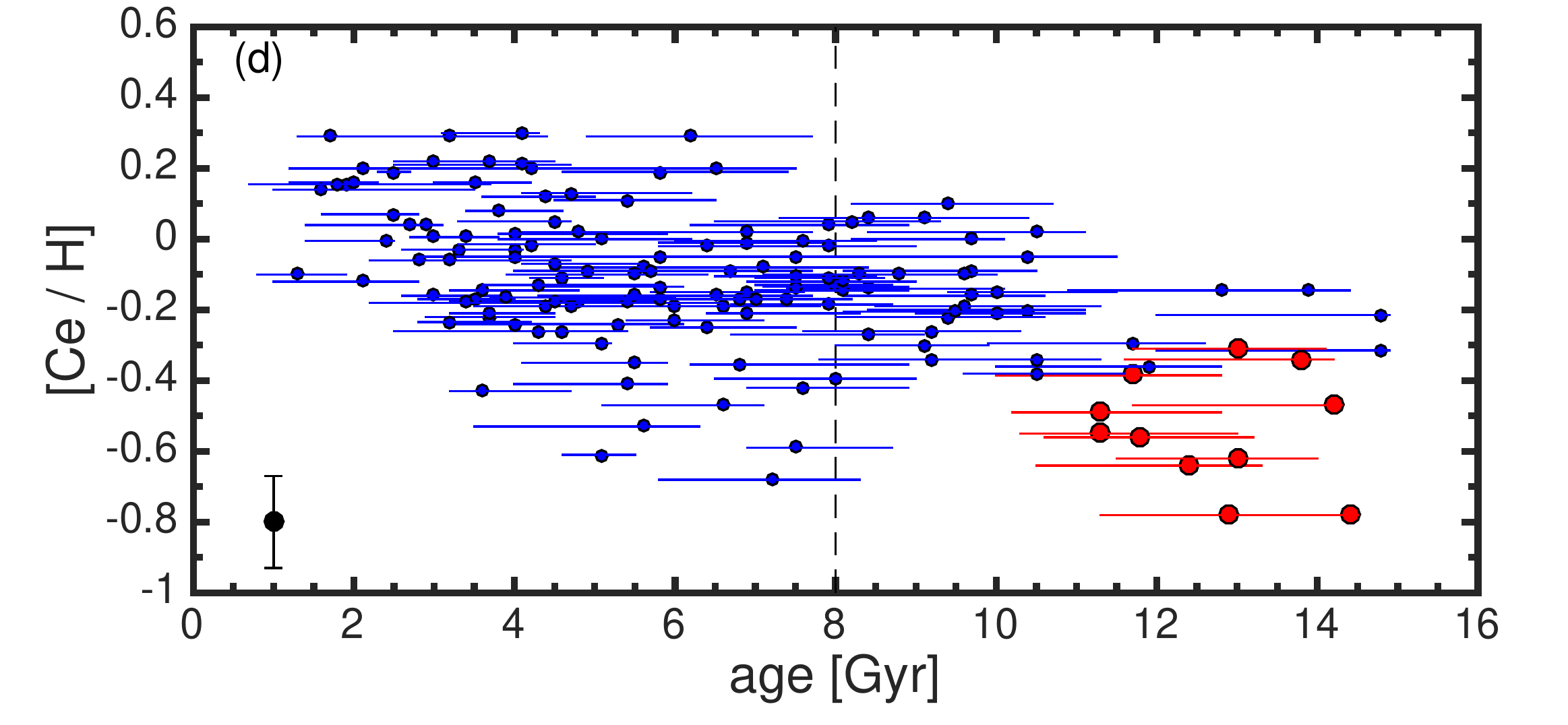}}
\resizebox{\hsize}{!}{
\includegraphics[viewport=0 250 650 560,clip]{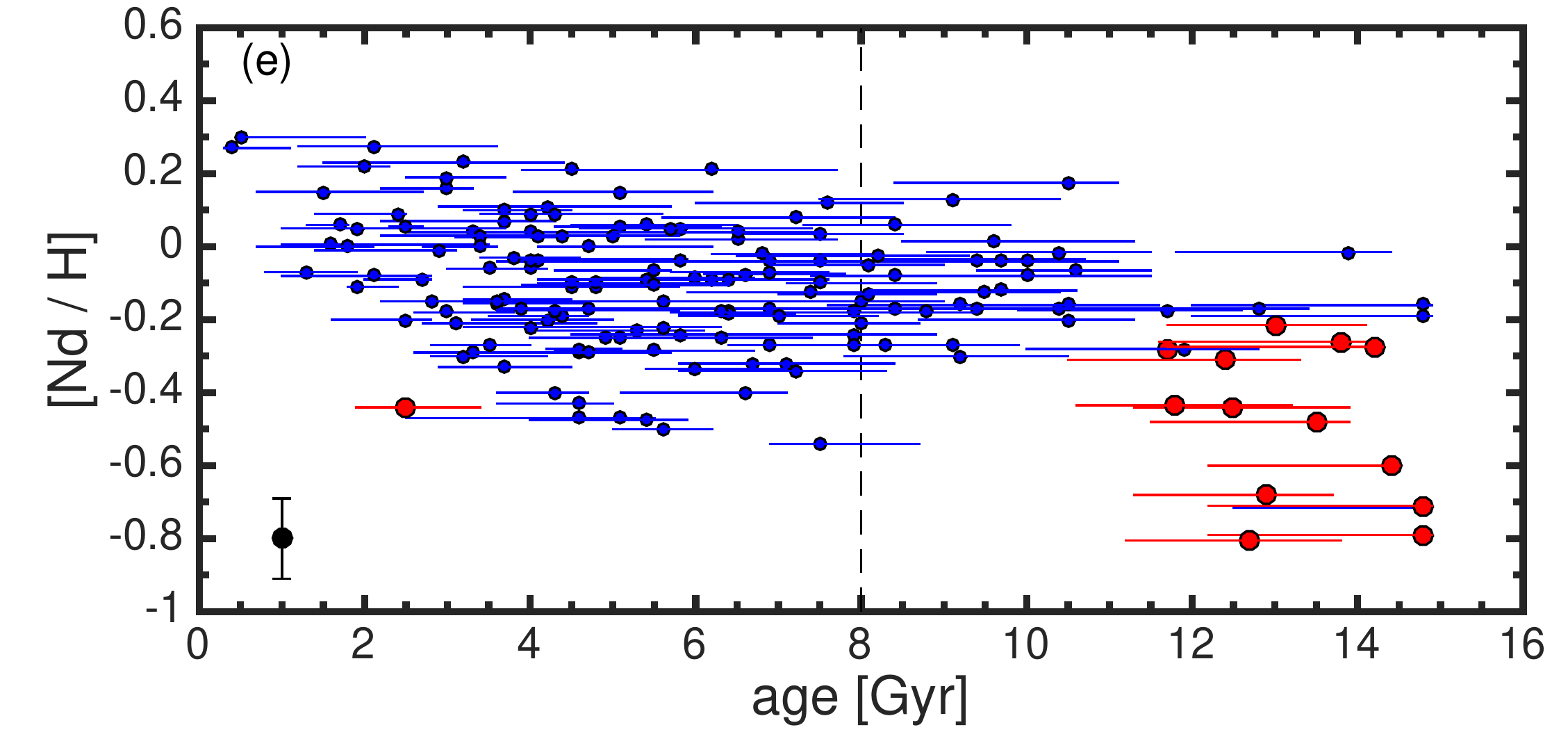}
\includegraphics[viewport=0 250 670 560,clip]{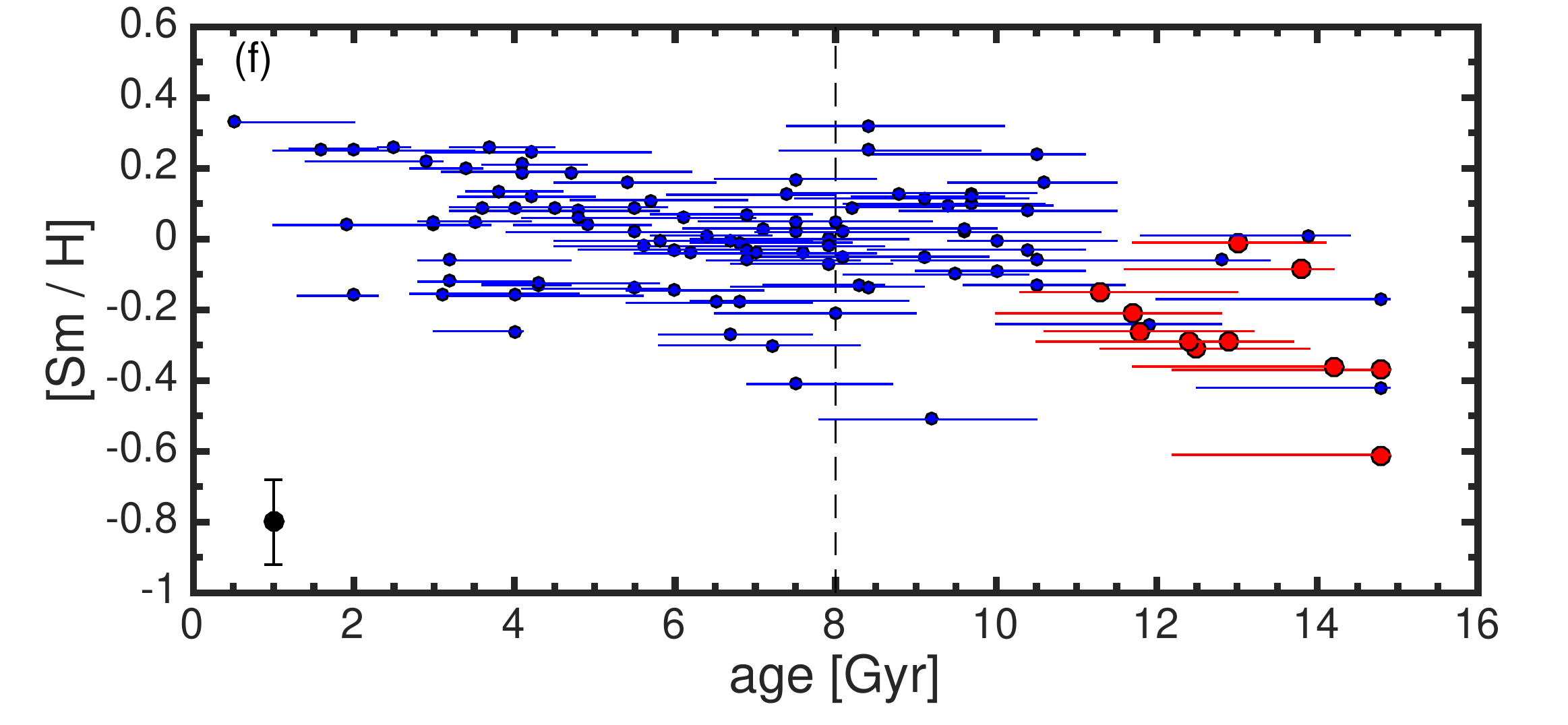}}
\resizebox{\hsize}{!}{
\includegraphics[viewport=-325 0 950 320,clip]{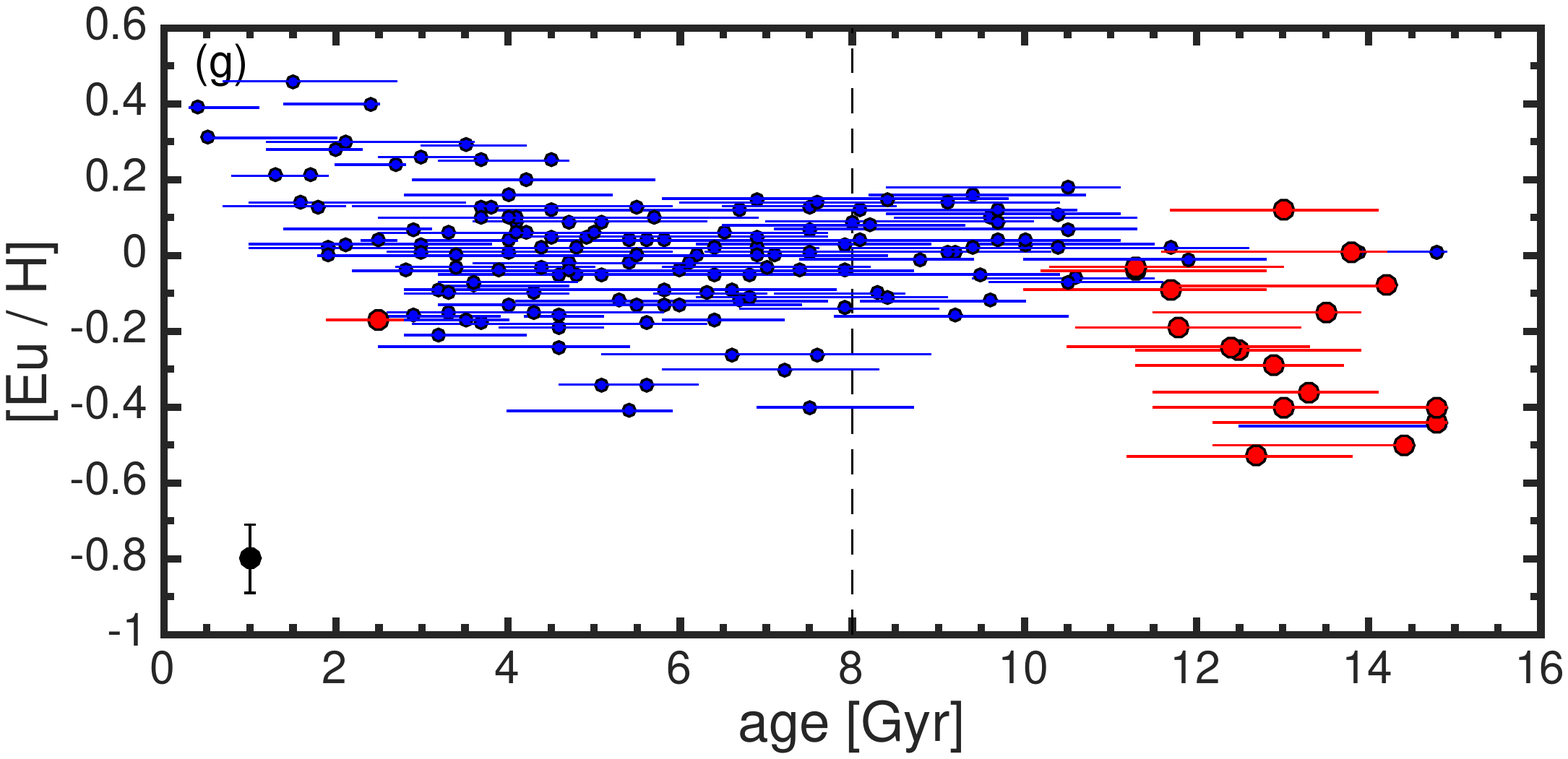}}
\caption{[$X$/H] for neutron-capture elements compared to age. Only stars with age uncertainties less than 3\,Gyr are plotted. The vertical dot line at 8 Gyr indicate an approximate age separation between thin and thick disk. Blue dots represent thin disk stars with [Ti/Fe] < 0.2 while red dots are for thick disk stars identified with [Ti/Fe] > 0.2. The average error on the abundance ratio is indicated in black.
 \label{fig:el_H_age_alpha}}
\end{figure*}

\section{Abundance evolution with age}

Since the r- and s-processes happen at different timescale in Galactic evolution, we now investigate how the [$X$/Fe] abundance ratios vary with time. Only stars with stellar ages with uncertainties less than 3\,Gyr are used (derived as the maximum age estimation minus the minimum one). As before, stars with ages less than 7\,Gyr are considered to be part of the thin disk, while stars older than 9\,Gyr are part of the thick disk. We stress here that the stellar ages are really accurate in relative terms, meaning that the relative age differences between stars is good.  

The results are presented in Fig.~\ref{fig:abundance_age}. The first thing that can be noticed is that in most of the cases there is a change in the slopes at an age of about 8\,Gyr, which happens coincide with the interface between the thin and thick disks. For the thick disk (stars older than 8 Gyr) the indication that we get from Fig.~\ref{fig:abundance_age} is that enrichment  of neutron-capture elements was higher at the early times of the Milky Way, so the r-process could produce large quantities of neutron-capture elements, even if only thanks to stars with masses in the 8-10 M$_{\odot}$ range. This could be explained assuming an intense and rapid star formation in the process of building up of the thick disk that ended around 8 Gyr ago. The decrease is then due to a reduction of r-process production with at the same time the beginning of s-process production. The basically flat [$X$/Fe] in thin disk for La, Ce, Nd, Sm and Eu can be explained as a combination of the fact that the contribution from low- and intermediate mass stars in AGB phase become more important, but this contribution is similar to Fe production from SN\,Ia, meaning a flattening in the ratio. In addition to this, since the s-process in the AGB phase requires seed nuclei of iron-group elements, the increase in metallicity of younger stars would produce an increase in the yields of neutron-capture elements. 

Interestingly, for Sr and Zr we see in Fig.~\ref{fig:abundance_age}a and Fig.~\ref{fig:abundance_age}b that abundances increase for younger thin disk stars. This phenomenon can be explained, as before, by the fact that the s-process is the main responsible for enrichment in thin disk, but with the addition that these two elements present the highest s-process rate among the other neutron-capture elements in this study (70\,\% for Sr and 65\,\% for Zr, as derived by \citep{Bisterzo2014}). Unfortunately, at this moment we cannot explain the high spread in [Zr/Fe] for the very old stars.

Considering the results in Figs.~\ref{fig:SrZr_vs_FeH},~\ref{fig:LaCe_vs_FeH}, and ~\ref{fig:NdSm_vs_FeH}, neutron-capture elements can be paired as they seem to share similar properties. For this reason we investigate how the abundance ratios of these element pairs change at different ages. The results are shown in Fig.~\ref{fig:abundance_age_ratio}, with the addition of the [Eu/Ba] ratio. 

As expected for elements that share the same production sites, the trends are almost flat, within the uncertainties. For [Sr/Zr] in Fig.~\ref{fig:abundance_age_ratio}a, there are indications for higher Zr production at early stage than Sr. A possible explanation could be that Zr is claimed to have an higher r- process production (15\,\% from classical r- process plus 18\,\% from LEPP, \citealt{Travaglio2004}) that will produce more Zr in the early stages of the evolution of the Galaxy. The [La/Ce] ratio in Fig.~\ref{fig:abundance_age_ratio}b presents a flat trend, indicating shared sites and mechanisms of production. For [Nd/Sm] in Fig.~\ref{fig:abundance_age_ratio}c the trend is basically flat, also in this case indicating shared production sites.

The trend of $\mathrm{[Eu/Ba]}$ can be explained in a different way. Eu and Ba have clearly different histories of productions and here are paired because they are typical representatives of the r- and s-processes, respectively. In Fig.~\ref{fig:abundance_age_ratio}d the decrease of [Eu/Ba] towards younger ages can be explained as an increase in production of Ba and a simultaneous decrease in Eu because AGB stars become more important than SN\,II after about 1\,Gyr from the bulk of star formation. 

We also investigate the trends of [$X$/H] with respect to age. This kind of investigation is not present in previous works since the age determination was not good enough. Results are presented in Fig.~\ref{fig:el_H_age}, using the same color coding as in Fig.~\ref{fig:abundance_age}, with stars with an age greater than 9\,Gyr considered as thick disk and stars younger than 7\,Gyr thin disk.

It is possible to see that in all the elements, apart for Sr probably due to the big scatter and the lower number of stars with a successful analysis, the trend is really similar. We checked for possible correlations with stellar parameters that could explain the trend in Fig.~\ref{fig:el_H_age} finding none. 

From older ages there is a steep increase that ends at 11\,Gyr, that is probably due to the production of these elements via SN\,II. This rise is then followed by a more flat part that in some cases ends in a increase in [$X$/H] for younger stars. This final increase can be explained considering the contribution from low-mass stars that polluted the gas from where the youngest stars were formed. However in the case of [Eu/H], this explanation has more problem to explain the results because Eu is almost completely produced via r-process meaning that the contribution from low-mass stars should be almost negligible.

In Fig.\,15 in \cite{Bensby2014}, it is possible to see the different abundance trends for different $\alpha$-elements. In particular Ti shows a clearer distinction between thin and thick disk compared to the other $\alpha$-elements, so we decided to use Ti to distinguish thin and thick disk stars in the [$X$/H] versus age comparison. We consider stars being thin disk stars when [Ti/Fe] < 0.2, while stars are thick disk stars when [Ti/Fe] > 0.2. The results are visible in Fig.~\ref{fig:el_H_age_alpha} for all the elements under investigation. Clearly basically all thick disk stars are located in the old age region and are responsible for the increase that is visible in Fig.~\ref{fig:el_H_age} up to 11\,Gyr. All thin disk stars are instead responsible for the flat and the uprising part at young age.

\section{Summary}\label{sect:summary}

We have performed a detailed chemical abundance analysis on a sample of F and G dwarf stars for several neutron-capture elements Sr, Zr, La, Ce, Nd, Sm and Eu in order to investigate their formation sites and their evolution in the Galactic disk. In the literature there are several works that focus on metal-poor stars due to their interesting highly enriched spectra, while for stars with $\rm [Fe/H] > -1$ the samples are usually smaller and not all the elements can be investigated due to the spectra that become extremely rich. In total we determine Sr abundance for 156 stars, Zr abundance for 311 stars, La abundance for 242 stars, Ce abundance for 365 stars, Nd abundance for 395 stars, Sm abundance for 280 stars and Eu abundance for 378 stars. In summary our findings and conclusions are:
\begin{itemize}
\item Strontium and zirconium are part of the same s-process peak production and they are intensively studied in theoretical works due to their position in the valley of stability due to the magic neutron number A = 50. They show similar abundance trends with a common flat part around solar metallicity. For Sr, the spread in thin and thick disk is too high to derive any conclusion. For Zr instead, the thin disk stars are grouped close to solar abundance values meaning a constant production in time, thanks to production in AGB stars in the mass range 1-8 M$_{\odot}$. Models from \cite{Travaglio2004} cannot explain our Sr trend since it is expected to be basically flat around solar [Sr/Fe] down to $\rm [Fe/H] \approx -2$. On the other hand, the same models can fit our Zr data, even if we observe sub-solar [Zr/Fe] at solar metallicity. The discrepancies could be due to uncertainties in the model on yields for stars in 1-8 M$_{\odot}$ range at different metallicities.

\item Lanthanum and cerium are basically flat, with solar [La,Ce/Fe]. This flat trend at solar value is basically conserved also when La and Ce are compared to Ba, a typical s-process element. When compared to Eu, some thick disk stars show pure r-process abundance and it is possible to see a turn in [Ce/Eu] when s-process become the more important enrichment process at $\rm [Fe/H] \approx -0.5$. In \cite{Travaglio1999}, r-process La and Ce productions come from SN\,II from stars of 8-10 M$_{\odot}$, while s-process production come from stars in the range 2-4 M$_{\odot}$. This gives a gap in the production of La and Ce and then a fast decrease as soon as AGB stars are actively involved in chemical enrichment, since Eu is basically not produce via this channel.

\item Neodymium and samarium are produced via both the s- and r-processes. For Nd the two processes are almost equally responsible for its enrichment while for Sm r-process is the main productive channel (70\,\%). Even if they are produced in different ways they share a similar trend as $\alpha$-element, derived by production from SN\,II. Considering the results from \cite{Travaglio1999} on different AGB sites for Nd, Sm and Ba, the smooth decrease in [Nd/Ba] is due to the fact that Nd has a quite high r-process production rate compared to Ba but at solar metallicities the flat [Nd/Ba] is due to the common production from stars in the mass range 2-4 M$_{\odot}$. Samarium, on the other hand, has a higher r-process production rate that makes it more similar to Eu, so [Sm/Eu] is almost flat with values around $\rm [Sm/Eu] = 0$.

\item Europium is the prototype of r-process element since its production is almost completely performed by rapid neutron-capture. It shows a really clean $\alpha$-trend when compared to metallicity, as Nd and Sm. This is expected since its high r-process production rate and both thin and thick disk stars present the same [Eu/Ba] decrease. Our data are in agreement with model from \cite{Travaglio1999} considering enrichment from SN\,II of mass 8-10 M$_{\odot}$.

\item The study of abundances as function of age shows different trend for thin and thick disk in almost all the elements. This is especially true for Sr, Zr, Nd, and Eu. In thick disk there is a decrease in the abundance as age decreases that can be explained with the decrease of SN\,II events while Fe production from SN\,Ia increases. In the thin disk most of the trends are basically flat because of the high production of neutron-capture elements via s-process from AGB stars. This is not the case for Sr and Zr that show increasing abundances for younger stars due to the high s-production rate, as derived by \cite{Bisterzo2014}. [Sr/Zr] shows an increasing trend with decreasing ages, probably due to higher Zr r- process production. The same production sites produce flat [La/Ce] and [Nd/Sm]. When [Eu/Ba] is related to age the trend shows a decrease with lower ages, due to the decrease in Eu production by SN\,II counterbalance by the increase in production by AGB stars of Ba.

When the [$X$/H] ratio is plotted as a function of the age, it is clear that around 11\,Gyr the trend changes clearly in almost all the elements, showing a flatter trend after a rise from older ages up to 6\,Gyr, followed in some cases by a clear increase in [$X$/H] at recent age. This is particularly relevant in Eu case, and the reason is not completely clear. In theory younger stars should not present such high abundance in Eu, since Eu is almost completely produced in SN\,II at the beginning of star formation. In theory if these stars are coming from an inner region of the Galaxy where the metallicity is higher compared to the Solar Neighborhood, the higher [Eu/H] also at younger stellar ages could be potentially explained. This possible explanation needs however to be studied in more detailed, taking into account also the orbit of the stars and possible membership with dynamical streams (for example Hercules streams, since it was studied in \cite{Bensby2014}).

\end{itemize}

\subsection*{Acknowledgements}
We would like to thank the referee Prof. Chris Sneden for the useful comments and suggestions in the improvement of this paper.
T.B. was supported by the project grant "The New Milky Way" from the Knut and Alice Wallenberg Foundation. 

\bibliographystyle{aa} 
 \bibliography{Chiara_references} 

\begin{thebibliography}{59}
\expandafter\ifx\csname natexlab\endcsname\relax\def\natexlab#1{#1}\fi

\bibitem[{{Adibekyan} {et~al.}(2012){Adibekyan}, {Sousa}, {Santos}, {Delgado
  Mena}, {Gonz{\'a}lez Hern{\'a}ndez}, {Israelian}, {Mayor}, \&
  {Khachatryan}}]{Adibekyan2012}
{Adibekyan}, V.~Z., {Sousa}, S.~G., {Santos}, N.~C., {et~al.} 2012, \aap, 545,
  A32

\bibitem[{{Andrievsky} {et~al.}(2011){Andrievsky}, {Spite}, {Korotin}, {Fran{\c
  c}ois}, {Spite}, {Bonifacio}, {Cayrel}, \& {Hill}}]{Andrievsky2011}
{Andrievsky}, S.~M., {Spite}, F., {Korotin}, S.~A., {et~al.} 2011, \aap, 530,
  A105

\bibitem[{{Aoki} {et~al.}(2001){Aoki}, {Ryan}, {Norris}, {Beers}, {Ando},
  {Iwamoto}, {Kajino}, {Mathews}, \& {Fujimoto}}]{Aoki2001}
{Aoki}, W., {Ryan}, S.~G., {Norris}, J.~E., {et~al.} 2001, \apj, 561, 346

\bibitem[{{Arlandini} {et~al.}(1999){Arlandini}, {K{\"a}ppeler}, {Wisshak},
  {Gallino}, {Lugaro}, {Busso}, \& {Straniero}}]{Arlandini1999}
{Arlandini}, C., {K{\"a}ppeler}, F., {Wisshak}, K., {et~al.} 1999, \apj, 525,
  886

\bibitem[{{Asplund} {et~al.}(2009){Asplund}, {Grevesse}, {Sauval}, \&
  {Scott}}]{Asplund2009}
{Asplund}, M., {Grevesse}, N., {Sauval}, A.~J., \& {Scott}, P. 2009, \araa, 47,
  481

\bibitem[{{Battistini} \& {Bensby}(2015)}]{Battistini2015}
{Battistini}, C. \& {Bensby}, T. 2015, \aap, 577, A9

\bibitem[{{Bautista} {et~al.}(2002){Bautista}, {Gull}, {Ishibashi}, {Hartman},
  \& {Davidson}}]{Bautista2002}
{Bautista}, M.~A., {Gull}, T.~R., {Ishibashi}, K., {Hartman}, H., \&
  {Davidson}, K. 2002, \mnras, 331, 875

\bibitem[{{Bedell} {et~al.}(2014){Bedell}, {Mel{\'e}ndez}, {Bean},
  {Ram{\'{\i}}rez}, {Leite}, \& {Asplund}}]{Bedell2014}
{Bedell}, M., {Mel{\'e}ndez}, J., {Bean}, J.~L., {et~al.} 2014, \apj, 795, 23

\bibitem[{{Bensby} {et~al.}(2003){Bensby}, {Feltzing}, \&
  {Lundstr{\"o}m}}]{Bensby2003}
{Bensby}, T., {Feltzing}, S., \& {Lundstr{\"o}m}, I. 2003, \aap, 410, 527

\bibitem[{{Bensby} {et~al.}(2005){Bensby}, {Feltzing}, {Lundstr{\"o}m}, \&
  {Ilyin}}]{Bensby2005}
{Bensby}, T., {Feltzing}, S., {Lundstr{\"o}m}, I., \& {Ilyin}, I. 2005, \aap,
  433, 185

\bibitem[{{Bensby} {et~al.}(2014){Bensby}, {Feltzing}, \& {Oey}}]{Bensby2014}
{Bensby}, T., {Feltzing}, S., \& {Oey}, M.~S. 2014, \aap, 562, A71

\bibitem[{{Bisterzo} {et~al.}(2011){Bisterzo}, {Gallino}, {Straniero},
  {Cristallo}, \& {K{\"a}ppeler}}]{Bisterzo2011}
{Bisterzo}, S., {Gallino}, R., {Straniero}, O., {Cristallo}, S., \&
  {K{\"a}ppeler}, F. 2011, \mnras, 418, 284

\bibitem[{{Bisterzo} {et~al.}(2014){Bisterzo}, {Travaglio}, {Gallino},
  {Wiescher}, \& {K{\"a}ppeler}}]{Bisterzo2014}
{Bisterzo}, S., {Travaglio}, C., {Gallino}, R., {Wiescher}, M., \&
  {K{\"a}ppeler}, F. 2014, \apj, 787, 10

\bibitem[{{Brewer} \& {Carney}(2006)}]{Brewer2006}
{Brewer}, M.-M. \& {Carney}, B.~W. 2006, \aj, 131, 431

\bibitem[{{Burbidge} {et~al.}(1957){Burbidge}, {Burbidge}, {Fowler}, \&
  {Hoyle}}]{Burbidge1957}
{Burbidge}, E.~M., {Burbidge}, G.~R., {Fowler}, W.~A., \& {Hoyle}, F. 1957,
  Reviews of Modern Physics, 29, 547

\bibitem[{{Burris} {et~al.}(2000){Burris}, {Pilachowski}, {Armandroff},
  {Sneden}, {Cowan}, \& {Roe}}]{Burris2000}
{Burris}, D.~L., {Pilachowski}, C.~A., {Armandroff}, T.~E., {et~al.} 2000,
  \apj, 544, 302

\bibitem[{{Cescutti} {et~al.}(2006){Cescutti}, {Fran{\c c}ois}, {Matteucci},
  {Cayrel}, \& {Spite}}]{Cescutti2006}
{Cescutti}, G., {Fran{\c c}ois}, P., {Matteucci}, F., {Cayrel}, R., \& {Spite},
  M. 2006, \aap, 448, 557

\bibitem[{{Demarque} {et~al.}(2004){Demarque}, {Woo}, {Kim}, \&
  {Yi}}]{Demarque2004}
{Demarque}, P., {Woo}, J.-H., {Kim}, Y.-C., \& {Yi}, S.~K. 2004, \apjs, 155,
  667

\bibitem[{{Frebel} \& {Norris}(2013)}]{Frebel2013}
{Frebel}, A. \& {Norris}, J.~E. 2013, {Metal-Poor Stars and the Chemical
  Enrichment of the Universe}, ed. T.~D. {Oswalt} \& G.~{Gilmore}, 55

\bibitem[{{Freeman} \& {Bland-Hawthorn}(2002)}]{Freeman2002}
{Freeman}, K. \& {Bland-Hawthorn}, J. 2002, \araa, 40, 487

\bibitem[{{Freiburghaus} {et~al.}(1999){Freiburghaus}, {Rosswog}, \&
  {Thielemann}}]{Freiburghaus1999}
{Freiburghaus}, C., {Rosswog}, S., \& {Thielemann}, F.-K. 1999, \apjl, 525,
  L121

\bibitem[{{Fuhrmann}(1998)}]{Fuhrmann1998}
{Fuhrmann}, K. 1998, \aap, 338, 161

\bibitem[{{Gustafsson} {et~al.}(2008){Gustafsson}, {Edvardsson}, {Eriksson},
  {J{\o}rgensen}, {Nordlund}, \& {Plez}}]{Gustafsson2008}
{Gustafsson}, B., {Edvardsson}, B., {Eriksson}, K., {et~al.} 2008, \aap, 486,
  951

\bibitem[{{Haywood} {et~al.}(2013){Haywood}, {Di Matteo}, {Lehnert}, {Katz}, \&
  {G{\'o}mez}}]{Haywood2013}
{Haywood}, M., {Di Matteo}, P., {Lehnert}, M.~D., {Katz}, D., \& {G{\'o}mez},
  A. 2013, \aap, 560, A109

\bibitem[{{Ishigaki} {et~al.}(2013){Ishigaki}, {Aoki}, \&
  {Chiba}}]{Ishigaki2013}
{Ishigaki}, M.~N., {Aoki}, W., \& {Chiba}, M. 2013, \apj, 771, 67

\bibitem[{{Ivans} {et~al.}(2006){Ivans}, {Simmerer}, {Sneden}, {Lawler},
  {Cowan}, {Gallino}, \& {Bisterzo}}]{Ivans2006}
{Ivans}, I.~I., {Simmerer}, J., {Sneden}, C., {et~al.} 2006, \apj, 645, 613

\bibitem[{{Karakas}(2014)}]{Karakas2014}
{Karakas}, A.~I. 2014, in IAU Symposium, Vol. 298, IAU Symposium, ed.
  S.~{Feltzing}, G.~{Zhao}, N.~A. {Walton}, \& P.~{Whitelock}, 142--153

\bibitem[{{Koch} \& {Edvardsson}(2002)}]{Koch2002}
{Koch}, A. \& {Edvardsson}, B. 2002, \aap, 381, 500

\bibitem[{{Korotin} {et~al.}(2011){Korotin}, {Mishenina}, {Gorbaneva}, \&
  {Soubiran}}]{Korotin2011}
{Korotin}, S., {Mishenina}, T., {Gorbaneva}, T., \& {Soubiran}, C. 2011,
  \mnras, 415, 2093

\bibitem[{{Kupka} {et~al.}(1999){Kupka}, {Piskunov}, {Ryabchikova}, {Stempels},
  \& {Weiss}}]{Kupka1999}
{Kupka}, F., {Piskunov}, N., {Ryabchikova}, T.~A., {Stempels}, H.~C., \&
  {Weiss}, W.~W. 1999, \aaps, 138, 119

\bibitem[{{Kupka} {et~al.}(2000){Kupka}, {Ryabchikova}, {Piskunov}, {Stempels},
  \& {Weiss}}]{Kupka2000}
{Kupka}, F.~G., {Ryabchikova}, T.~A., {Piskunov}, N.~E., {Stempels}, H.~C., \&
  {Weiss}, W.~W. 2000, Baltic Astronomy, 9, 590

\bibitem[{{Lambert}(1989)}]{Lambert1989}
{Lambert}, D.~L. 1989, in American Institute of Physics Conference Series, Vol.
  183, Cosmic Abundances of Matter, ed. C.~J. {Waddington}, 168--199

\bibitem[{{Lawler} {et~al.}(2006){Lawler}, {Den Hartog}, {Sneden}, \&
  {Cowan}}]{Lawler2006}
{Lawler}, J.~E., {Den Hartog}, E.~A., {Sneden}, C., \& {Cowan}, J.~J. 2006,
  \apjs, 162, 227

\bibitem[{{Lawler} {et~al.}(2009){Lawler}, {Sneden}, {Cowan}, {Ivans}, \& {Den
  Hartog}}]{Lawler2009}
{Lawler}, J.~E., {Sneden}, C., {Cowan}, J.~J., {Ivans}, I.~I., \& {Den Hartog},
  E.~A. 2009, \apjs, 182, 51

\bibitem[{{Lawler} {et~al.}(2001){Lawler}, {Wickliffe}, {den Hartog}, \&
  {Sneden}}]{Lawler2001}
{Lawler}, J.~E., {Wickliffe}, M.~E., {den Hartog}, E.~A., \& {Sneden}, C. 2001,
  \apj, 563, 1075

\bibitem[{{Ljung} {et~al.}(2006){Ljung}, {Nilsson}, {Asplund}, \&
  {Johansson}}]{Ljung2006}
{Ljung}, G., {Nilsson}, H., {Asplund}, M., \& {Johansson}, S. 2006, \aap, 456,
  1181

\bibitem[{{Mashonkina} \& {Gehren}(2001)}]{Mashonkina2001}
{Mashonkina}, L. \& {Gehren}, T. 2001, \aap, 376, 232

\bibitem[{{Mashonkina} {et~al.}(2004){Mashonkina}, {Kamaeva}, {Samotoev}, \&
  {Sakhibullin}}]{Mashonkina2004}
{Mashonkina}, L.~I., {Kamaeva}, L.~A., {Samotoev}, V.~A., \& {Sakhibullin},
  N.~A. 2004, Astronomy Reports, 48, 185

\bibitem[{{Mashonkina} {et~al.}(2007){Mashonkina}, {Vinogradova}, {Ptitsyn},
  {Khokhlova}, \& {Chernetsova}}]{Mashonkina2007}
{Mashonkina}, L.~I., {Vinogradova}, A.~B., {Ptitsyn}, D.~A., {Khokhlova},
  V.~S., \& {Chernetsova}, T.~A. 2007, Astronomy Reports, 51, 903

\bibitem[{{Mishenina} {et~al.}(2013){Mishenina}, {Pignatari}, {Korotin},
  {Soubiran}, {Charbonnel}, {Thielemann}, {Gorbaneva}, \&
  {Basak}}]{Mishenina2013}
{Mishenina}, T.~V., {Pignatari}, M., {Korotin}, S.~A., {et~al.} 2013, \aap,
  552, A128

\bibitem[{{Nishimura} {et~al.}(2006){Nishimura}, {Kotake}, {Hashimoto},
  {Yamada}, {Nishimura}, {Fujimoto}, \& {Sato}}]{Nishimura2006}
{Nishimura}, S., {Kotake}, K., {Hashimoto}, M.-a., {et~al.} 2006, \apj, 642,
  410

\bibitem[{{Pignatari} {et~al.}(2010){Pignatari}, {Gallino}, {Heil}, {Wiescher},
  {K{\"a}ppeler}, {Herwig}, \& {Bisterzo}}]{Pignatari2010}
{Pignatari}, M., {Gallino}, R., {Heil}, M., {et~al.} 2010, \apj, 710, 1557

\bibitem[{{Piskunov} {et~al.}(1995){Piskunov}, {Kupka}, {Ryabchikova}, {Weiss},
  \& {Jeffery}}]{Piskunov1995}
{Piskunov}, N.~E., {Kupka}, F., {Ryabchikova}, T.~A., {Weiss}, W.~W., \&
  {Jeffery}, C.~S. 1995, \aaps, 112, 525

\bibitem[{{Prochaska} {et~al.}(2000){Prochaska}, {Naumov}, {Carney},
  {McWilliam}, \& {Wolfe}}]{Prochaska2000}
{Prochaska}, J.~X., {Naumov}, S.~O., {Carney}, B.~W., {McWilliam}, A., \&
  {Wolfe}, A.~M. 2000, \aj, 120, 2513

\bibitem[{{Reddy} {et~al.}(2006){Reddy}, {Lambert}, \& {Allende
  Prieto}}]{Reddy2006}
{Reddy}, B.~E., {Lambert}, D.~L., \& {Allende Prieto}, C. 2006, \mnras, 367,
  1329

\bibitem[{{Reddy} {et~al.}(2003){Reddy}, {Tomkin}, {Lambert}, \& {Allende
  Prieto}}]{Reddy2003}
{Reddy}, B.~E., {Tomkin}, J., {Lambert}, D.~L., \& {Allende Prieto}, C. 2003,
  \mnras, 340, 304

\bibitem[{{Roederer} {et~al.}(2008){Roederer}, {Lawler}, {Sneden}, {Cowan},
  {Sobeck}, \& {Pilachowski}}]{Roederer2008}
{Roederer}, I.~U., {Lawler}, J.~E., {Sneden}, C., {et~al.} 2008, \apj, 675, 723

\bibitem[{{Rosswog} {et~al.}(2014){Rosswog}, {Korobkin}, {Arcones},
  {Thielemann}, \& {Piran}}]{Rosswog2014}
{Rosswog}, S., {Korobkin}, O., {Arcones}, A., {Thielemann}, F.-K., \& {Piran},
  T. 2014, \mnras, 439, 744

\bibitem[{{Ryabchikova} {et~al.}(1997){Ryabchikova}, {Piskunov}, {Kupka}, \&
  {Weiss}}]{Ryabchikova1997}
{Ryabchikova}, T.~A., {Piskunov}, N.~E., {Kupka}, F., \& {Weiss}, W.~W. 1997,
  Baltic Astronomy, 6, 244

\bibitem[{{Sackmann} {et~al.}(1993){Sackmann}, {Boothroyd}, \&
  {Kraemer}}]{Sackmann1993}
{Sackmann}, I.-J., {Boothroyd}, A.~I., \& {Kraemer}, K.~E. 1993, \apj, 418, 457

\bibitem[{{Simmerer} {et~al.}(2004){Simmerer}, {Sneden}, {Cowan}, {Collier},
  {Woolf}, \& {Lawler}}]{Simmerer2004}
{Simmerer}, J., {Sneden}, C., {Cowan}, J.~J., {et~al.} 2004, \apj, 617, 1091

\bibitem[{{Sneden} {et~al.}(2008){Sneden}, {Cowan}, \& {Gallino}}]{Sneden2008}
{Sneden}, C., {Cowan}, J.~J., \& {Gallino}, R. 2008, \araa, 46, 241

\bibitem[{{Surman} {et~al.}(2008){Surman}, {McLaughlin}, {Ruffert}, {Janka}, \&
  {Hix}}]{Surman2008}
{Surman}, R., {McLaughlin}, G.~C., {Ruffert}, M., {Janka}, H.-T., \& {Hix},
  W.~R. 2008, \apjl, 679, L117

\bibitem[{{Travaglio} {et~al.}(1999){Travaglio}, {Galli}, {Gallino}, {Busso},
  {Ferrini}, \& {Straniero}}]{Travaglio1999}
{Travaglio}, C., {Galli}, D., {Gallino}, R., {et~al.} 1999, \apj, 521, 691

\bibitem[{{Travaglio} {et~al.}(2004){Travaglio}, {Gallino}, {Arnone}, {Cowan},
  {Jordan}, \& {Sneden}}]{Travaglio2004}
{Travaglio}, C., {Gallino}, R., {Arnone}, E., {et~al.} 2004, \apj, 601, 864

\bibitem[{{Valenti} \& {Fischer}(2005)}]{Valenti2005}
{Valenti}, J.~A. \& {Fischer}, D.~A. 2005, \apjs, 159, 141

\bibitem[{{Valenti} \& {Piskunov}(1996)}]{Valenti1996}
{Valenti}, J.~A. \& {Piskunov}, N. 1996, \aaps, 118, 595

\bibitem[{{Winckler} {et~al.}(2006){Winckler}, {Dababneh}, {Heil},
  {K{\"a}ppeler}, {Gallino}, \& {Pignatari}}]{Winckler2006}
{Winckler}, N., {Dababneh}, S., {Heil}, M., {et~al.} 2006, \apj, 647, 685

\bibitem[{{Woosley} {et~al.}(1994){Woosley}, {Wilson}, {Mathews}, {Hoffman}, \&
  {Meyer}}]{Woosley1994}
{Woosley}, S.~E., {Wilson}, J.~R., {Mathews}, G.~J., {Hoffman}, R.~D., \&
  {Meyer}, B.~S. 1994, \apj, 433, 229

\end{thebibliography}
 
\begin{appendix}

\section{Complete linelists}
\label{sec:linelists}

In this section we listed the hfs components for the lines that suffer from hyperfine splitting.

\begin{table*}[ht]\tiny
\begin{center}
\caption{Linelists for the synthesis of Sr. The lists include atomic data for Sr from \cite{Bautista2002} as well as additional lines nearby. Columns represent element line, wavelength and log(gf) values, respectively.\label{tab:hfs1}}
\setlength{\tabcolsep}{1 mm}
\begin{tabular}{c c c }
\hline
\multicolumn{3}{c} {\ion{Sr}{ii}  4607  } \\
Element & $\lambda$ ($\AA$) & log(gf)  \\
\hline
\ion{Fe}{i} & 4607.0856 & -3.530 \\
\ion{Ni}{i} & 4607.1378 & -2.759 \\
\ion{Sr}{i} & 4607.3310 & 0.283 \\
\ion{Nd}{ii} & 4607.3800 & -1.100 \\
\ion{Mn}{i} & 4607.5880 & -1.900\\
\hline
\end{tabular}
\end{center}
\end{table*}

\begin{table*}[ht]\tiny
\begin{center}
\caption{Linelists for the synthesis of Zr. The lists include atomic data for Zr from \cite{Ljung2006} as well as additional lines nearby. Columns represent element line, wavelength and log(gf) values, respectively.\label{tab:hfs2}}
\setlength{\tabcolsep}{1 mm}
\begin{tabular}{c c c | c c c | c c c | c c c}
\hline
\multicolumn{3}{c|} {\ion{Zr}{ii}  4208  } & \multicolumn{3}{c|}{\ion{Zr}{i} 4687 }  & \multicolumn{3}{c|}{\ion{Zr}{i} 4739} &  \multicolumn{3}{c}{\ion{Zr}{ii} 5112}\\
Element & $\lambda$ ($\AA$) & log(gf)  & Element & $\lambda$ ($\AA$) & log(gf) & Element & $\lambda$ ($\AA$) & log(gf) & Element & $\lambda$ ($\AA$) & log(gf) \\
\hline
\ion{Th}{ii} & 4208.8910 & -0.738 & \ion{Si}{i} & 4687.6210 & -2.944 & \ion{Mn}{i} & 4739.0870 & -0.490 & \ion{Fe}{i} & 5111.8550 & -1.993 \\
\ion{Cr}{i} & 4208.9529 & -0.528 & \ion{Fe}{i} & 4687.6700 & -3.780 & \ion{Ce}{ii} & 4739.1470 & -0.810 &  \ion{K}{i} & 5112.2460 & -2.080 \\
\ion{Zr}{ii} & 4208.9800 & -0.510 & \ion{Zr}{i} & 4687.8090 & 0.550 & \ion{Zr}{i} & 4739.4800 & 0.230  & \ion{Zr}{ii} & 5112.2700 & -0.850 \\
\ion{Cr}{ii} & 4209.0941 & -2.379 & \ion{Ti}{i} & 4687.8090 & -2.399 & \ion{Ce}{ii} & 4739.5190 & -1.020  & \ion{Sm}{ii} & 5112.2930 & -1.550 \\
\ion{Cr}{i} & 4209.1730 & -1.388 & \ion{Fe}{i} & 4688.1770 & -1.538 & \ion{V}{i} & 4739.5750 & -0.849  & \ion{Cr}{i} & 5112.4850 & -3.700 \\

\hline
\end{tabular}
\end{center}
\end{table*}

\begin{table*}[ht]\tiny
\begin{center}
\caption{Linelists for the synthesis of La. The lists include the hfs (when available) of the La in question taken from \cite{Ivans2006} as well as additional lines nearby. Columns represent element line, wavelength and log(gf) values, respectively.\label{tab:hfs3}}
\setlength{\tabcolsep}{1 mm}
\begin{tabular}{c c c | c c c | c c c | c c c }
\hline
\multicolumn{3}{c|} {\ion{La}{ii}  4662  } & \multicolumn{3}{c|}{\ion{La}{ii} 4748 }  & \multicolumn{3}{c|}{\ion{La}{ii} 5122} & \multicolumn{3}{c}{\ion{La}{ii} 6390} \\
Element & $\lambda$ ($\AA$) & log(gf)  & Element & $\lambda$ ($\AA$) & log(gf) & Element & $\lambda$ ($\AA$) & log(gf) & Element & $\lambda$ ($\AA$) & log(gf) \\
\hline
\ion{Fe}{i} & 4662.0970 & -4.751 & \ion{Fe}{i} & 4748.0123 & -2.544 & \ion{C2}{i} & 5122.8253 & 0.183 & \ion{Fe}{i} & 6388.4047 & -4.476\\
\ion{Cr}{i} & 4662.4020 & -2.735 & \ion{V}{i} & 4748.5030 & -0.620 & \ion{C2}{i} & 5122.8948 & 0.175 & \ion{Nd}{ii} & 6389.9800 & -0.770 \\
\ion{La}{ii} & 4662.4773 & -2.951 & \ion{La}{ii} & 4748.7300 & -0.540 & \ion{La}{ii} & 5122.9798 & -1.476 & \ion{La}{ii} & 6390.4551 & -2.012 \\
\ion{La}{ii} & 4662.4814 & -2.511 & \ion{Fe}{i} & 4749.0097 & -3.058 & \ion{La}{ii} & 5122.9799 & -2.046 & \ion{La}{ii} & 6390.4683 & -2.752\\
\ion{La}{ii} & 4662.4852 & -2.240 & \ion{Cr}{i} & 4749.2091 & -1.643 & \ion{La}{ii} & 5122.9859 & -2.046 & \ion{La}{ii} & 6390.4685 & -2.183 \\
\ion{La}{ii} & 4662.4901 & -2.252 & & & & \ion{La}{ii} & 5122.9864 & -1.888 & \ion{La}{ii} & 6390.4794 & -3.752 \\
\ion{La}{ii} & 4662.4914 & -2.136 & & & &\ion{La}{ii} & 5122.9865 & -1.874 & \ion{La}{ii} & 6390.4796 & -2.570 \\
\ion{La}{ii} & 4662.4924 & -2.256 & & & &\ion{La}{ii} & 5122.9911 & -1.888 & \ion{La}{ii} & 6390.4802 & -2.390 \\
\ion{La}{ii} & 4662.5024 & -2.511 & & & &\ion{La}{ii} & 5122.9915 & -2.556 & \ion{La}{ii} & 6390.4887 & -3.330 \\
\ion{La}{ii} & 4662.5044 & -2.056 & & & &\ion{La}{ii} & 5122.9919 & -1.899 & \ion{La}{ii} & 6390.4887 & -2.536 \\
\ion{Ce}{ii} & 4662.6950 & -0.850 & & & &\ion{La}{ii} & 5122.9954 & -1.899 & \ion{La}{ii} & 6390.4900 & -2.661\\
\ion{Ti}{ii} & 4662.7410 & -3.429 & & & &\ion{La}{ii} & 5122.9958 & -3.999 & \ion{La}{ii} & 6390.4964 & -3.100\\
& & & & & &\ion{La}{ii} & 5122.9962 & -2.072  & \ion{La}{ii} & 6390.4971 & -2.595 \\
& & & & & & \ion{La}{ii} & 5122.9986 & -2.072 & \ion{La}{ii} & 6390.4977 & -3.079 \\
& & & & & & \ion{La}{ii} & 5122.9990 & -2.248 & \ion{La}{ii} & 6390.5021 & -2.954 \\
& & & & & & \ion{C2}{i} & 5123.1242 & -0.410 & \ion{La}{ii} & 6390.5021 & -2.857 \\
& & & & & & \ion{C2}{i} & 5123.1796 & -0.426 & \ion{La}{ii} & 6390.5058 & -2.778 \\
& & & & & & & & & \ion{Si}{i} & 6391.0205 & -2.065\\
& & & & & & & & & \ion{Mn}{i} & 6391.2109 & -1.491\\
\hline
\end{tabular}
\end{center}
\end{table*}

\begin{table*}[ht]\tiny
\begin{center}
\caption{Linelists for the synthesis of Ce. The lists include atomic data for Ce from \cite{Lawler2009} as well as additional lines nearby. Columns represent element line, wavelength and log(gf) values, respectively.\label{tab:hfs4}}
\setlength{\tabcolsep}{1 mm}
\begin{tabular}{c c c | c c c | c c c | c c c }
\hline
\multicolumn{3}{c|} {\ion{Ce}{ii}  4523  } & \multicolumn{3}{c|}{\ion{Ce}{ii} 4572 }  & \multicolumn{3}{c|}{\ion{Ce}{ii} 4628} & \multicolumn{3}{c}{\ion{Ce}{ii} 5187} \\
Element & $\lambda$ ($\AA$) & log(gf)  & Element & $\lambda$ ($\AA$) & log(gf) & Element & $\lambda$ ($\AA$) & log(gf) & Element & $\lambda$ ($\AA$) & log(gf) \\
\hline
\ion{Gd}{ii} & 4522.8360 & -0.310 & \ion{Ti}{i} & 4572.0658 & -1.361 & \ion{Co}{i} & 4627.4250 & -1.420 & \ion{Ni}{i} & 5186.5527 & -1.124\\
\ion{Sm}{ii} & 4523.0300 & -0.990 & \ion{Cr}{ii} & 4572.1870 & -1.247 & \ion{Fe}{i} & 4627.5471 & -3.057 & \ion{Fe}{i} & 5186.7543 & -2.867\\
\ion{Ce}{ii} & 4523.0750 & -0.080 & \ion{Ce}{ii} & 4572.2780 & 0.220 & \ion{Ce}{ii} & 4628.1609 & 0.140 & \ion{Ce}{ii} & 5187.4578 & 0.170 \\
\ion{Ti}{i} & 4523.2453 & -1.848 & \ion{Co}{ii} & 4572.4616 & -3.078 & \ion{Ti}{i} & 4628.1842 & -1.287 & \ion{Ni}{i} & 5187.8199 & -1.910 \\
\ion{Mn}{i} & 4523.3952 & -1.360 & \ion{Ti}{i} & 4572.7676 & -1.560 & \ion{Ce}{ii} & 4628.2391 & -0.430 & \ion{Fe}{i} & 5187.9142 & -1.371\\
\hline
\end{tabular}
\end{center}
\end{table*}

\begin{table*}[ht]\tiny
\begin{center}
\caption{Linelists for the synthesis of Nd. The lists include the hfs (when available) of the Nd in question taken from \cite{Roederer2008} as well as additional lines nearby. Columns represent element line, wavelength and log(gf) values, respectively.\label{tab:hfs4}}
\setlength{\tabcolsep}{1 mm}
\begin{tabular}{c c c | c c c | c c c | c c c | c c c}
\hline
\multicolumn{3}{c|} {\ion{Nd}{ii}  4177  } & \multicolumn{3}{c|}{\ion{Nd}{ii} 4358 }  & \multicolumn{3}{c|}{\ion{Nd}{ii} 4446} & \multicolumn{3}{c}{\ion{Nd}{ii} 5130} & \multicolumn{3}{c}{\ion{Nd}{ii} 5319} \\
Element & $\lambda$ ($\AA$) & log(gf)  & Element & $\lambda$ ($\AA$) & log(gf) & Element & $\lambda$ ($\AA$) & log(gf) & Element & $\lambda$ ($\AA$) & log(gf) & Element & $\lambda$($\AA$) & log(gf) \\
\hline
\ion{Fe}{i} & 4177.0789 & -3.369 & \ion{CH}{i} & 4358.0000 & -1.517 & \ion{Fe}{ii} & 4446.2433 & -2.776 & \ion{Cr}{i} & 5130.4239 & -1.051 & \ion{Fe}{i} & 5319.6228 & -0.140 \\
\ion{Fe}{ii} & 4177.2014 & -8.294 & \ion{Fe}{i} & 4358.0763 & -2.960 & \ion{Fe}{i} & 4446.2747 & -2.116 & \ion{C2}{i} & 5130.5777 & -0.262 & \ion{Fe}{i} & 5319.7629 & -2.594\\
\ion{Nd}{ii} & 4177.2885 & -4.315 & \ion{Nd}{ii} & 4358.1255 & -3.132 & \ion{Nd}{ii} & 4446.3635 & -4.140 & \ion{Nd}{ii} & 5130.5900 & 0.450 & \ion{Nd}{ii} & 5319.8100 & -0.140 \\
\ion{Nd}{ii} & 4177.2948 & -2.865 & \ion{Nd}{ii} & 4358.1346 & -1.821 & \ion{Nd}{ii} & 4446.3647 & -3.935 & \ion{Fe}{i} & 5130.9095 & -2.818 & \ion{Fe}{i} & 5320.0356 & -2.540 \\
\ion{Nd}{ii} & 4177.2984 & -3.927 & \ion{Nd}{ii} & 4358.1351 & -2.908 & \ion{Nd}{ii} & 4446.3648 & -3.141 & \ion {Ce}{ii} & 5131.1908 & -2.470 & \ion{Cr}{i} & 5320.4280 & -1.527 \\
\ion{Nd}{ii} & 4177.3021 & -1.717 & \ion{Nd}{ii} & 4358.1433 & -1.909 & \ion{Nd}{ii} & 4446.3657 & -2.690 & & & & & \\
\ion{Nd}{ii} & 4177.3025 & -4.487 & \ion{Nd}{ii} & 4358.1438 & -2.820 & \ion{Nd}{ii} & 4446.3664 & -2.930 & & &  & & \\
\ion{Nd}{ii} & 4177.3039 & -2.647 & \ion{Nd}{ii} & 4358.1455 & -3.299 & \ion{Nd}{ii} & 4446.3665 & -3.908 & & &  & & \\
\ion{Nd}{ii} & 4177.3064 & -3.032 & \ion{Nd}{ii} & 4358.1501 & -0.725 & \ion{Nd}{ii} & 4446.3677 & -2.571 &  & &  & & \\
\ion{Nd}{ii} & 4177.3074 & -2.736 & \ion{Nd}{ii} & 4358.1511 & -1.989 & \ion{Nd}{ii} & 4446.3686 & -2.835 & & &  & & \\
\ion{Nd}{ii} & 4177.3087 & -4.096 & \ion{Nd}{ii} & 4358.1512 & -1.977 & \ion{Nd}{ii} & 4446.3690 & -3.986 & & &  & & \\
\ion{Nd}{ii} & 4177.3102 & -1.802 & \ion{Nd}{ii} & 4358.1514 & -3.075 & \ion{Nd}{ii} & 4446.3703 & -2.453 & & &  & & \\
\ion{Nd}{ii} & 4177.3109 & -1.884 & \ion{Nd}{ii} & 4358.1517 & -2.799 & \ion{Nd}{ii} & 4446.3714 & -2.799 &  & &  & & \\
\ion{Nd}{ii} & 4177.3120 & -2.564 & \ion{Nd}{ii} & 4358.1524 & -3.132 & \ion{Nd}{ii} & 4446.3722 & -4.177 &  & &  & & \\
\ion{Nd}{ii} & 4177.3121 & -2.814 & \ion{Nd}{ii} & 4358.1566 & -2.076 & \ion{Nd}{ii} & 4446.3735 & -2.342 & & &  & & \\
\ion{Nd}{ii} & 4177.3143 & -3.904 & \ion{Nd}{ii} & 4358.1569 & -2.987 & \ion{Nd}{ii} & 4446.3749 & -2.814 & & &  & & \\
\ion{Nd}{ii} & 4177.3154 & -3.658 & \ion{Nd}{ii} & 4358.1581 & -2.086 & \ion{Nd}{ii} & 4446.3750 & -4.289 & & &  & & \\
\ion{Nd}{ii} & 4177.3160 & -1.969 & \ion{Nd}{ii} & 4358.1586 & -2.830 & \ion{Nd}{ii} & 4446.3758 & -4.102 & & &  & & \\
\ion{Nd}{ii} & 4177.3161 & -0.665 & \ion{Nd}{ii} & 4358.1594 & -2.908 & \ion{Nd}{ii} & 4446.3758 & -4.289 & & &  & & \\
\ion{Nd}{ii} & 4177.3172 & -2.732 & \ion{Nd}{ii} & 4358.1597 & -0.783 & \ion{Nd}{ii} & 4446.3762 & -5.565 & & &  & & \\
\ion{Nd}{ii} & 4177.3175 & -1.892 & \ion{Nd}{ii} & 4358.1614 & -2.165 & \ion{Nd}{ii} & 4446.3763 & -2.857 & & &  & & \\
\ion{Nd}{ii} & 4177.3192 & -0.723 & \ion{Nd}{ii} & 4358.1617 & -2.966 & \ion{Nd}{ii} & 4446.3768 & -4.098 & & &  & & \\
\ion{Nd}{ii} & 4177.3193 & -2.549 & \ion{Nd}{ii} & 4358.1622 & -3.299 & \ion{Nd}{ii} & 4446.3769 & -4.054 &  & &  & & \\
\ion{Nd}{ii} & 4177.3194 & -3.824 & \ion{Nd}{ii} & 4358.1642 & -2.172 & \ion{Nd}{ii} & 4446.3773 & -2.238 & & &  & & \\
\ion{Nd}{ii} & 4177.3196 & -0.864 & \ion{Nd}{ii} & 4358.1647 & -2.925 & \ion{Nd}{ii} & 4446.3774 & -0.915 & & &  & & \\
\ion{Nd}{ii} & 4177.3206 & -2.059 & \ion{Nd}{ii} & 4358.1655 & -2.820 & \ion{Nd}{ii} & 4446.3777 & -2.739 & & &  & & \\
\ion{Nd}{ii} & 4177.3216 & -2.716 & \ion{Nd}{ii} & 4358.1658 & -2.254 & \ion{Nd}{ii} & 4446.3782 & -3.002 & & &  & & \\
\ion{CN}{i} & 4177.3222 & -1.091 & \ion{Nd}{ii} & 4358.1661 & -2.997 & \ion{Nd}{ii} & 4446.3785 & -4.154 &  & &  & & \\
\ion{Nd}{ii} & 4177.3223 & -3.685 & \ion{Nd}{ii} & 4358.1666 & -3.075 & \ion{Nd}{ii} & 4446.3791 & -2.897 & & &  & & \\
\ion{Nd}{ii} & 4177.3235 & -3.852 & \ion{Nd}{ii} & 4358.1686 & -0.924 & \ion{Nd}{ii} & 4446.3793 & -2.621 & & &  & & \\
\ion{Nd}{ii} & 4177.3251 & -1.344 & \ion{Nd}{ii} & 4358.1693 & -2.250 & \ion{Nd}{ii} & 4446.3800 & -2.966 & & &  & & \\
\ion{Nd}{ii} & 4177.3238 & -1.988 & \ion{Nd}{ii} & 4358.1695 & -2.370 & \ion{Nd}{ii} & 4446.3805 & -4.346 & & &  & & \\
\ion{Nd}{ii} & 4177.3245 & -2.155 & \ion{Nd}{ii} & 4358.1699 & -3.093 & \ion{Nd}{ii} & 4446.3813 & -2.509 & & &  & & \\
\ion{Nd}{ii} & 4177.3253 & -2.585 & \ion{Nd}{ii} & 4358.1700 & -3.149 & \ion{Nd}{ii} & 4446.3817 & -2.142 & & &  & & \\
\ion{Nd}{ii} & 4177.3254 & -2.572 & \ion{Nd}{ii} & 4358.1703 & -2.987 & \ion{Nd}{ii} & 4446.3821 & -2.982 & & &  & & \\
\ion{Nd}{ii} & 4177.3271 & -4.059 & \ion{Nd}{ii} & 4358.1706 & -2.799 & \ion{Nd}{ii} & 4446.3829 & -0.973 &  & &  & & \\
\ion{Nd}{ii} & 4177.3278 & -2.259 & \ion{Nd}{ii} & 4358.1727 & -2.417 & \ion{Nd}{ii} & 4446.3829 & -4.737 & & &  & & \\
\ion{Nd}{ii} & 4177.3281 & -3.890 & \ion{Nd}{ii} & 4358.1731 & -3.316 & \ion{Nd}{ii} & 4446.3836 & -2.405 & & &  & & \\
\ion{Nd}{ii} & 4177.3283 & -1.352 & \ion{Nd}{ii} & 4358.1735 & -2.966 & \ion{Nd}{ii} & 4446.3837 & -2.051 &  & &  & & \\
\ion{Nd}{ii} & 4177.3285 & -2.848 & \ion{Nd}{ii} & 4358.1736 & -2.311 & \ion{Nd}{ii} & 4446.3840 & -3.115 &  & &  & & \\
\ion{Nd}{ii} & 4177.3291 & -2.092 & \ion{Nd}{ii} & 4358.1749 & -2.830 & \ion{Nd}{ii} & 4446.3847 & -3.064 & & &  & & \\
\ion{Nd}{ii} & 4177.3303 & -2.680 & \ion{Nd}{ii} & 4358.1754 & -2.478 & \ion{Nd}{ii} & 4446.3864 & -2.309 & & &  & & \\
\ion{Nd}{ii} & 4177.3304 & -2.371 & \ion{Nd}{ii} & 4358.1762 & -2.997 & \ion{Nd}{ii} & 4446.3878 & -3.282 & & &  & & \\
\ion{Nd}{ii} & 4177.3308 & -3.059 & \ion{Nd}{ii} & 4358.1771 & -2.341 & \ion{Nd}{ii} & 4446.3882 & -1.114 & & &  & & \\
\ion{Nd}{ii} & 4177.3322 & -2.489 & \ion{Nd}{ii} & 4358.1775 & -2.508 & \ion{Nd}{ii} & 4446.3896 & -2.219 & & &  & & \\
\ion{Nd}{ii} & 4177.3333 & -2.607 & \ion{Nd}{ii} & 4358.1783 & -2.925 & \ion{Nd}{ii} & 4446.3928 & -1.967 & & &  & & \\
\ion{Nd}{ii} & 4177.3334 & -2.203 & \ion{Nd}{ii} & 4358.1783 & -3.093 & \ion{Nd}{ii} & 4446.3932 & -2.134 & & &  & & \\
\ion{Nd}{ii} & 4177.3341 & -2.891 & \ion{Nd}{ii} & 4358.1784 & -1.404 & \ion{Nd}{ii} & 4446.3937 & -1.594 & & &  & & \\
\ion{Nd}{ii} & 4177.3363 & -2.321 & \ion{Nd}{ii} & 4358.1799 & -3.316 & \ion{Nd}{ii} & 4446.4000 & -1.602 &  & &  & & \\
\ion{Nd}{ii} & 4177.3382 & -2.440 & \ion{Nd}{ii} & 4358.1808 & -3.149 & \ion{Fe}{i} & 4446.4786 & -3.963 & & &  & & \\
\ion{CN}{i} & 4177.3480 & -1.123 & \ion{Nd}{ii} & 4358.1912 & -1.412 & \ion{Gd}{ii} & 4446.4786 & -1.280 & & &  & & \\
\ion{Ti}{i} & 4177.3790 & -1.900 & \ion{CH}{i} & 4358.2170 & -1.412 & & & & &  & & \\
& & & \ion{Fe}{ii} & 4358.3674 & -7.807 & & & & &  & & \\
\hline
\end{tabular}
\end{center}
\end{table*}

\begin{table*}[ht]\tiny
\begin{center}
\caption{Linelists for the synthesis of Sm. The lists include the hfs (when available) of the Sm in question taken from \cite{Lawler2006} and \cite{Roederer2008} as well as additional lines nearby. Columns represent element line, wavelength and log(gf) values, respectively.\label{tab:hfs4}}
\setlength{\tabcolsep}{1 mm}
\begin{tabular}{c c c | c c c | c c c | c c c }
\hline
\multicolumn{3}{c|} {\ion{Sm}{ii}  4467  } & \multicolumn{3}{c|}{\ion{Sm}{ii} 4523 }  & \multicolumn{3}{c|}{\ion{Sm}{ii} 4577} & \multicolumn{3}{c}{\ion{Sm}{ii} 4669} \\
Element & $\lambda$ ($\AA$) & log(gf)  & Element & $\lambda$ ($\AA$) & log(gf) & Element & $\lambda$ ($\AA$) & log(gf) & Element & $\lambda$ ($\AA$) & log(gf) \\
\hline
\ion{Gd}{ii} & 4467.2260 & -0.848 & \ion{Nd}{ii} & 4522.8220 & -1.080 & \ion{V}{i} & 4577.1741 & -1.048 & \ion{Sm}{ii} & 4669.3900 & -0.600\\
\ion{Cr}{i} & 4467.3154 & -2.193 & \ion{Gd}{ii} & 4522.8360 & -0.310 & \ion{Fe}{i} & 4577.6736 & -3.465 & \ion{Ce}{ii} & 4669.4997 & -0.410 \\
\ion{Sm}{ii} & 4467.3135 & -1.957 & \ion{Sm}{ii} & 4523.0300 & -0.990 & \ion{Sm}{ii} & 4577.6900 & -0.650 & \ion{Sm}{ii} & 4669.6400 & -0.530 \\
\ion{Sm}{ii} & 4467.3146 & -2.682 & \ion{Ce}{ii} & 4523.0750 & -0.080 & \ion{Fe}{ii} & 4577.7803 & -5.237 & \ion{Cr}{i} & 4669.6642 & -2.268 \\ 
\ion{Sm}{ii} & 4467.3160 & -3.896 & \ion{Ti}{i} & 4523.2453 & -1.848 & \ion{Mn}{i} & 4578.0555 & -1.185 & \ion{Fe}{i} & 4669.9606 & -2.108 \\
\ion{Sm}{ii} & 4467.3173 & -1.874 & & & & & &  & & & \\
\ion{Sm}{ii} & 4467.3188 & -2.464 & & &  & & & & & & \\
\ion{Sm}{ii} & 4467.3189 & -1.993 && &  & & & & & & \\
\ion{Sm}{ii} & 4467.3200 & -2.717 & & &  & & & & & & \\
\ion{Sm}{ii} & 4467.3207 & -3.662 & & &  & & & & & & \\
\ion{Sm}{ii} & 4467.3215 & -3.931 & & &  & & & & & & \\
\ion{Sm}{ii} & 4467.3222 & -1.788 & & &  & & & & & & \\
\ion{Sm}{ii} & 4467.3223 & -1.909 & & &  & & & & & & \\
\ion{Sm}{ii} & 4467.3236 & -2.499 & & &  & & & & & & \\
\ion{Sm}{ii} & 4467.3241 & -2.370 & & &  & & & & & & \\
\ion{Sm}{ii} & 4467.3255 & -3.698 &  & &  & & & & & & \\
\ion{Sm}{ii} & 4467.3263 & -1.823 & & &  & & & & & & \\
\ion{Sm}{ii} & 4467.3265 & -3.614 & & &  & & & & & & \\
\ion{Sm}{ii} & 4467.3281 & -1.704 & & &  & & & & & & \\
\ion{Sm}{ii} & 4467.3282 & -2.405 & & &  & & & & & & \\
\ion{Sm}{ii} & 4467.3303 & -3.651 & & &  & & & & & & \\
\ion{Sm}{ii} & 4467.3305 & -2.338 & & &  & & & & & & \\
\ion{Sm}{ii} & 4467.3314 & -1.739 & & &  & & & & & & \\
\ion{Sm}{ii} & 4467.3335 & -3.683 & & &  & & & & & & \\
\ion{Sm}{ii} & 4467.3336 & -2.373 & & &  & & & & & & \\
\ion{Sm}{ii} & 4467.3352 & -1.623 & & &  & & & & & & \\
\ion{Sm}{ii} & 4467.3360 & -3.716 & & &  & & & & & & \\
\ion{Sm}{ii} & 4467.3373 & -1.658 & & &  & & & & & & \\
\ion{Sm}{ii} & 4467.3382 & -2.348 &  & &  & & & & & & \\
\ion{Sm}{ii} & 4467.3382 & -1.363 & & &  & & & & & & \\
\ion{Sm}{ii} & 4467.3398 & -2.394 & & &  & & & & & & \\
\ion{Sm}{ii} & 4467.3406 & -0.799 & & &  & & & & & & \\
\ion{Sm}{ii} & 4467.3408 & -0.982 & & &  & & & & & & \\
\ion{Sm}{ii} & 4467.3410 & -0.423 & & &  & & & & & & \\
\ion{Sm}{ii} & 4467.3418 & -3.863 & & &  & & & & & & \\
\ion{Sm}{ii} & 4467.3420 & -0.493 & & &  & & & & & & \\
\ion{Sm}{ii} & 4467.3426 & -3.901 & & &  & & & & & & \\
\ion{Sm}{ii} & 4467.3434 & -1.545 & & &  & & & & & & \\
\ion{Sm}{ii} & 4467.3441 & -1.581 & & &  & & & & & & \\
\ion{Sm}{ii} & 4467.3469 & -2.480 & & &  & & & & & & \\
\ion{Sm}{ii} & 4467.3471 & -2.445 &  & &  & & & & & & \\
\ion{Sm}{ii} & 4467.3500 & -4.282 & & &  & & & & & & \\
\ion{Sm}{ii} & 4467.3514 & -4.248 & & &  & & & & & & \\
\ion{Sm}{ii} & 4467.3518 & -1.507 & & &  & & & & & & \\
\ion{Sm}{ii} & 4467.3529 & -1.471 & & &  & & & & & & \\
\ion{Sm}{ii} & 4467.3549 & -2.703 & & &  & & & & & & \\
\ion{Sm}{ii} & 4467.3572 & -2.668 & & &  & & & & & & \\
\ion{Sm}{ii} & 4467.3603 & -1.437 & & &  & & & & & & \\
\ion{Sm}{ii} & 4467.3636 & -1.401 & & &  & & & & & & \\
\ion{Fe}{i} & 4467.4254 & -2.655 & & &  & & & & & & \\
\ion{Co}{i} & 4467.4773 & -1.412 & & &  & & & & & & \\
\hline
\end{tabular}
\end{center}
\end{table*}

\begin{table*}[ht]\tiny
\begin{center}
\caption{Linelists for the synthesis of \ion{Eu}{ii}. The lists include the hfs of the \ion{Eu}{ii} in question taken from \cite{Lawler2001} as well as additional lines nearby. Columns represent element line, wavelength and log(gf) values, respectively.\label{tab:hfs1}}
\setlength{\tabcolsep}{1 mm}
\begin{tabular}{c c c}
\hline
\multicolumn{3}{c} {\ion{Eu}{ii}  4129  } \\
Element & $\lambda$ ($\AA$) & log(gf)  \\
\hline
\ion{CN}{i} & 4129.6006 & -0.585  \\
\ion{Eu}{ii} & 4129.5966 & -1.833  \\
\ion{Eu}{ii} & 4129.6001 & -1.356 \\
\ion{Eu}{ii} & 4129.6137 & -1.637  \\
\ion{Eu}{ii} & 4129.6185 & -1.298  \\
\ion{Eu}{ii} & 4129.6220 & -1.833  \\
\ion{Eu}{ii} & 4129.6387 & -1.578  \\
\ion{Ti}{i}   & 4129.6429 & -1.424 \\
\ion{Eu}{ii} & 4129.6444 & -1.167 \\ 
\ion{Eu}{ii} & 4129.6492 & -1.637  \\
\ion{Eu}{ii} & 4129.6716 & -1.614 \\
\ion{Eu}{ii} & 4129.6774 & -1.795  \\ 
\ion{Eu}{ii} & 4129.6781 & -1.017  \\
\ion{Eu}{ii} & 4129.6801 & -1.318  \\
\ion{Eu}{ii} & 4129.6838 & -1.598  \\
\ion{Eu}{ii} & 4129.6839 & -1.578  \\
\ion{Eu}{ii} & 4129.6871 & -1.260  \\
\ion{Eu}{ii} & 4129.6898 & -1.795 \\ 
\ion{Eu}{ii} & 4129.6941 & -1.539 \\
\ion{Eu}{ii} & 4129.6977 & -1.129 \\
\ion{Eu}{ii} & 4129.7007 & -1.598\\
\ion{Eu}{ii} & 4129.7117 & -0.979  \\
\ion{Eu}{ii} & 4129.7130 & -1.801  \\
\ion{Eu}{ii} & 4129.7150 & -1.539  \\
\ion{Eu}{ii} & 4129.7198 & -0.866  \\
\ion{Eu}{ii} & 4129.7240 & -1.576 \\
\ion{Eu}{ii} & 4129.7263 & -1.614  \\
\ion{Eu}{ii} & 4129.7295 & -1.763  \\
\ion{Eu}{ii} & 4129.7305 & -0.828  \\
\ion{Eu}{ii} & 4129.7331 & -1.576  \\
\ion{Eu}{ii} & 4129.7548 & -0.683  \\
\ion{Eu}{ii} & 4129.7558 & -1.763  \\
\ion{Eu}{ii} & 4129.7769 & -1.801 \\
\ion{Eu}{ii} & 4129.7700 & -0.721 \\
\ion{Nd}{ii} & 4129.8370 & 0.180 \\
\ion{Co}{i} & 4129.8421 & -1.587 \\
\hline
\end{tabular}
\end{center}
\end{table*}

\section{Solar abundance plots}
\label{sec:solarabundanceplots}

\begin{figure*}[ht]
\centering
\resizebox{\hsize}{!}{
\includegraphics[bb=0 50 792 612,clip]{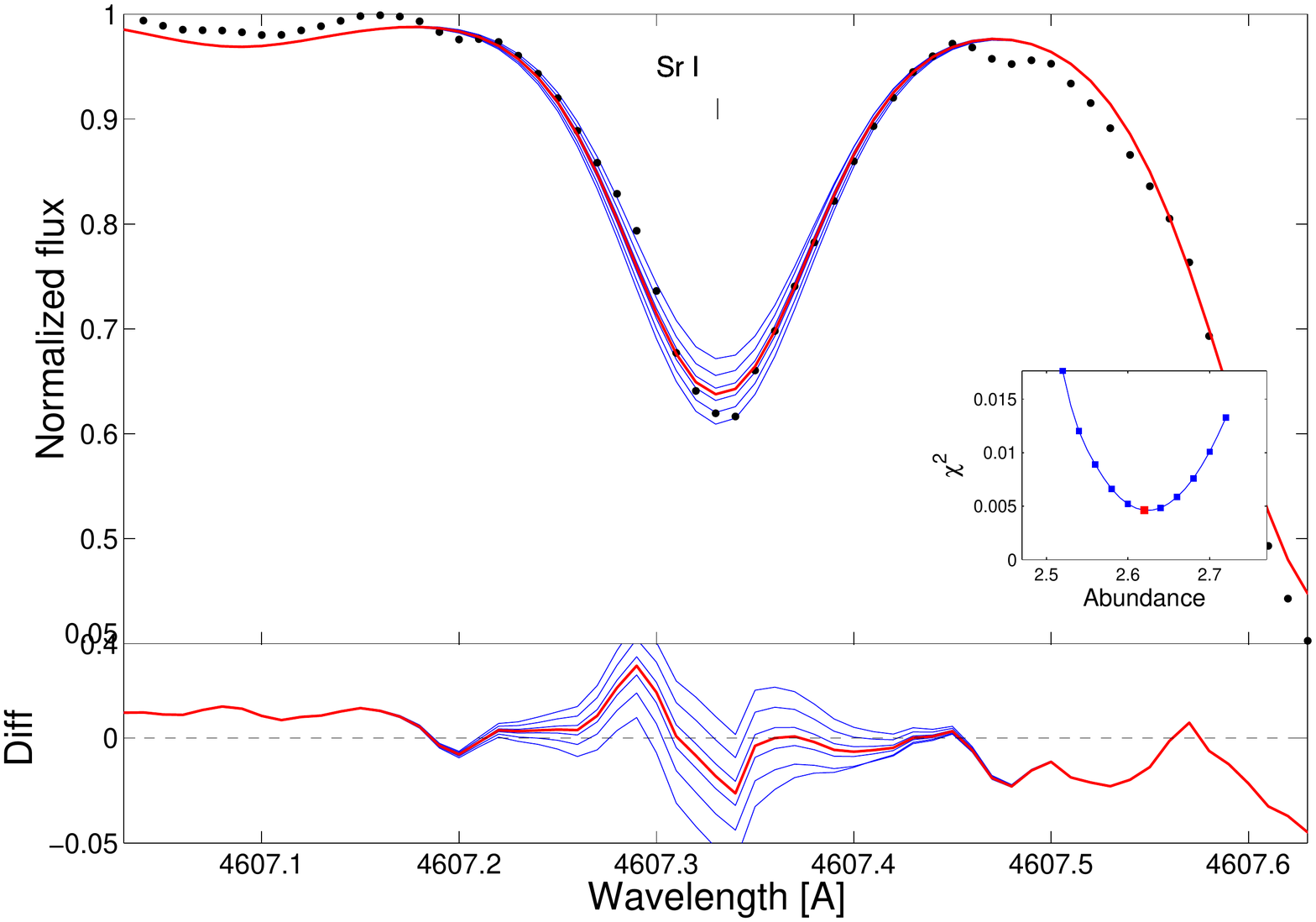} 
\includegraphics[bb=0 50 792 612,clip]{Zr4208abhipVesta.pdf} }
\resizebox{\hsize}{!}{
\includegraphics[bb=0 50 792 612,clip]{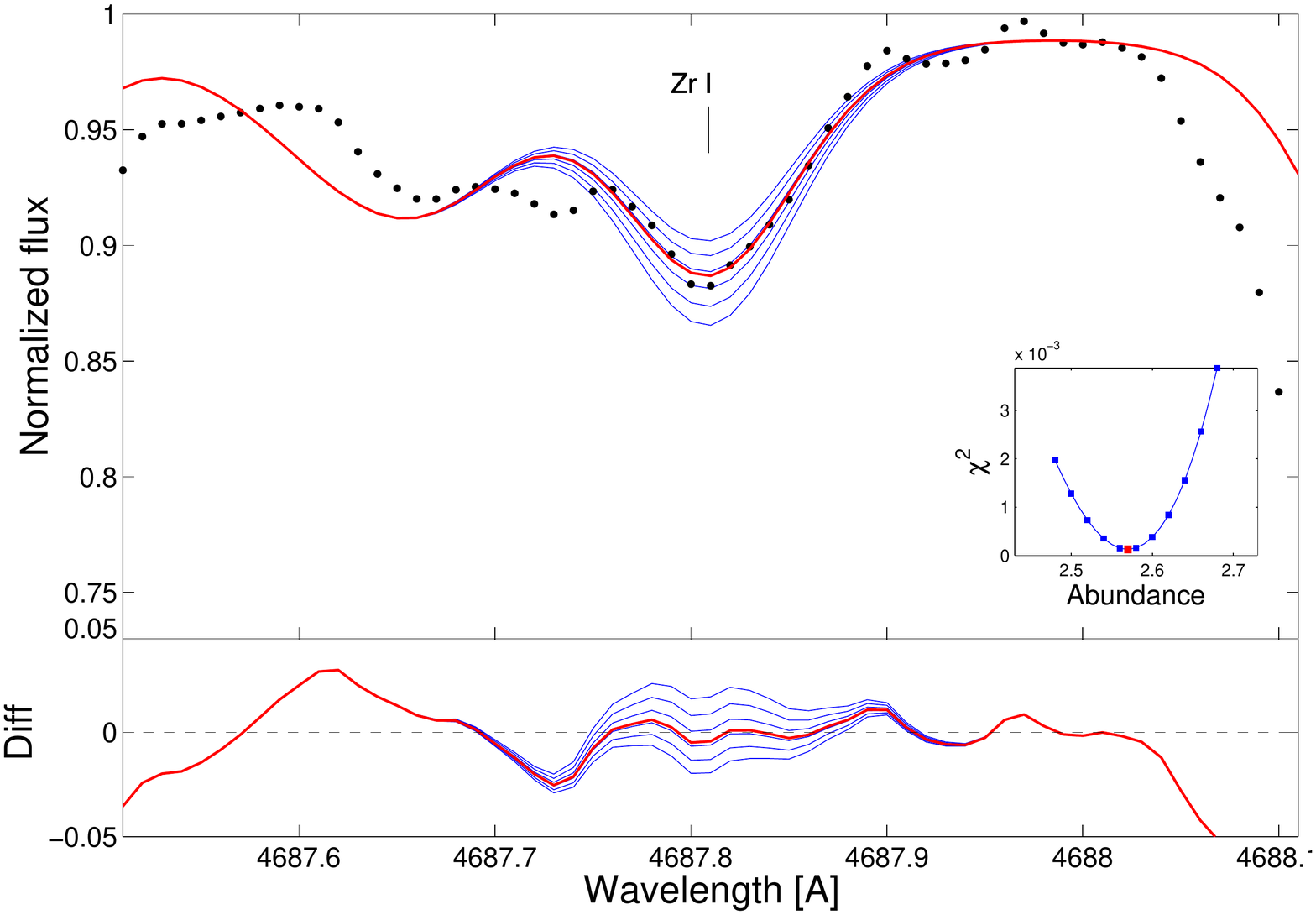} 
\includegraphics[bb=0 50 792 612,clip]{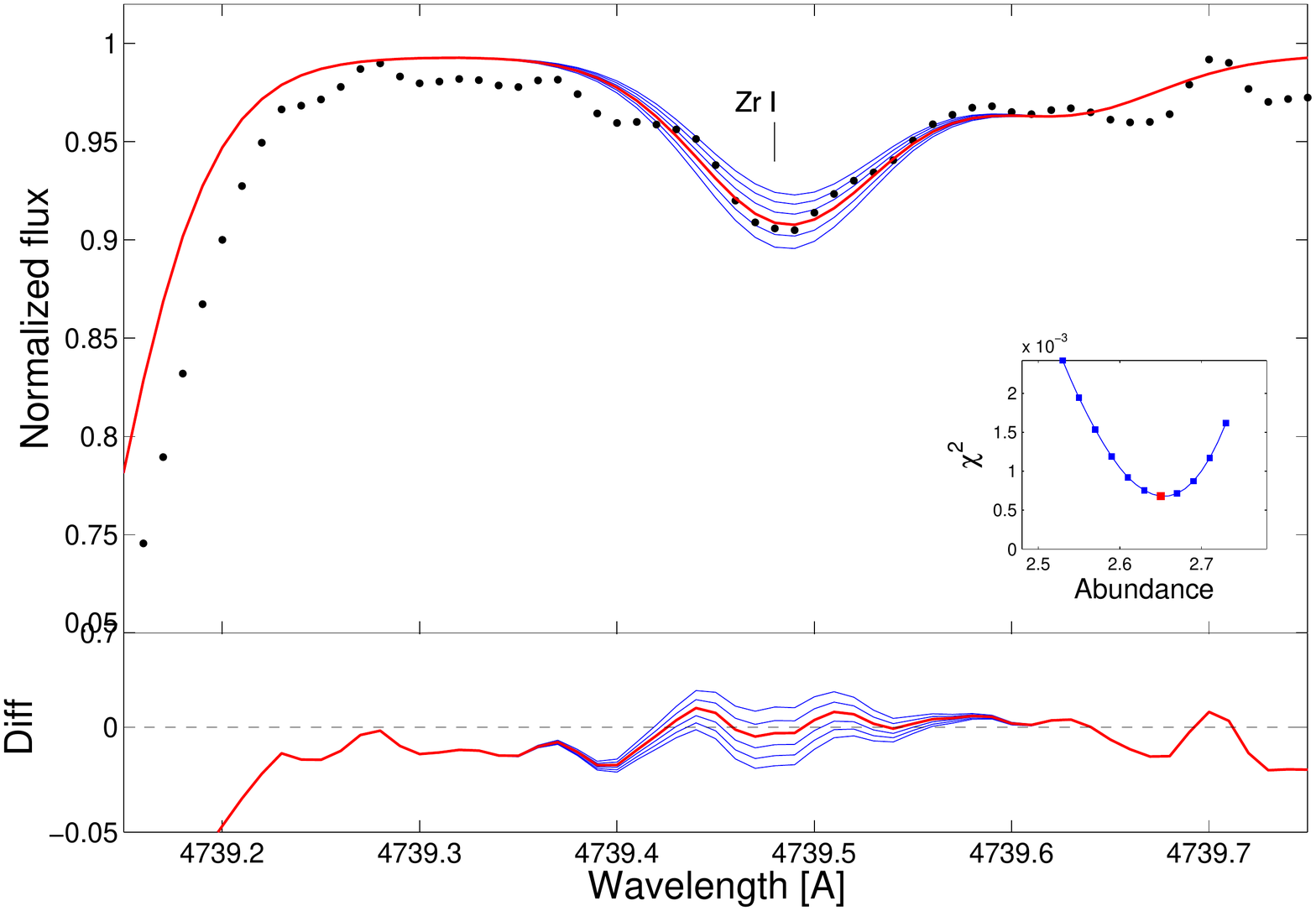} }
\resizebox{\hsize}{!}{
\includegraphics[bb=-396 50 1188 612,clip]{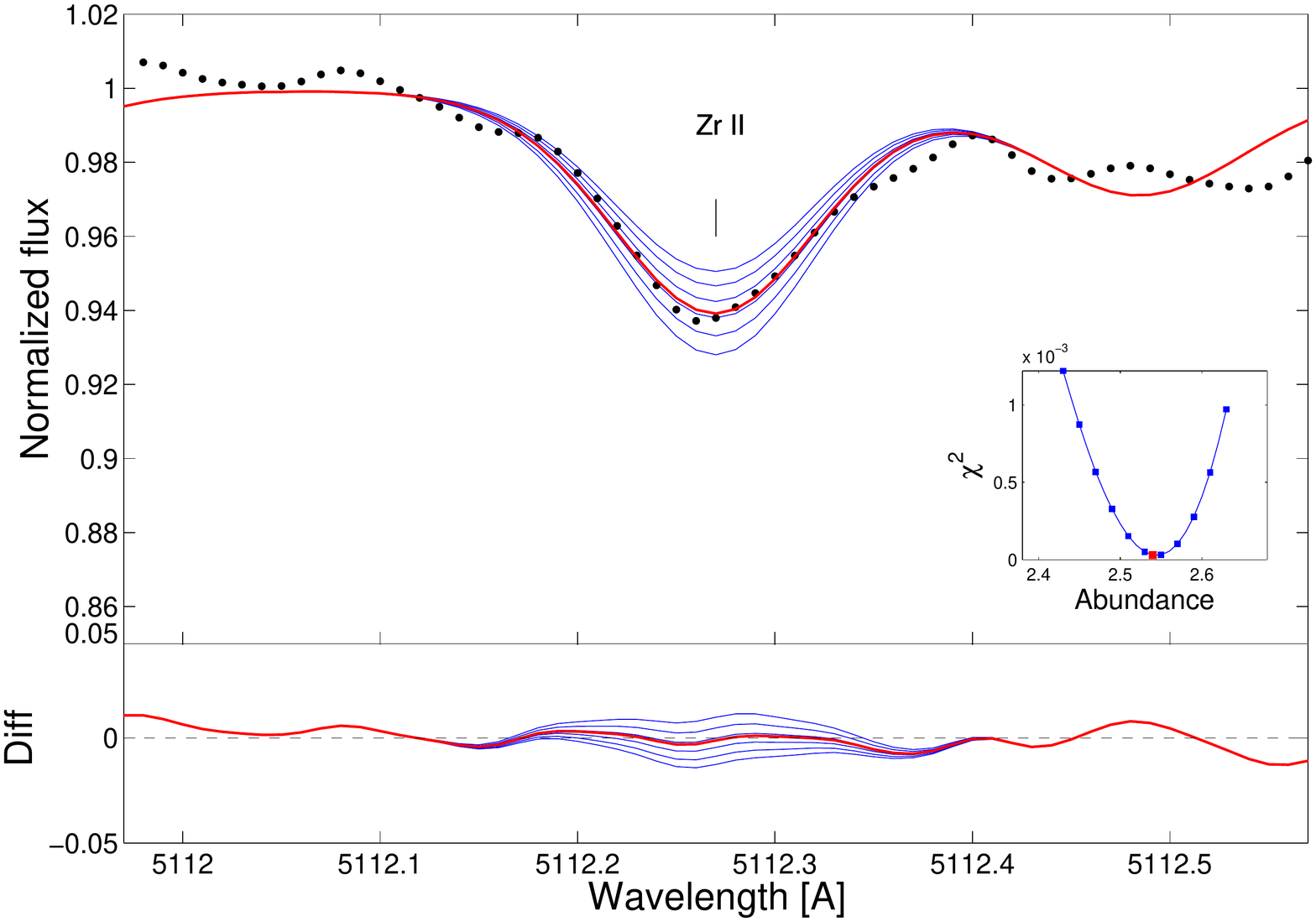}}
\caption{Solar spectrum taken using asteroid Vesta during run at Magellan in January 2006. Spectral lines are listed in order of element and wavelength. Chemical elements with all the hfs lines are indicated inside the lines. The different colored lines are the different synthetic spectra with different abundances in steps of 0.04\,dex, while the dots are the observed spectra. In the lower panels are the values of differences between the real and synthetic spectra for the synthetic spectra plotted above. The red lines represent the best fit derived from unnormalized $\chi^{2}$,   visible in the small plot as a red dot. Here the fits for the Sr line and the five Zr lines are shown.\label{fig:solar_spectrum1} }
\end{figure*}

\begin{figure*}[!ht]
\resizebox{\hsize}{!}{
\includegraphics[bb=0 50 792 612,clip]{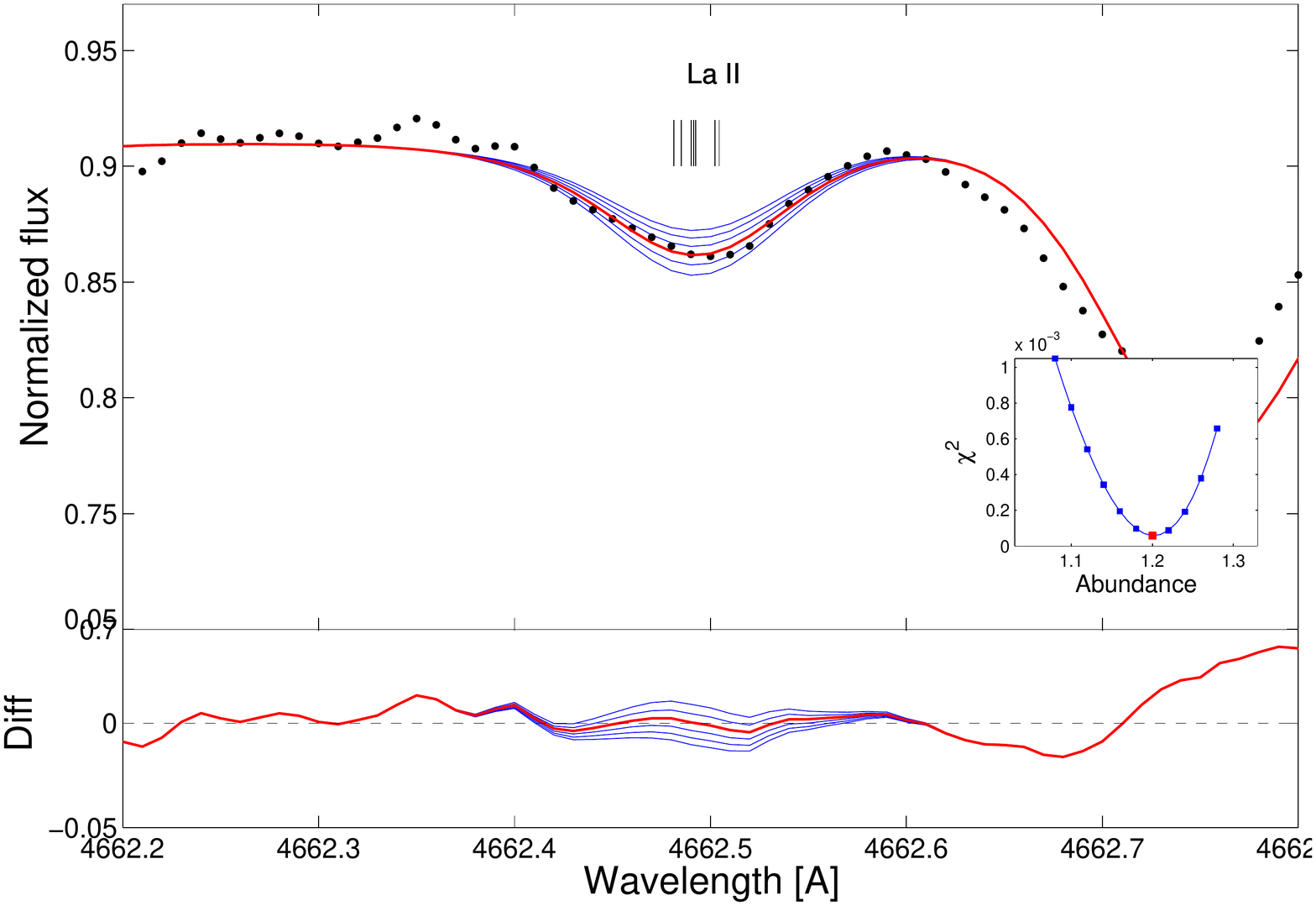}
\includegraphics[bb=0 50 792 612,clip]{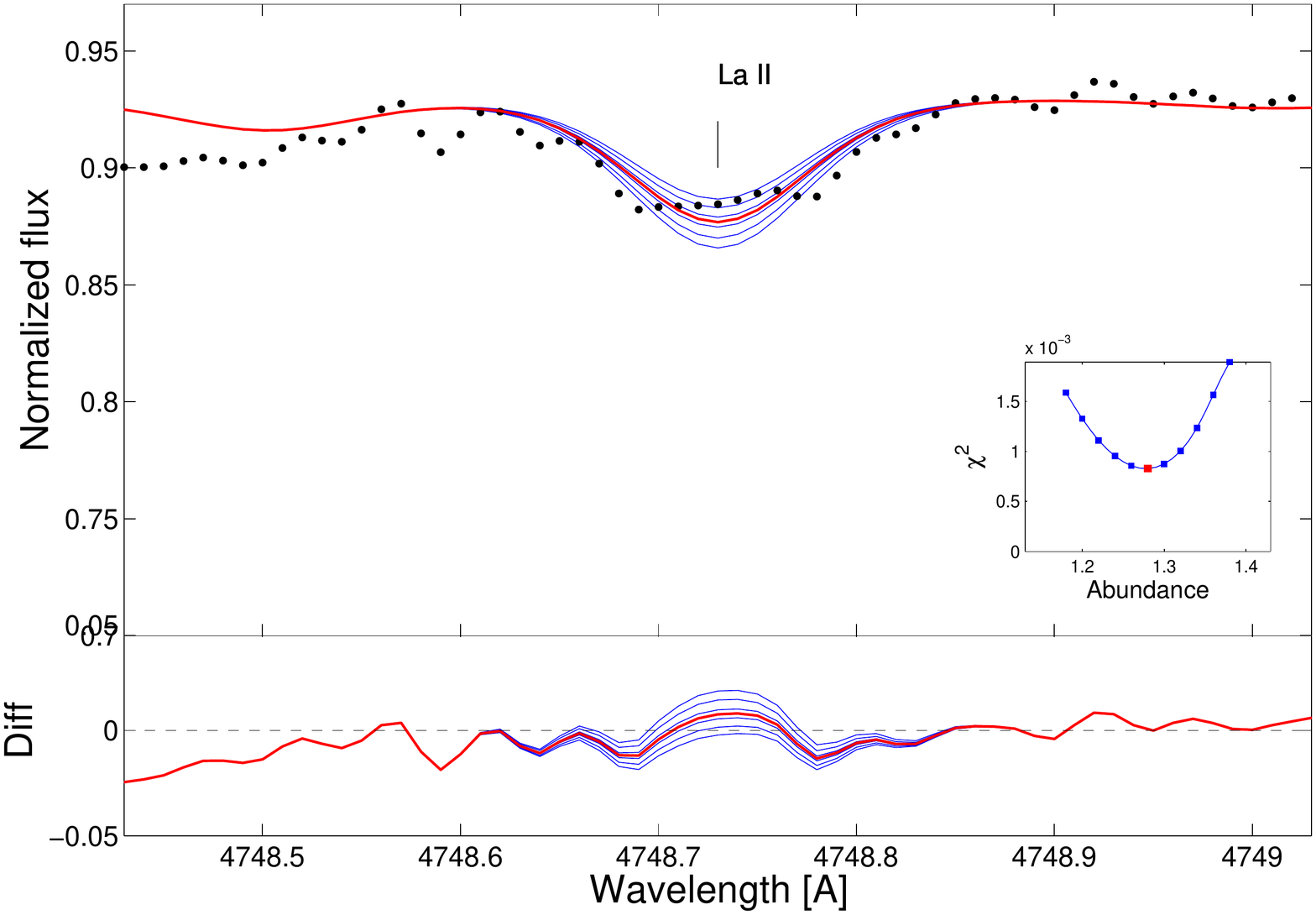}}
\resizebox{\hsize}{!}{
\includegraphics[bb=0 50 792 612,clip]{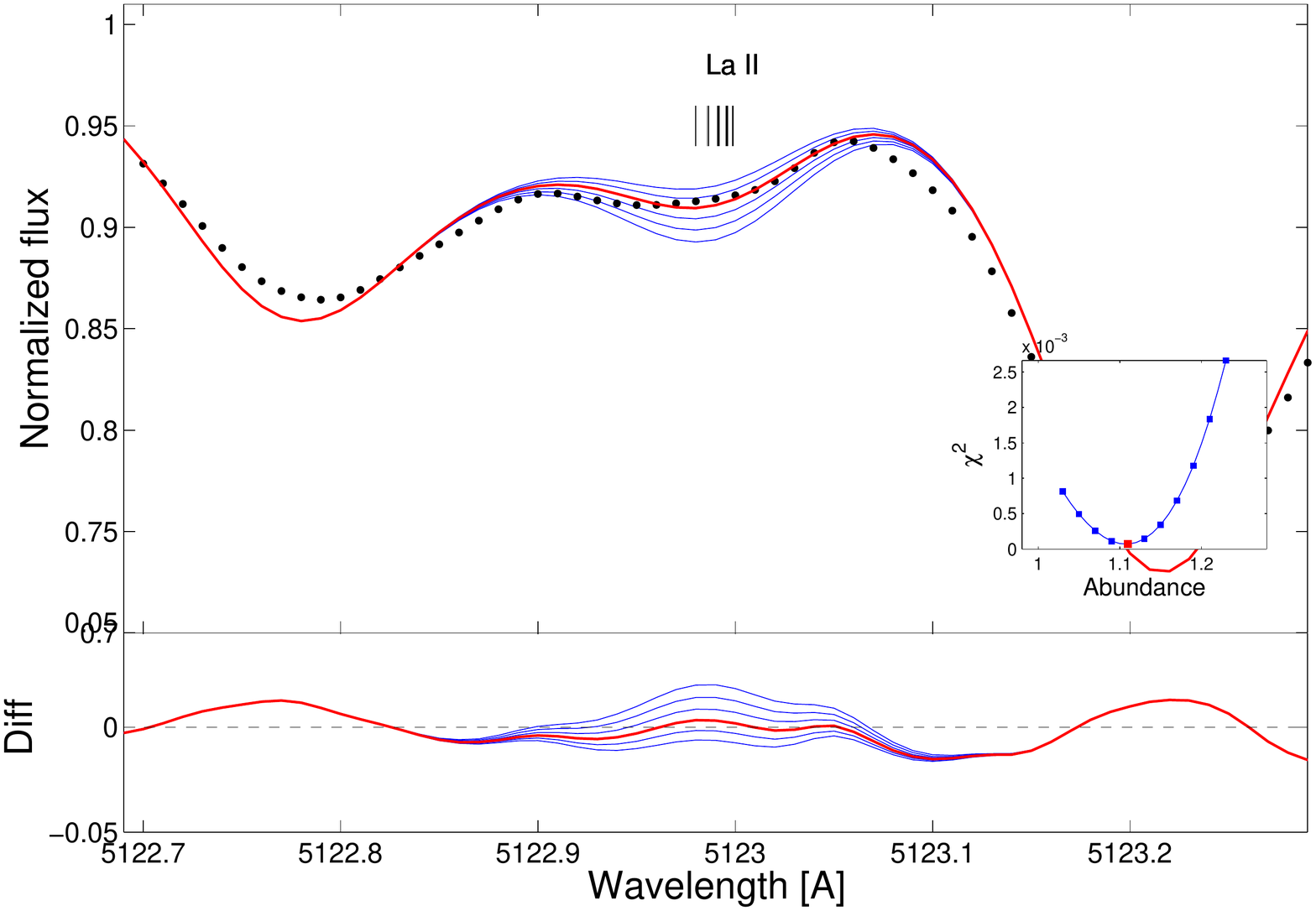}
\includegraphics[bb=0 50 792 612,clip]{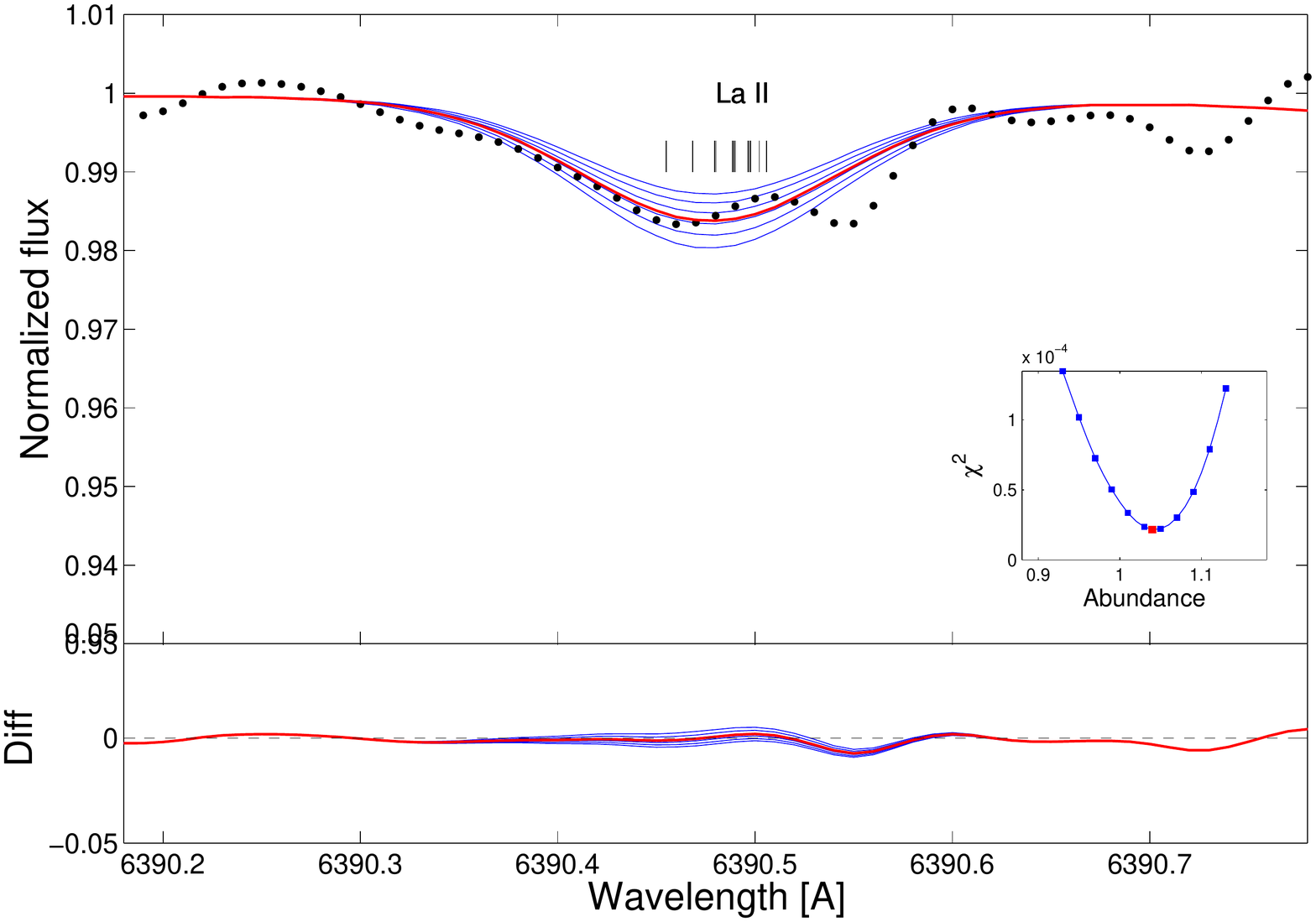}}
\caption{As in Fig.~\ref{fig:solar_spectrum1} but for the four La lines. \label{fig:solar_spectrum2}}
\end{figure*}

\begin{figure*}[!ht]
\resizebox{\hsize}{!}{
\includegraphics[bb=0 50 792 612,clip]{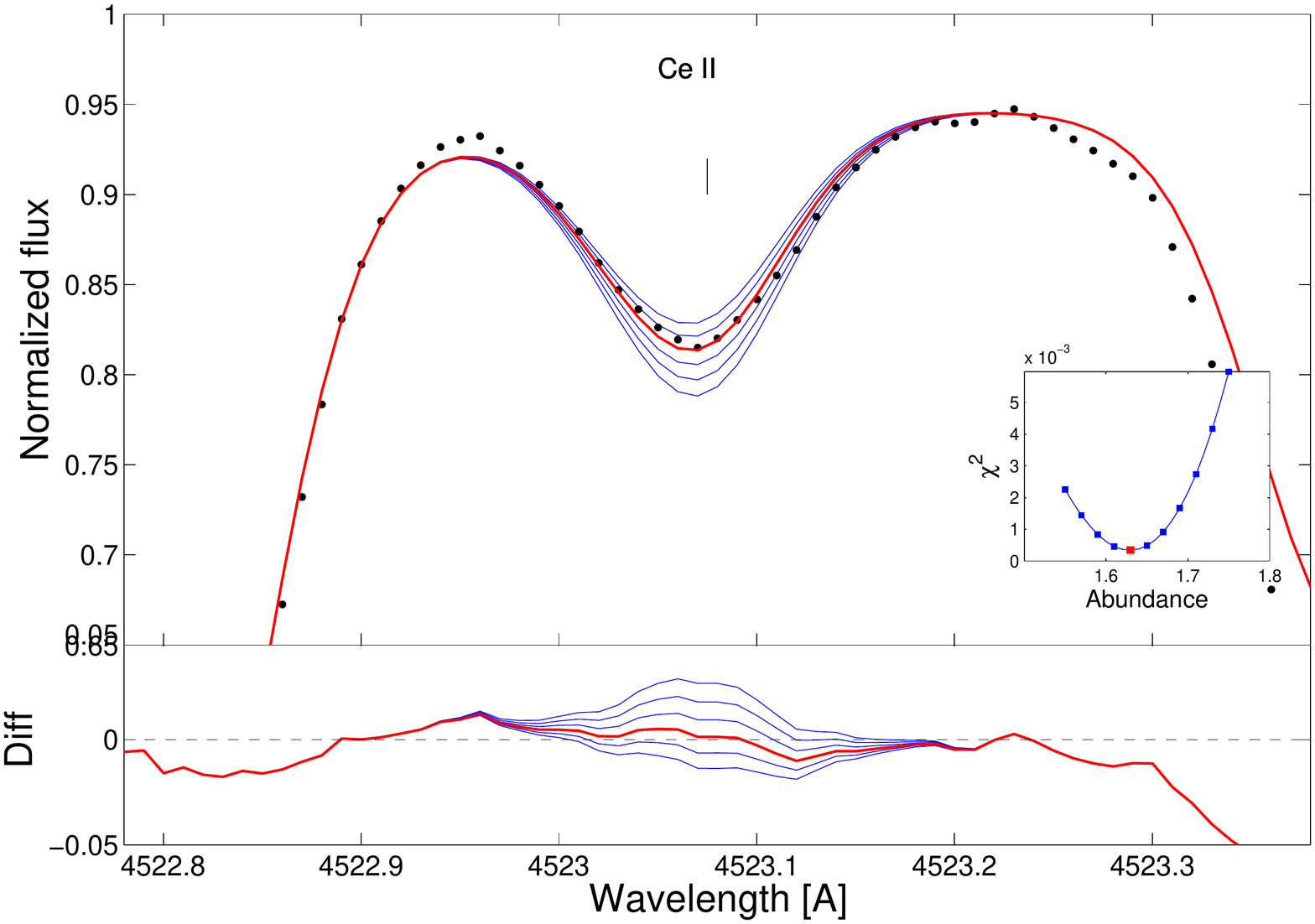}
\includegraphics[bb=0 50 792 612,clip]{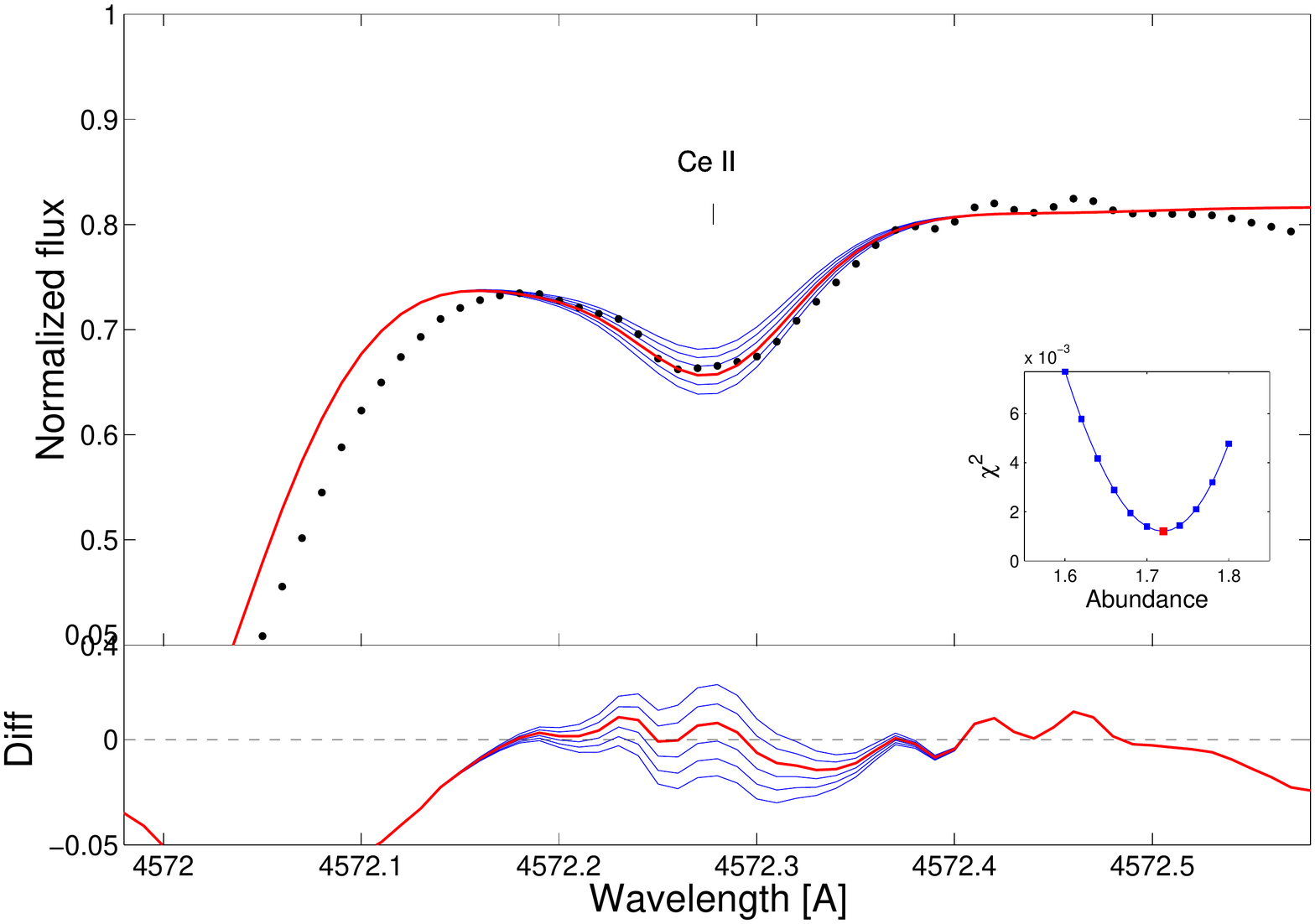}}
\resizebox{\hsize}{!}{
\includegraphics[bb=0 50 792 612,clip]{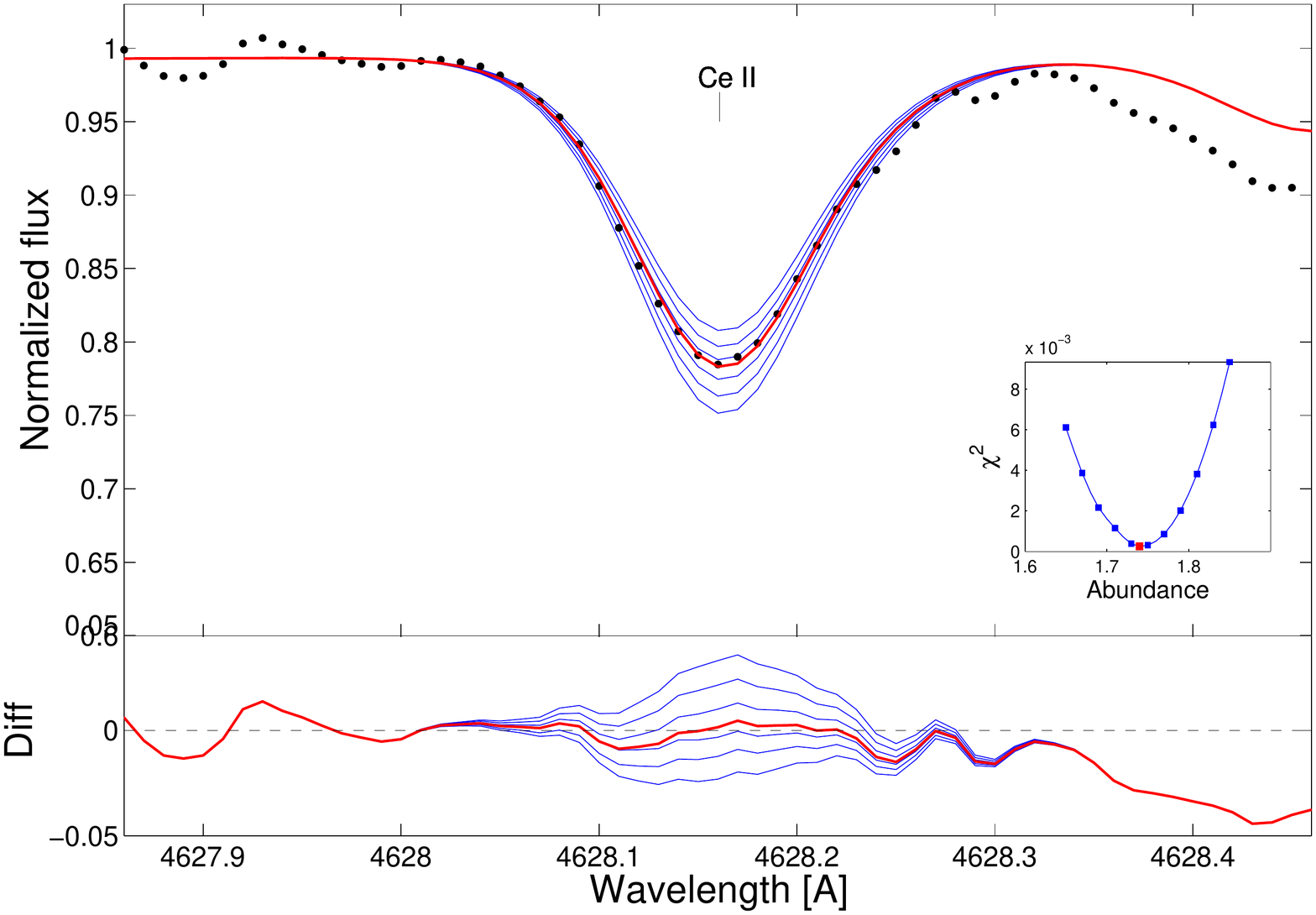}
\includegraphics[bb=0 50 792 612,clip]{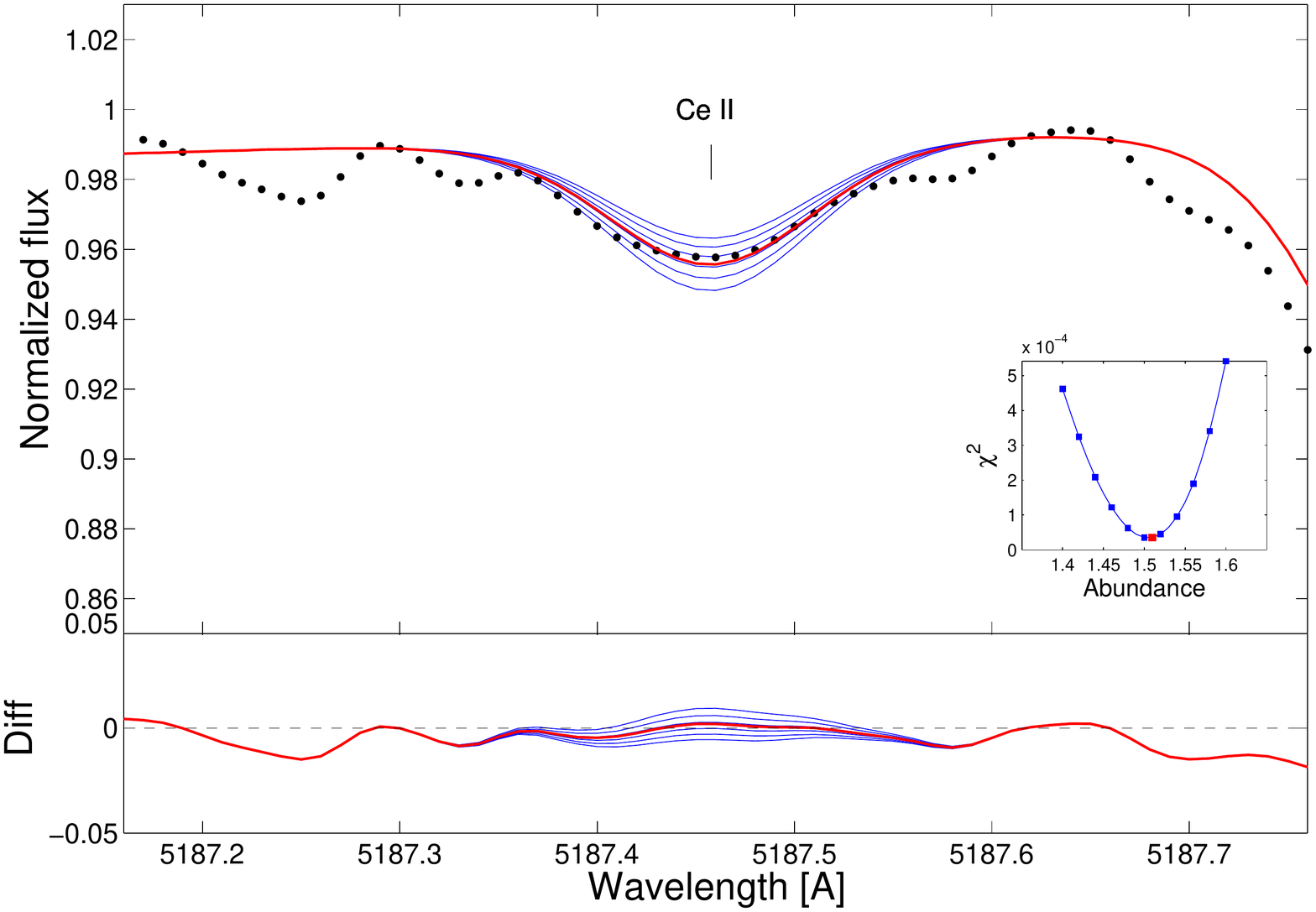}}
\caption{As in Fig.~\ref{fig:solar_spectrum1} but for the four Ce lines. \label{fig:solar_spectrum3}}
\end{figure*}

\begin{figure*}[!ht]
\resizebox{\hsize}{!}{
\includegraphics[bb=0 50 792 612,clip]{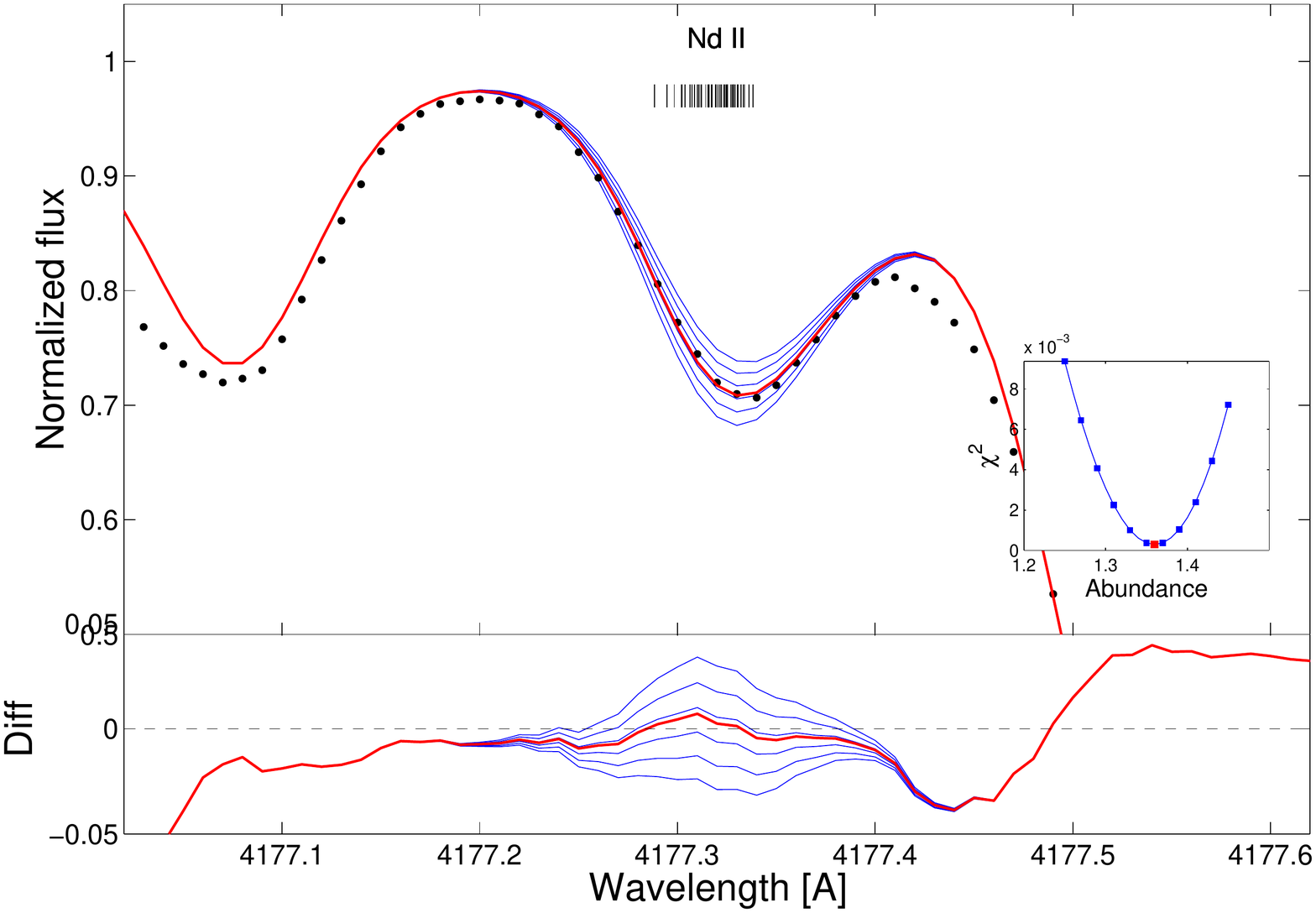}
\includegraphics[bb=0 50 792 612,clip]{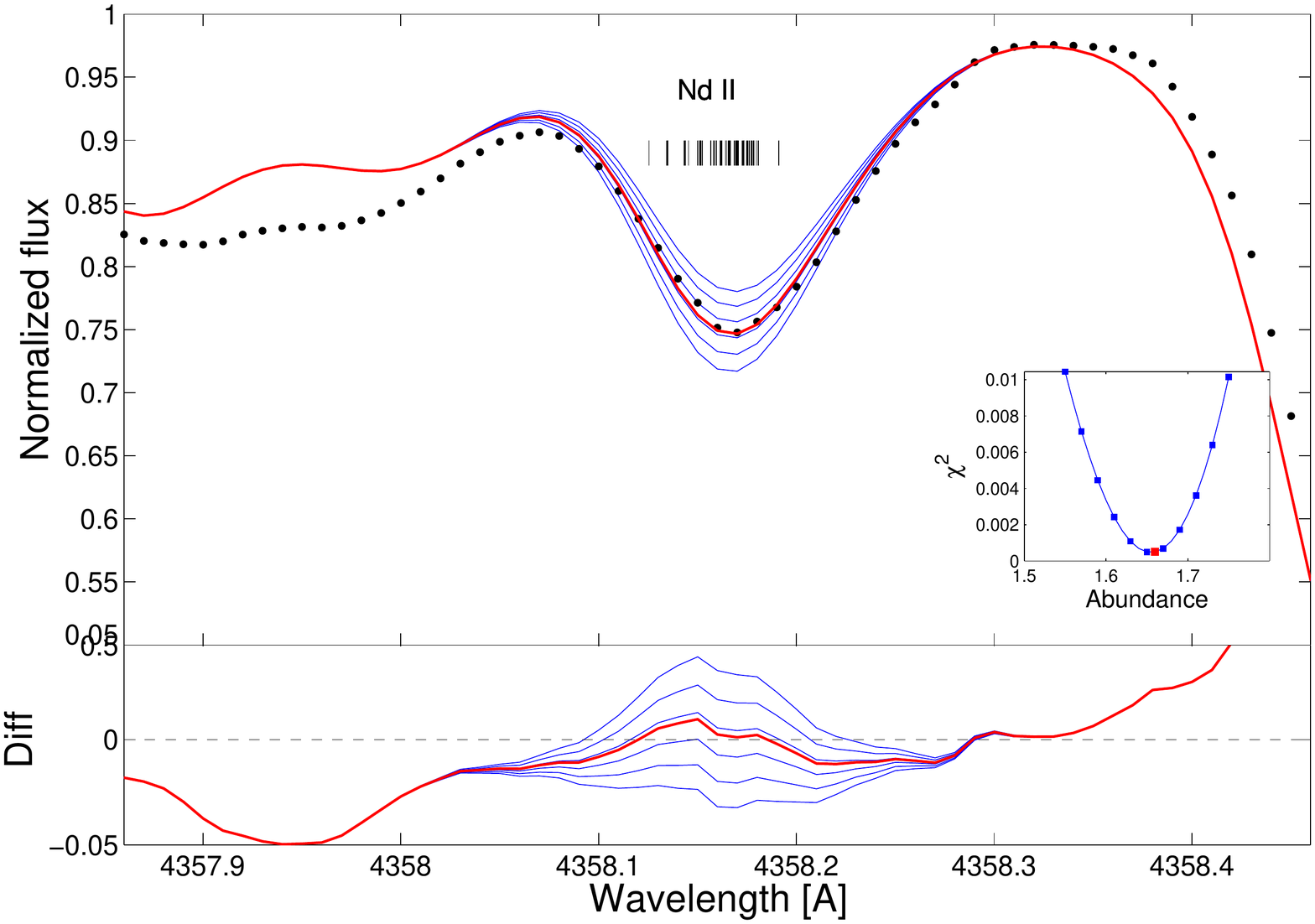}}
\resizebox{\hsize}{!}{
\includegraphics[bb=0 50 792 612,clip]{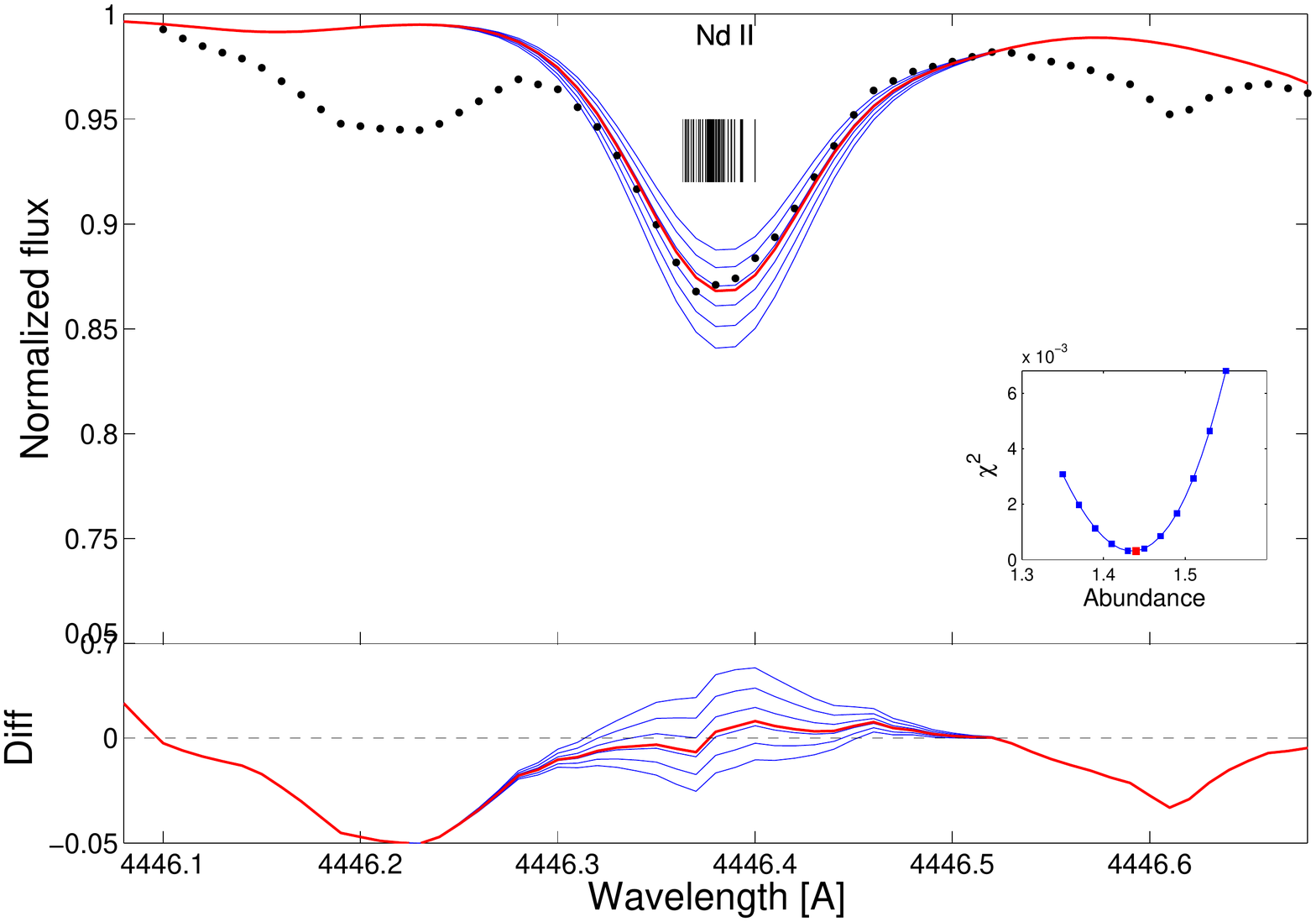}
\includegraphics[bb=0 50 792 612,clip]{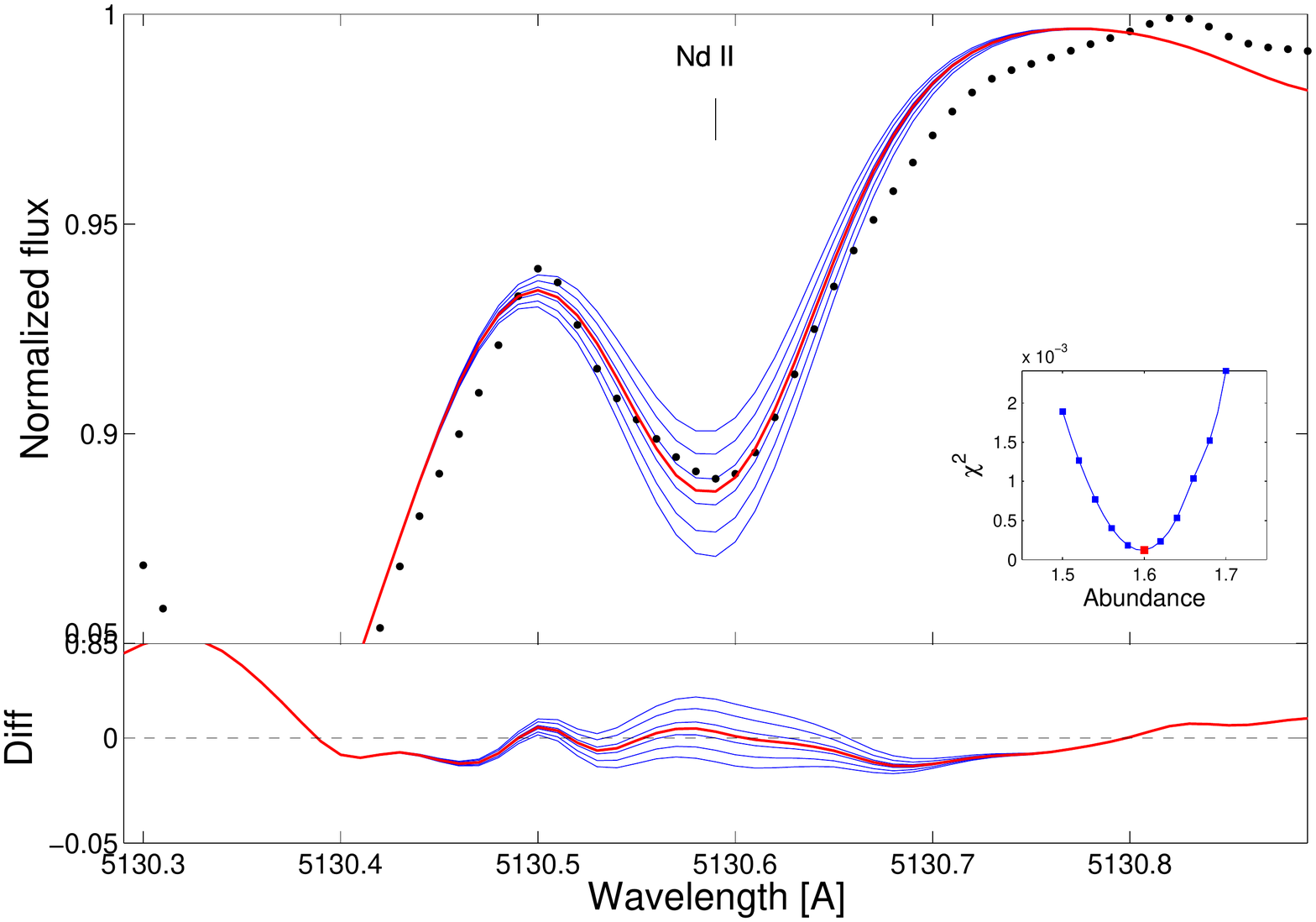}}
\resizebox{\hsize}{!}{
\includegraphics[bb=-396 50 1188 612,clip]{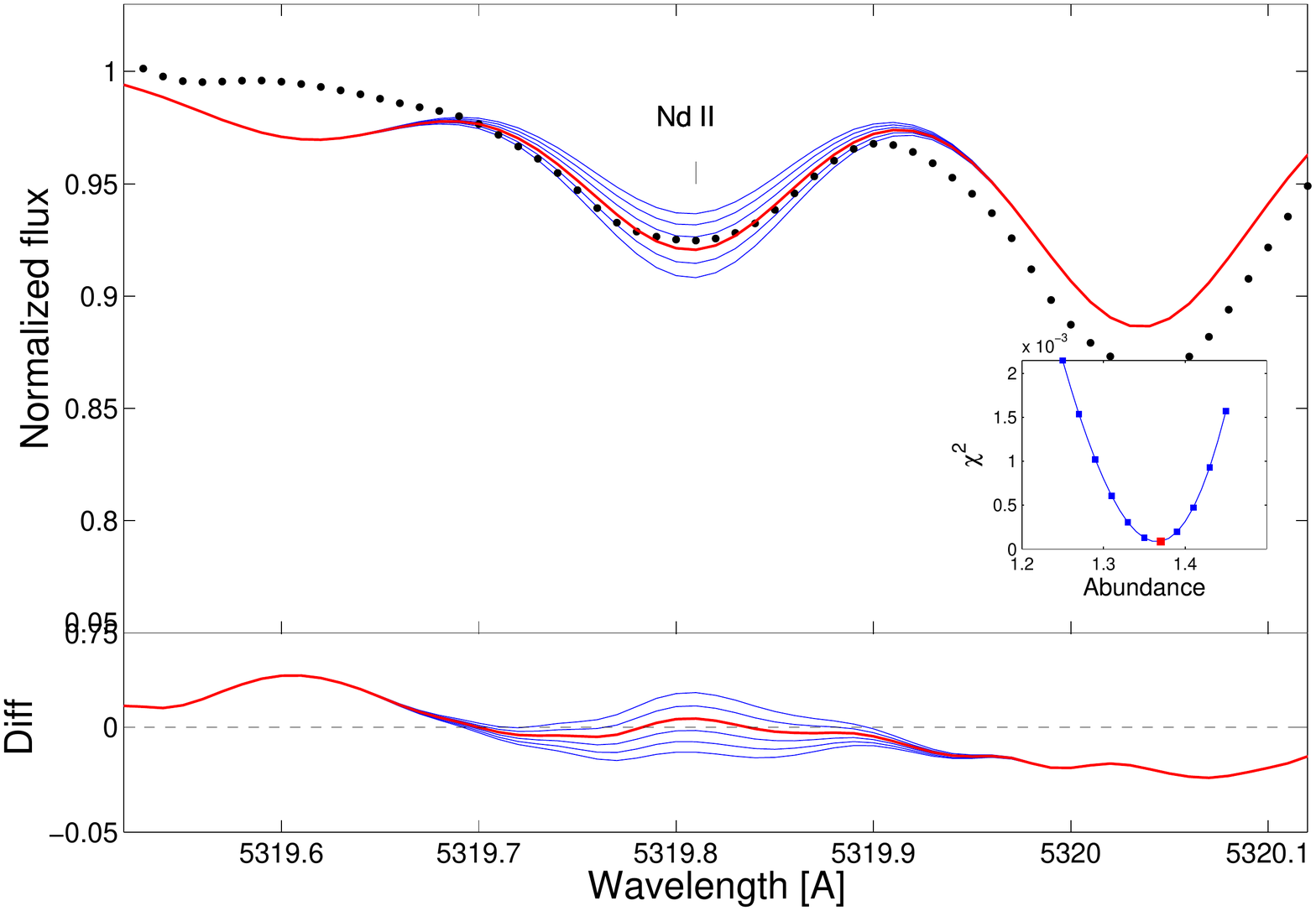}}
\caption{As in Fig.~\ref{fig:solar_spectrum1} but for the five Nd lines. \label{fig:solar_spectrum4}}
\end{figure*}

\begin{figure*}[!ht]
\resizebox{\hsize}{!}{
\includegraphics[bb=0 50 792 612,clip]{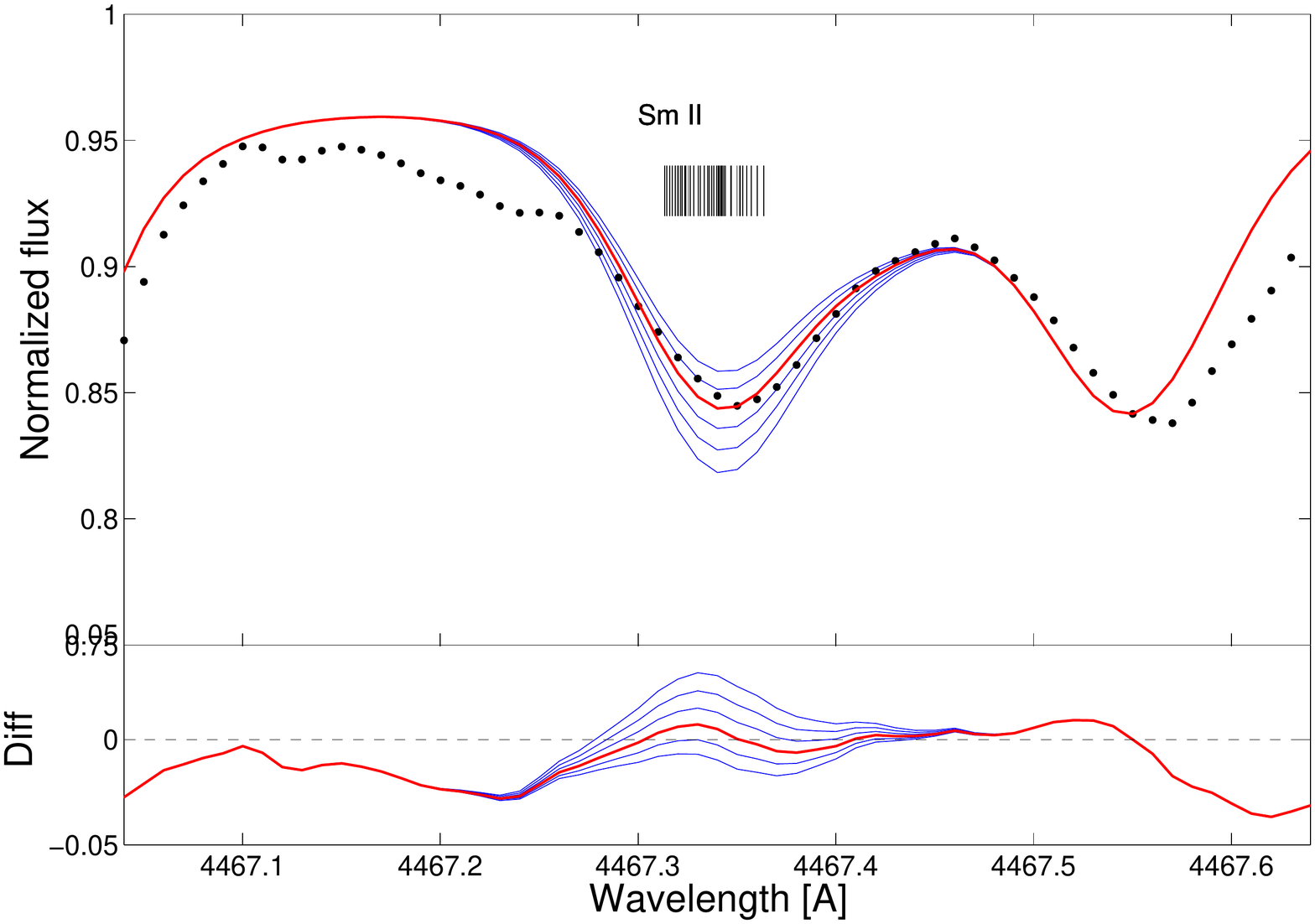}
\includegraphics[bb=0 50 792 612,clip]{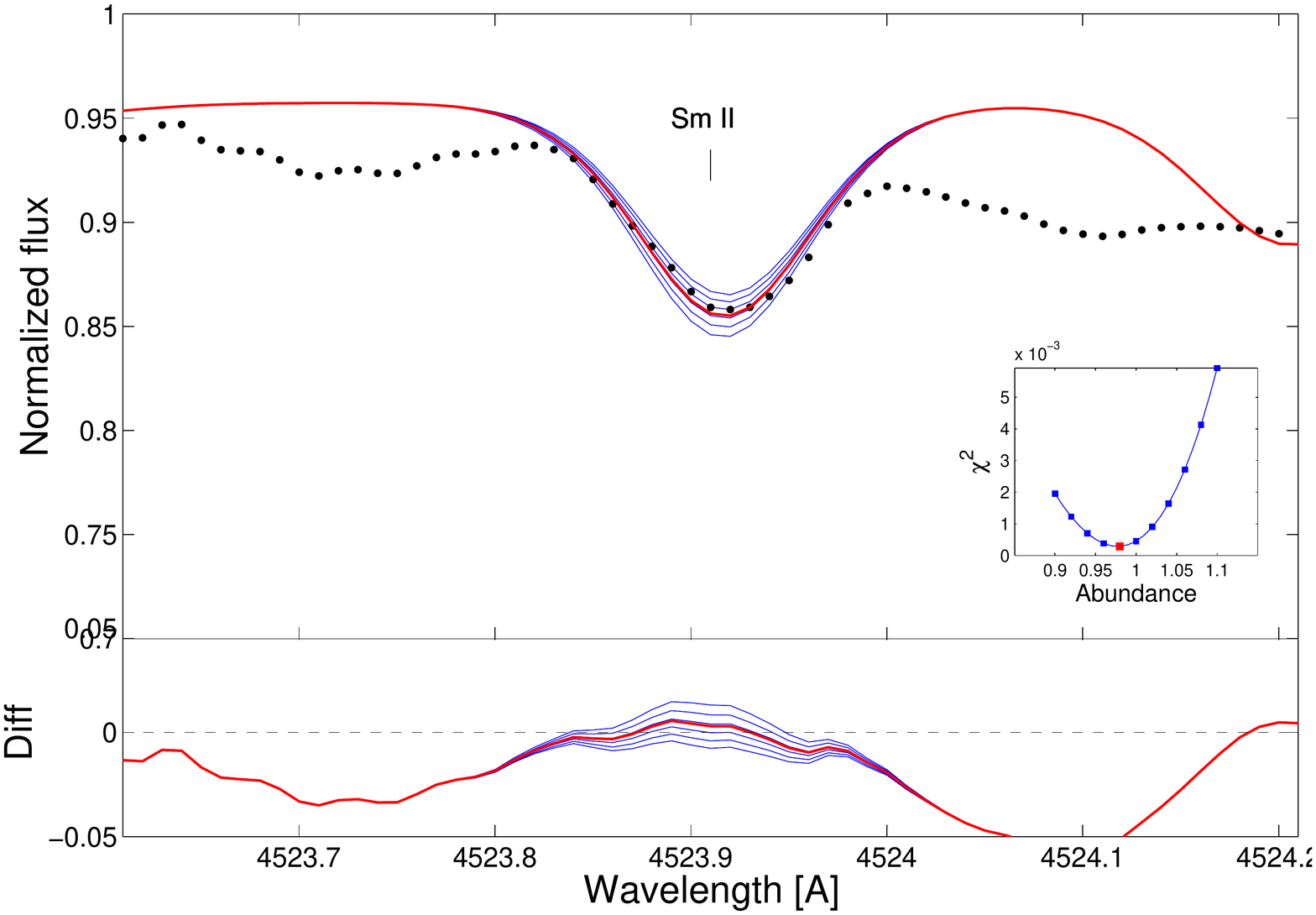}}
\resizebox{\hsize}{!}{
\includegraphics[bb=0 50 792 612,clip]{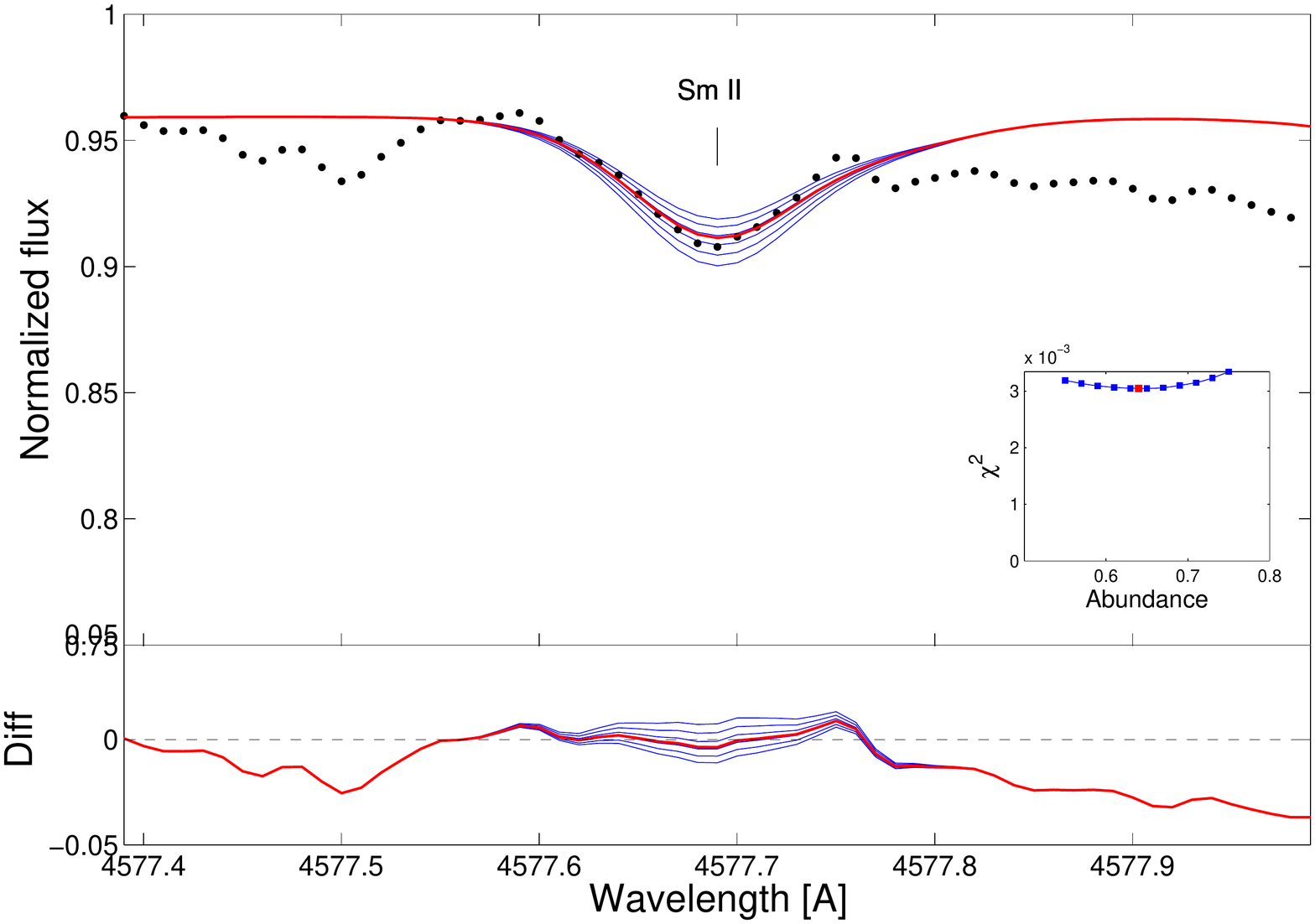}
\includegraphics[bb=0 50 792 612,clip]{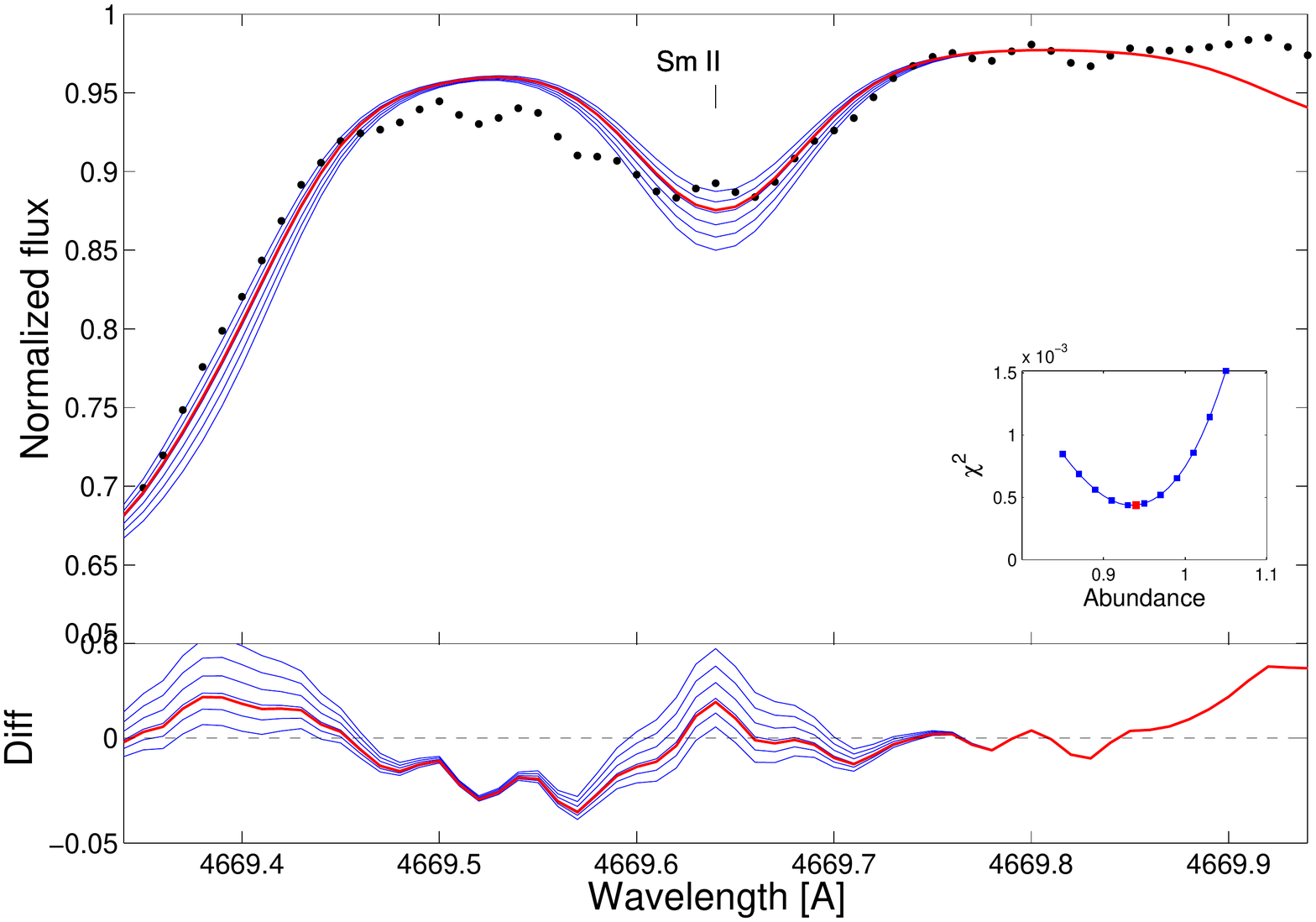}}
\resizebox{\hsize}{!}{
\includegraphics[bb=-396 50 1188 612,clip]{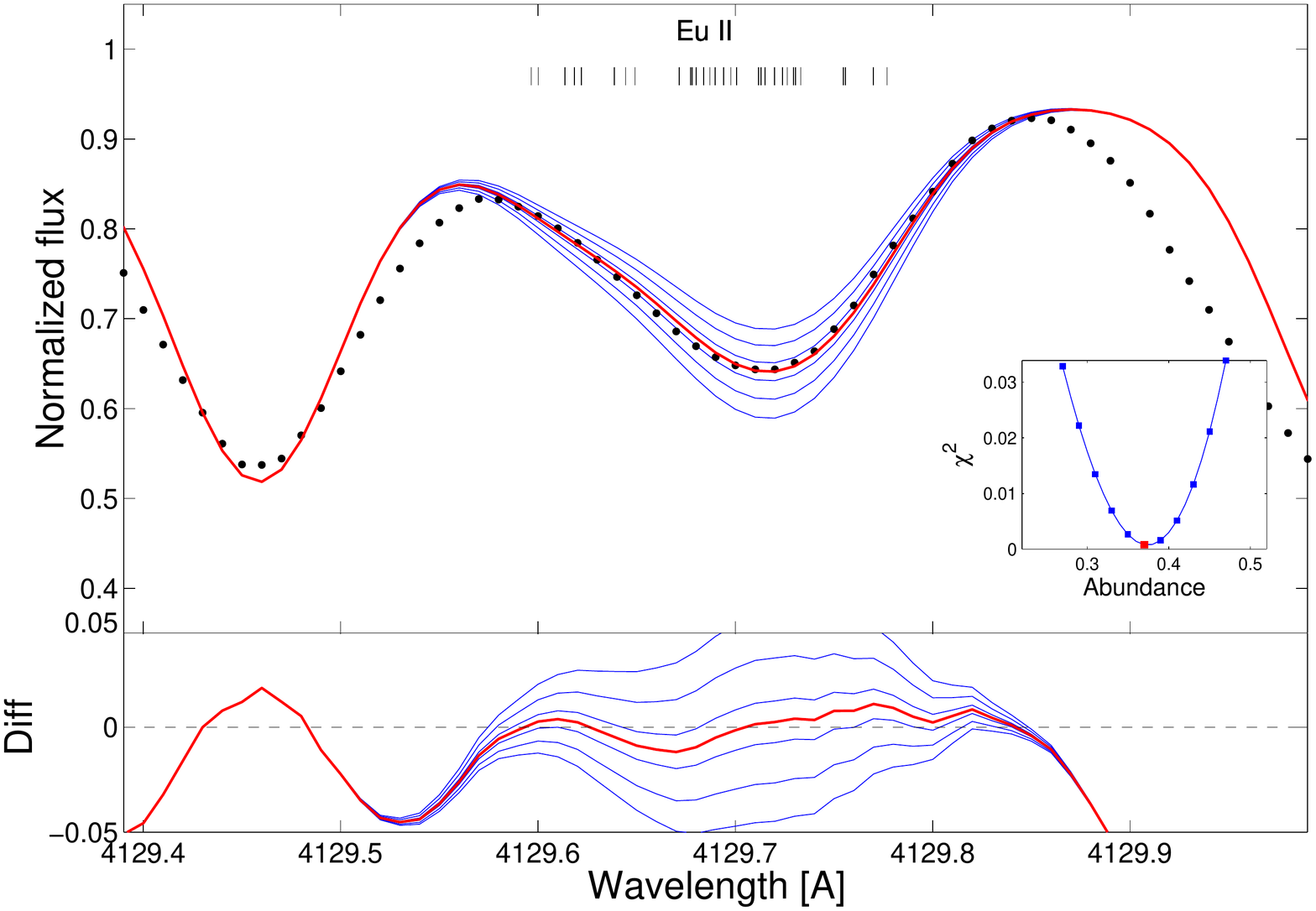}}
\caption{As in Fig.~\ref{fig:solar_spectrum1} but for the four Sm lines and the two Eu lines. \label{fig:solar_spectrum5}}
\end{figure*}

In Figs.~\ref{fig:solar_spectrum1}--\ref{fig:solar_spectrum5}  the synthesis for all the lines analyzed in this work for  the solar spectrum from Vesta observed at Magellan in January 2006 is shown. Each plot shows (in the bigger panel) the line fitting with the best fit  to the observed one  and the different abundances in steps of 0.04\,dex. In the lower panel the differences between the observed spectrum and the synthetic one are shown, with the difference with the best fit highlighted. In the small square panel  the $\chi^{2}$ values for the different abundances are represented, with the red dot representing the minimum and therefore the best fit value for the elemental abundance.

\end{appendix}

\end{document}